\documentclass[useAMS,usenatbib,usegraphicx,usefloat]{mn2e}
\usepackage{textcomp}
\usepackage{amssymb}

\newcommand{\aap}{A\&A}          
\newcommand{\aj}{AJ}             
\newcommand{\apj}{ApJ}           
\newcommand{\apjs}{ApJS}         
\newcommand{\apjl}{ApJL}         
\newcommand{\mnras}{MNRAS}       
\newcommand{\aaps}{A\&AS}        
\newcommand{\pasp}{PASP}         
\newcommand{\araa}{ARA\&A}       

\title[A unified picture of breaks and truncations]{A unified picture of breaks and truncations in spiral galaxies from SDSS and S$^4$G imaging}
\author[Mart\'in-Navarro et al.]
{Ignacio Mart\'in-Navarro$^{1,2}$\thanks{E-mail:imartin@iac.es}, Judit Bakos$^{1,2}$, Ignacio Trujillo$^{1,2}$, Johan H. Knapen$^{1,2}$,
\newauthor 
E. Athanassoula$^{3}$, Albert Bosma$^{3}$, S\'ebastien Comer\'on$^{4}$, Bruce G. Elmegreen$^{5}$, 
\newauthor
Santiago Erroz-Ferrer$^{1,2}$, Dimitri A. Gadotti$^{6}$, Armando Gil de Paz$^{7}$,
\newauthor
Joannah L. Hinz$^{8}$, Luis C. Ho$^{9}$, Benne W. Holwerda$^{10}$, Taehyun Kim$^{11,6,12}$,
\newauthor
Jarkko Laine$^{4}$, Eija Laurikainen$^{4}$, Kar\'in Men\'endez-Delmestre$^{9}$,
\newauthor
Trisha Mizusawa$^{11,13,14}$, Juan-Carlos Mu\~noz-Mateos$^{11}$, Michael W. Regan$^{15}$,
\newauthor
Heikki Salo$^{4}$, Mark Seibert$^{9}$ and Kartik Sheth$^{11,13,14}$
\\
$^{1}${Instituto de Astrof\'isica de Canarias, E-38200 La Laguna, Tenerife, Spain}\\
$^{2}${Departamento de Astrof\'isica, Universidad de La Laguna, E-38205 La Laguna, Tenerife, Spain}\\
$^{3}${Laboratoire d'Astrophysique de Marseille (LAM), Marseille, France}\\
$^{4}${Division of Astronomy, Department of Physical Sciences, University of Oulu, Oulu, FIN-90014, Finland}\\
$^{5}${IBM T. J. Watson Research Center, P.O. Box 218, Yorktown Heights, NY 10598, USA}\\
$^{6}${European Southern Observatory, Casilla 19001, Santiago 19, Chile}\\
$^{7}${Departamento de Astrof\'isica, Universidad Complutense Madrid, Madrid, Spain}\\
$^{8}${University of Arizona, 933 N. Cherry Ave, Tucson, AZ 85721}\\
$^{9}${The Observatories of the Carnegie Institution for Science, Pasadena, CA, USA}\\
$^{10}${European Space Agency, ESTEC, 2200 AG Noordwijk, The Netherlands}\\
$^{11}${National Radio Astronomy Observatory / NAASC, 520 Edgemont Road, Charlottesville, VA 22903}\\
$^{12}${Astronomy Program, Department of Physics and Astronomy, Seoul National University, Seoul 151-742, Korea}\\
$^{13}${Spitzer Science Center, 1200 East California Boulevard, Pasadena, CA 91125}\\
$^{14}${California Institute of Technology, 1200 East California Boulevard, Pasadena, CA 91125}\\
$^{15}${Space Telescope Science Institute, Baltimore, MD, USA}
}
\begin{document}

\date{June 2012}

\pagerange{\pageref{firstpage}--\pageref{lastpage}} \pubyear{2012}

\maketitle
\label{firstpage}

\begin{abstract}
The mechanism causing breaks in the radial surface brightness distribution of spiral galaxies is not yet well known. Despite theoretical efforts, there is not a unique explanation for these features and the observational results are not conclusive. In an attempt to address this problem, we have selected a sample of 34 highly inclined spiral galaxies present both in the Sloan Digital Sky Survey and in the \textit{Spitzer} Survey of Stellar Structure in Galaxies. We have measured the surface brightness profiles in the five Sloan optical bands and in the  3.6$\mu m$ \textit{Spitzer} band. We have also calculated the color and stellar surface mass density profiles using the available photometric information, finding two differentiated features: an innermost \textit{break radius} at distances of $\sim 8 \pm 1$ kpc [$0.77 \pm 0.06$ $R_{25}$] and a second characteristic radius, or \textit{truncation radius}, close to the outermost optical extent ($\sim 14 \pm 2$ kpc [$1.09 \pm 0.05$ $R_{25}$]) of the galaxy. We propose in this work that the breaks might be a phenomena related to a threshold in the star formation, while truncations are more likely a real drop in the stellar mass density of the disk associated with the maximum angular momentum of the stars.
\end{abstract}


\section{Introduction}\label{sec:intro}

The vast majority of spiral galaxies do not follow a radial surface brightness decline mimicking a perfect exponential law as proposed by \citet{patt40,devau58} \& \citet{free70}. Depending on the shape of the surface brightness distribution, a classification for face-on galaxies has been developed by \citet{erw05} and \citet{pt06}. It distinguishes three different types of profiles. Type I (TI) is the classical case, with a single exponential describing the entire profile; Type II (TII) profiles have a downbending brightness beyond the break. Type III (TIII) profiles are characterized by an upbending brightness beyond the break radius. Relative frequencies for each type are 10\%, 60\% and 30\% \citep{pt06} in the case of late-type spirals.

Photometric studies of TII galaxies reveal that the radial scalelength of the surface brightness profiles changes when a characteristic radius ($\sim$10~kpc) is reached. This so called break radius is described in several studies of face-on galaxies \citep{erw05,pt06,erw08} and is also found if galaxies are observed in edge-on projections \citep{van82,grijs01,kruit07}. Using faint magnitude stars, \citet{fer07} found also a break in the surface brightness profile of M~33. On the contrary, no break was detected by star counting neither in NGC~300 \citep{bh05,vla09} nor in NGC~7793 \citep{vla11}. Works on galaxies beyond the nearby Universe \citep{per04,tp05,azzo08} show the presence of a break at redshifts up to z~$\sim$~1. These results suggest that breaks, once formed, must have been stable for the last 8 Gyrs of galaxy evolution. Cosmological simulations \citep{gov07,mar09} support the idea of a break in the light distribution of disk galaxies as well.

Different mechanisms explaining the origin of TII breaks have been proposed. All those theories can be sorted into two families. A possible scenario is that the break could be located at a position where a threshold in the star formation occurs \citep{fall80,kenn89,elme94,scha04,elme06}. A change in the stellar population would thus be expected at the break radius, but not necessarily a downbending of the surface mass density profile as shown by some simulations \citep{pat09,mar09}. Supporting this, \citet{bakos08} found that in a sample of 85 face-on galaxies (see Pohlen \& Trujilo, 2006), the ($g'-r'$) Sloan Digital Sky Survey (SDSS) color profile of the TII galaxies is, in general, U-shaped, with a minimum at the break radius, hinting at a minimum also in the mean age of the stellar population at the break radius. This is in agreement with studies on resolved stellar populations across the break \citep{jong07,rad12}. Furthermore, the surface mass density profile recovered from the photometry shows a much smoother behavior, in which the break is almost absent compared to the surface brightness profile. Numerical simulations \citep{deba06,ros08,pat09,mar09} reveal how the secular redistribution of the angular momentum through stellar migration can drive the formation of a break. In the papers of \citet{ros08} and \citet{pat09}, a minimum in the age of the stellar population is found at the break radius, in agreement with the results of \citet{bakos08}. It is also remarkable that the mean break radius in the simulations of \citet{ros08} is very close (2.6 $h_\mathrm{r}$) to the observational result (2.5$\pm 0.6$ $h_\mathrm{r}$) measured by \citet{pt06}, where the values are given in radial scalelength units ($h_\mathrm{r}$). However, \citet{van82} showed that the Goldreich-Lynden-Bell criterion for stability of gas layers offered a poor prediction for the truncation radius in their sample of edge-on galaxies. Focusing on these high inclined galaxies, \citet{kruit87,kruit88} proposed an alternative scenario where the break could be related to the maximum angular momentum of the protogalactic cloud, leading to a break radius of around four or five times the radial scalelength. This mechanism would also lead to a fast drop in the density of stars beyond the break. Observations of edge-on galaxies, such as those by \citet{van82}, \citet{bart94} or \citet{kreg02}, place the break at a radius around four times the radial scalelength, supporting this second explanation.

There is a clear discrepancy between the positions of the breaks located in the face-on view compared to those breaks found in the edge-on perspective. In the face-on view, the breaks are located at closer radial distances of the center than in the edge-on cases. Are the two types of breaks the same phenomenon? On the one hand, the edge-on observations tend to support the idea of the maximum angular momentum as the main actor in the break formation. On the other hand, face-on galaxies, supported by numerical simulations, favor a scenario where the break is associated with some kind of threshold in the star formation along the disk. Only a few studies have attempted to give a more global vision of the problem \citep{poh04,poh07} by deprojecting edge-on surface brightness profiles. \citet{poh07} and \citet{seb12} found that the classification exposed at the beginning of the current paper into TI, TII and TIII profiles is basically independent of the geometry of the problem. Consequently, the difference between the break radius obtained from the edge-on galaxies and the one from the face-on galaxies can not be related to the different inclination angle. More details on the current understanding of breaks and truncations can be found in the recent review by \citet[$\S$3.8]{van11}.

Despite truncations in edge-on galaxies and breaks in face-on galaxies have been traditionally considered equivalent features, our aim in the current paper is to understand the already noticed observational differences, proposing a global and self-consistent understanding of the breaks. We use images in six different filters (five SDSS bands and the S$^4$G 3.6$\mu m$ band) to study the radial surface brightness distribution in a sample of 34 edge-on galaxies. We also measure the color and the stellar surface mass density profiles for each object, trying to constrain the most plausible mechanism for the break formation.

The layout of this work is as follows: in Section \ref{sec:data} the characteristics of the sample are presented. Section \ref{sec:prof} describes how the profiles are measured. The analysis is shown in Section \ref{sec:ana} and then we discuss the results in Section \ref{sec:discu}. The main conclusions listed in Section \ref{sec:summ}. Tables referenced in the paper are in Appendix \ref{sec:tables}.

Throughout, we adopt a standard $\Lambda$CDM set of cosmological parameters ($H_0 = 70$ km  s$^{-1}$ Mpc$^{-1}$; $\Omega_\mathrm{M} = 0.30$;  $\Omega_\mathrm{\Lambda} = 0.70$) to calculate the redshift dependent quantities. We have used AB magnitudes unless otherwise stated.

\section{Sample and data} \label{sec:data}
Our sample of galaxies has been selected to cover edge-on objects present both in the Sloan Digital SDSS Data Release 7 \citep{aba09} and in the \textit{Spitzer} Survey of Stellar Structure in Galaxies (S$^4$G) \citep{seth10}. We find 49 highly inclined (inclination $\gtrsim 88$\textdegree) objects satisfying this criterion. In this sense our sample is neither volume limited nor flux limited.

For each object, the morphological type, the corrected $B$-band absolute magnitude ($M_\mathrm{B}$) and the maximum velocity of the rotation curve were obtained from the HyperLeda database \citep{patu03}. Using the NASA/IPAC Extragalactic Database (NED) \citep{helu91} we obtained most distances from primary indicators, except for three cases (PGC~029466, UGC~5347 and UGC~9345) where the distance was obtained using the redshift.

Finally, from the original 49 galaxies,  we keep only those objects brighter than $M_\mathrm{B} = -17$ mag and with morphological type Sc or later. Table \ref{tab:data} summarizes the general information available for each galaxy in the final sample. We focus our studies on late-type spiral galaxies. These late Hubble-types are known to have the most quiescent evolution during the cosmic time, hence the fossil records imprinted by the early galaxy formation and evolution processes likely survive to present day.

\subsection{SDSS data}
For all the galaxies in the sample, the SDSS data were downloaded in all the five available bands \{$u',g',r',i',z'$\}. SDSS observations are stored in the format of 2048$\times$1489 pixel ($\sim$~13.51'$\times$9.83') frames that cover the same piece of sky in all bands. These frames are bias subtracted, flat-fielded and purged of bright stars, but still contain the sky background. These frames might not be large enough to contain extended objects completely. For this reason, to assemble mosaics that are large enough to study extended objects, we needed to download several adjacent frames from the SDSS archive. We estimate the sky background of these frames independently (see \S\ref{sec:sky} for a more detailed description) and then, we assemble the final mosaics in all five bands, centered at the target galaxy, by using SWARP \citep{swarp}.

The pixel size in the SDSS survey is 0.396~\textquotedblright, and the exposure time is $\sim$ 54 sec. From this, we calculated the surface brightness in the AB system as:
\begin{eqnarray*}
\mu_\mathrm{SDSS} &=& -2.5\times\log(counts)+ZP \\
ZP &=& -2.5\times 0.4 \times (aa+kk+airmass) \\
& & +2.5 \times \log(53.907456\times0.396^2)
\end{eqnarray*}
where \textit{aa}, \textit{kk} and \textit{airmass} are the photometric zero point, the extinction term and the air mass, respectively. These parameters are stored in the header of each SDSS image. 

\subsection{S$^4$G data}
The S$^4$G project is a survey of a representative sample of nearby Universe galaxies (D$<$40 Mpc) that collects images of more than 2000 galaxies with the \textit{Spitzer} Infrared Array Camera (IRAC); \citet{faz04}. In particular, we are interested in the 3.6$\mu m$ band (channel 1) which is less affected by dust than the SDSS optical bands and is a good tracer for the stellar mass in our  galaxies. The fact that the optical depth in this band is lower than in the SDSS images is crucial as we are dealing with edge-on objects, with a lot of dust along the line of sight. The IRAC 3.6$\mu m$ images were reduced by the S$^4$G team. They have a pixel scale of 0.75~\textquotedblright/pix and a spatial resolution is $\sim$ 100 pc at the median distance of the survey volume. Also, the azimuthally-averaged surface brightness profile typically traces isophotes down to a (1$\sigma$) surface brightness limit $\sim$ 27 mag arcsec$^{-2}$ (see Sheth et al. 2010). The conversion to AB surface magnitudes is given by:
\begin{displaymath}
\mu_\mathrm{S^4G} = -2.5\times\log(counts)+20.472
\end{displaymath}

\section{Data handling and results} \label{sec:prof}
\subsubsection{Sky subtraction}\label{sec:sky}
Since the breaks are expected to happen at low surface brightness (a typical value of $\sim$ 24 mag arcsec$^{-2}$ in the SDSS $g'$ band was found by Pohlen \& Trujillo 2006 for face-on galaxies), a very careful sky estimation is needed to obtain reliable results. In this context, working with data taken with different instruments at different wavelengths, requires a different treatment for the SDSS and S$^4$G data. 

In the case of the SDSS images, after removing the 1000 counts of the SOFTBIAS added to all SDSS chips, we measure the fluxes in about ten thousand randomly placed, five pixel wide apertures. We apply a resistant mean\footnote{The resistant mean trims away outliers using the median and the median absolute deviation. It is implemented as an Interactive Data Language (IDL) routine.} to the distribution of the aperture fluxes, and carry out several iterations to get rid of the fluxes biased by stars, and other background objects. The mean of the bias-free distribution provides accurate measurement of the sky background, and is then subtracted from the chips. 

\begin{figure*}
\begin{center}
\includegraphics[width=145pt]{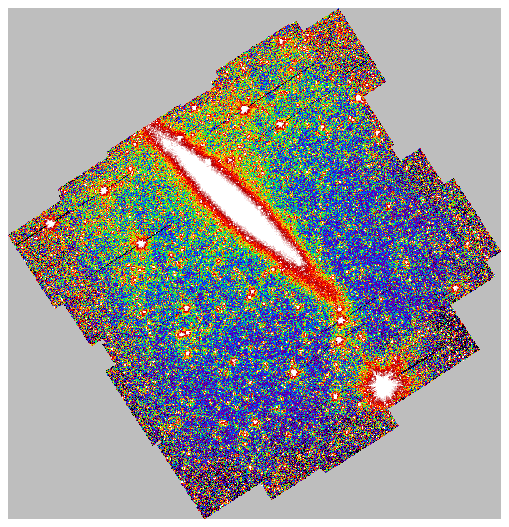} 
\includegraphics[width=145pt]{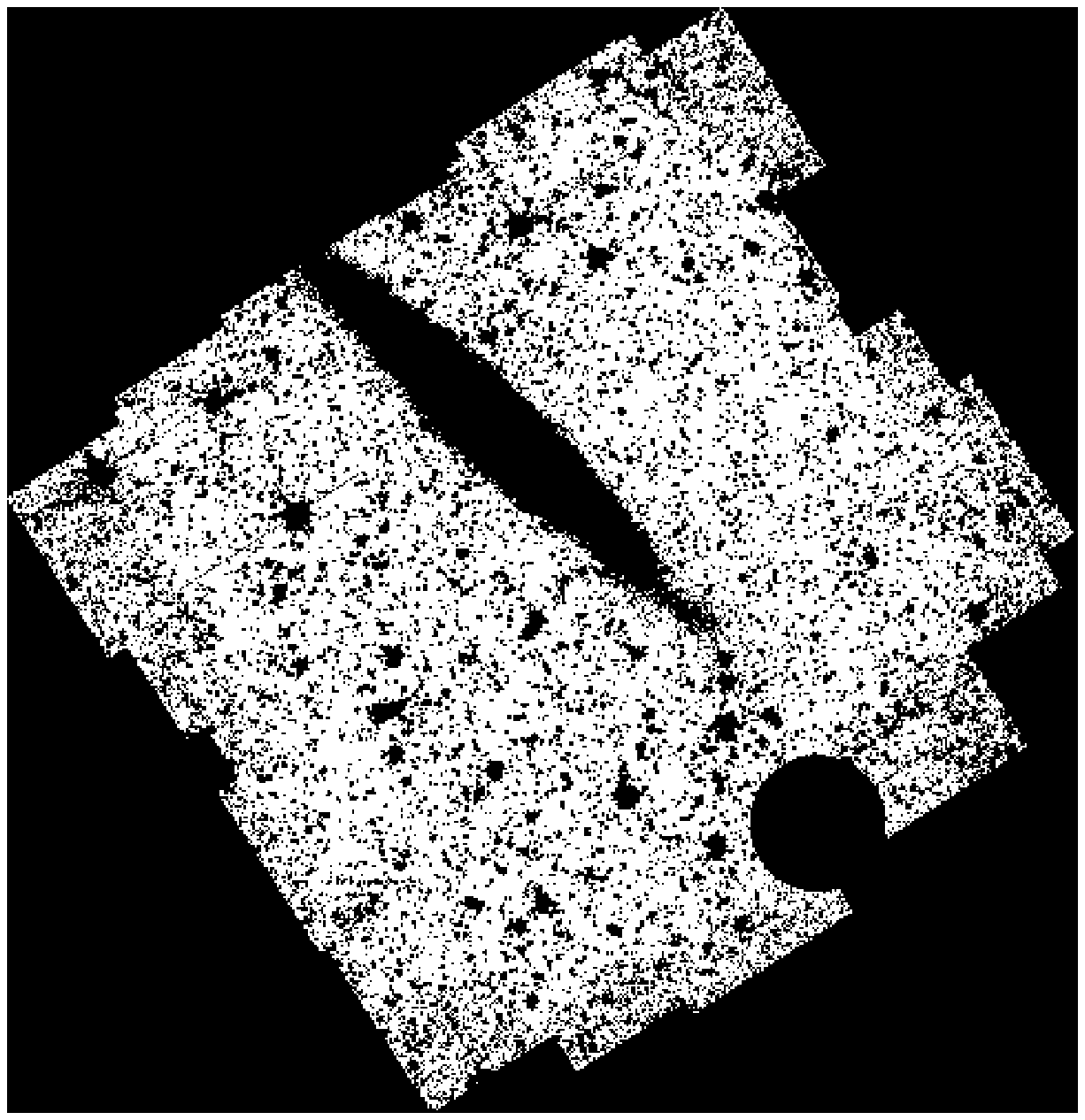}
\includegraphics[width=145pt]{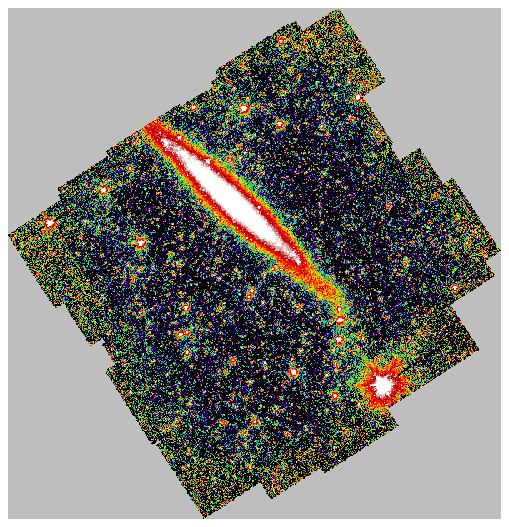} 
\end{center}
\caption{Sky subtraction steps for NGC 4244 (3.6$\mu m$ band). From left to right: 1- Original image. 2- Image mask where white pixels are those used to estimate the background shape. 3- Image after the subtraction of one of the polynomial fittings (3rd order).}
\label{img:mask}
\end{figure*}

For the S$^4$G data, the sky treatment was completely different because of the presence of important gradients across the images (see Comer\'on et al. 2011). The adopted strategy was to mask every object in the image and then make a low order polynomial fit of the background in the same way that \citet{seb11} did in their study of NGC 4244. In order to establish a threshold between sky/non-sky pixels, we built the histogram of the image and, with the assumption that this histogram is dominated by the sky pixels, we calculated the maximum and the FWHM of this distribution. The FWHM measured in that way can be associated with an effective sigma ($\sigma_\mathrm{e}$) that we supposed representative of the standard deviation of the background. After this, we performed five two-dimensional fits with five different polynomial orders ($\mathcal{O} = \{0,1,2,3,4\}$), including only those pixels with a value between the maximum of the histogram $\pm 3\sigma_\mathrm{e}$. The outcome of this stage was five sky-subtracted images, each one derived from one of the fits. The difference between these images was used (see \S\ref{sec:s4glim}) to establish the surface brightness limit in the 3.6$\mu m$ surface brightness profile. In Fig. \ref{img:mask} we show a raw S$^4$G image (left) where the gradients across the image are obvious; the mask image (center) with the pixels used to fit the background represented in white and, as an example, the 3rd order sky subtracted image (right). 

\subsubsection{Estimating the position angle of the slit}
Our aim is to extract surface brightness profiles along the disk in a slit that represents the same physical distance (1.2  kpc) vertically in all galaxies. For this reason, this slit has to be aligned accurately with the disk of our galaxies. The position angle (PA) was defined as the angle that maximizes the total number of non-sky pixels within the previously fixed slit. This definition is consistent with the idea followed in the surface brightness profile calculation. The measurement will be robust if the size of the galaxy is large enough compared to the field of view. As in \S \ref{sec:sky}, we distinguished sky pixels from non-sky pixels, setting them to 0 and 1 respectively. We rotated the image by a certain angular step and then sum the value of all the pixels within the slit. The desired angle corresponds to the maximum of this distribution. 

\begin{figure}
\begin{center}
\includegraphics[width=215pt]{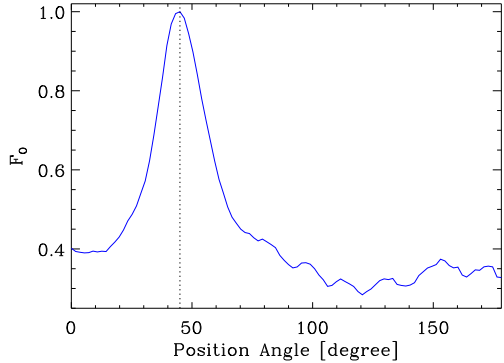} 
\end{center}
\caption{The fraction of non-background pixels within the slit aperture (F$_\mathrm{O}$) as a function of the position angle of the slit for NGC 4244 ($r'$-band). The vertical dashed line marks the center of the angular interval used to refine the measurement of the final PA.}
\label{img:rot}
\end{figure}

In order to obtain the PA with enough accuracy, but with an acceptable computational cost, the process was divided into two phases. In the first one, a low resolution profile was obtained by rotating the galaxy between $0$\degr \ and $180$\degr \ in 100 consecutive steps. We then defined an angular interval around the maximum of the resulting low resolution profile and used a high resolution step of 0.2\degr. The maximum found in this second phase is the PA of the galaxy. Fig. \ref{img:rot} shows the normalized fraction of flux within the aperture vs the slit position angle profile for the case of NGC 4244, with a peaked distribution around the maximum (PA = 47.2\degr).

As the galactic plane is expected to be dust attenuated, we employed the 3.6$\mu m$ S$^4$G images to calculate the PA. The PA derived form the SDSS data was in all the cases compatible with the S$^4$G measurements taking into account the associated errors. In addition, we repeated the measurement of the PA changing the threshold used to distinguish the sky pixels, taking at the end the average of the resulting values.

The PA calculated in this way is intuitive but is sensitive to the presence of extended objects in the field and to asymmetries in the light distribution of the galaxy. For that reason, we had to manually set the PA in seven cases to completely align the galactic plane with the slit, with a typical deviation with respect to the automatic measurement $\left\langle  PA_\mathrm{final}-PA_\mathrm{auto} \right\rangle  =$ 0.5 $\pm$ 0.1\degr. The PA here calculated is compared in Table \ref{tab:rota} with the position angle from HyperLeda\footnote{By definition, the PA increases (from 0\degr \ to 180\degr) from the North to the East, i.e., counterclockwise.} for each galaxy in the sample, showing a mean difference between our measurements and the PA from HyperLeda $\sim 0.7$\degr. It is worth noting here that the PA given by HyperLeda is the result of averaging all entries in the database for each object, sometimes with very different values (e.g. for NGC 4244, the HyperLeda PA (42.2\degr) is the average of four significantly different measurements: 48\degr, 48\degr, 27\degr \  and 45.2\degr).

\subsubsection{Image masking}
To study the outskirts of the galaxies with the desired precision it was necessary to mask objects that clearly do not belong to the galaxy (foreground stars, background galaxies). This masking process allowed us to obtain cleaner surface brightness profiles and also to avoid the contamination of the faintest parts by external objects.

For the S$^4$G images, the masks were supplied by the S4G team and are described in \citet{seth10}. In the case of the SDSS images, we generated the masks using the package SEXTRACTOR \citep{bert96}. Background objects and foreground stars were detected using a master image composed of all five SDSS bands, scaled to the $r'$-band flux. Then, over-sized elliptical masks were place onto the detected sources using SEXTRACTOR parameters such measured flux, elongation and similar. To increase the accuracy in this process, the masking was done in two consecutive steps. In the the first one, the bigger and brighter background objects were masked. In the second step, we performed a more detailed detection process to mask any possible remaining object in the field. By doing this, we could successfully mask background objects within a wide range of sizes.

\subsection{Surface brightness profiles}\label{sec:asy}
\begin{figure}
\begin{center}
\includegraphics[width=190pt]{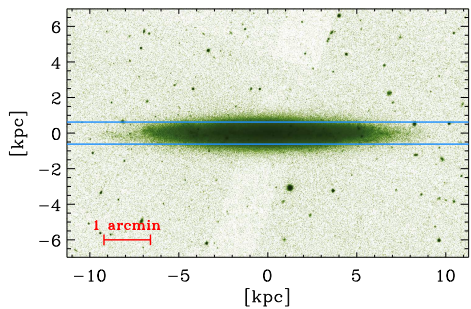} 
\end{center}
\caption{The galaxy NGC 5023 is shown with the region (slit) used to calculate the surface brightness profile delimited by the solid blue lines.} 
\label{img:show}
\end{figure}

To extract the profile along the radial distance of the galaxies, a slit of constant physical width is placed over the galactic plane and then used to calculate how the flux varies along the slit. The width of the slit was fixed to 1.2 kpc following the assumption of a vertical scalelength equal to 0.6 kpc \citep{kreg02}. For illustrative purposes, in Fig. \ref{img:show} the slit position (in blue) is shown over the $r'$-band image of NGC~5023.

In general, the process to obtain the surface brightness profiles was the same for the SDSS and for the S$^4$G data. The first step was to divide the slit along the major axis in cells of 0.3" radial width. The averaged value of the flux within each cell was calculated using a $3\sigma$ rejection mean. This first stage yields a crude estimate of the surface brightness profile. The center of the galaxy was supposed to correspond to the brightest cell of this profile and, to avoid any kind of misalignment between the different filters, we took the right ascension and  declination of the S$^4$G brightest cell as the center in all the other profiles. The sky was again measured on the left and right sides of the galaxies in order to, later, remove any possible residual contribution after the initial sky subtraction (\S \ref{sec:sky}). 

After that, we calculated the \textit{right} and \textit{left} surface brightness profiles with the size of the bins linearly increasing with a factor 1.03 between consecutive cells to improve the signal to noise ratio in the outskirts of the galaxies. The right/left residual sky value measured at the beginning of the process was also subtracted for the right/left profiles to obtain the final surface brightness distributions. The surface brightnesses of this residual sky corrections were typically $\sim 30$ mag~arcsec$^{-2}$ for the $g'$-band and $\sim 29$ mag~arcsec$^{-2}$ for the $3.6\mu m$ band.

\begin{figure}
\begin{center}
\includegraphics[width=215pt]{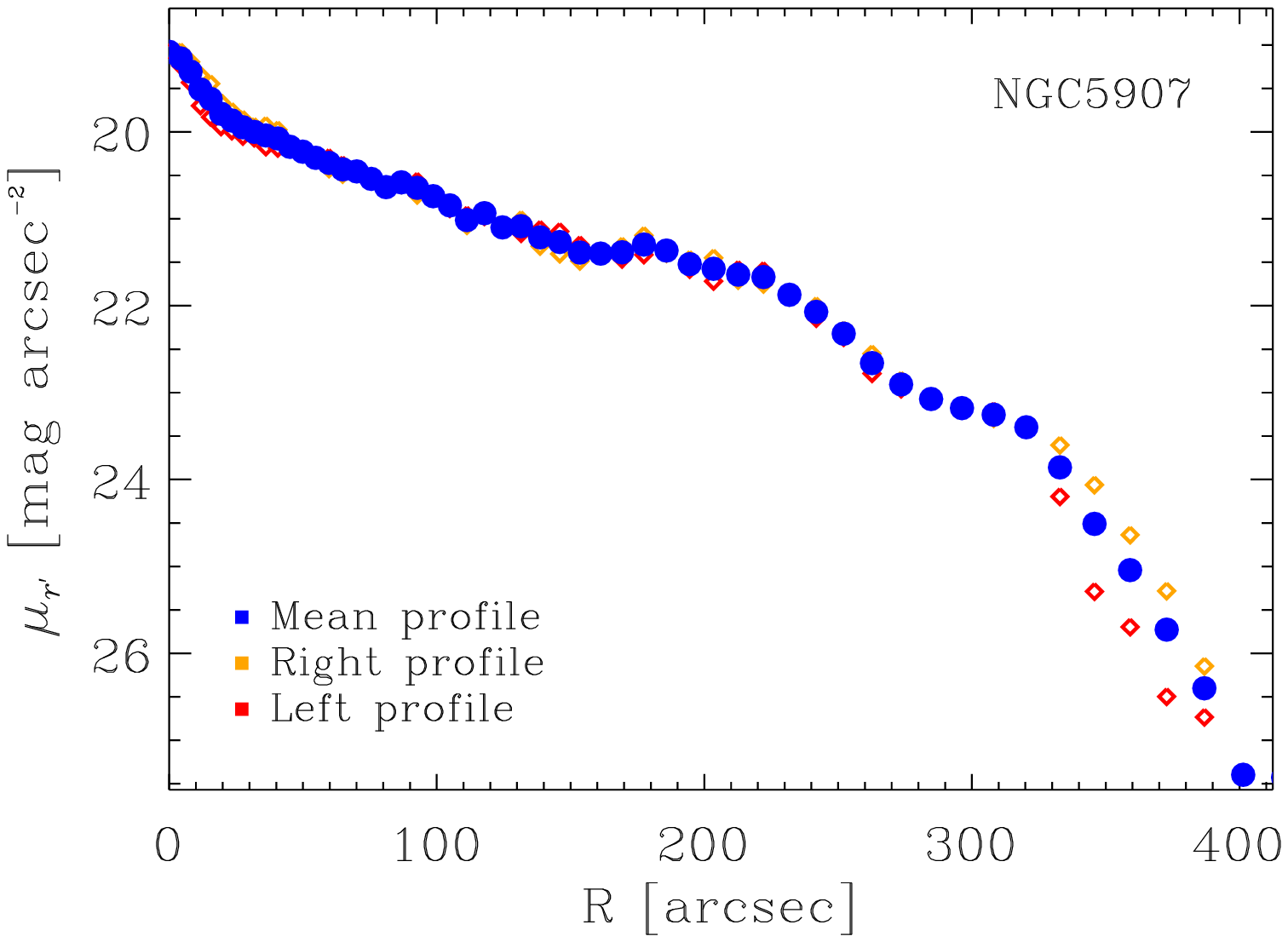}
\includegraphics[width=215pt]{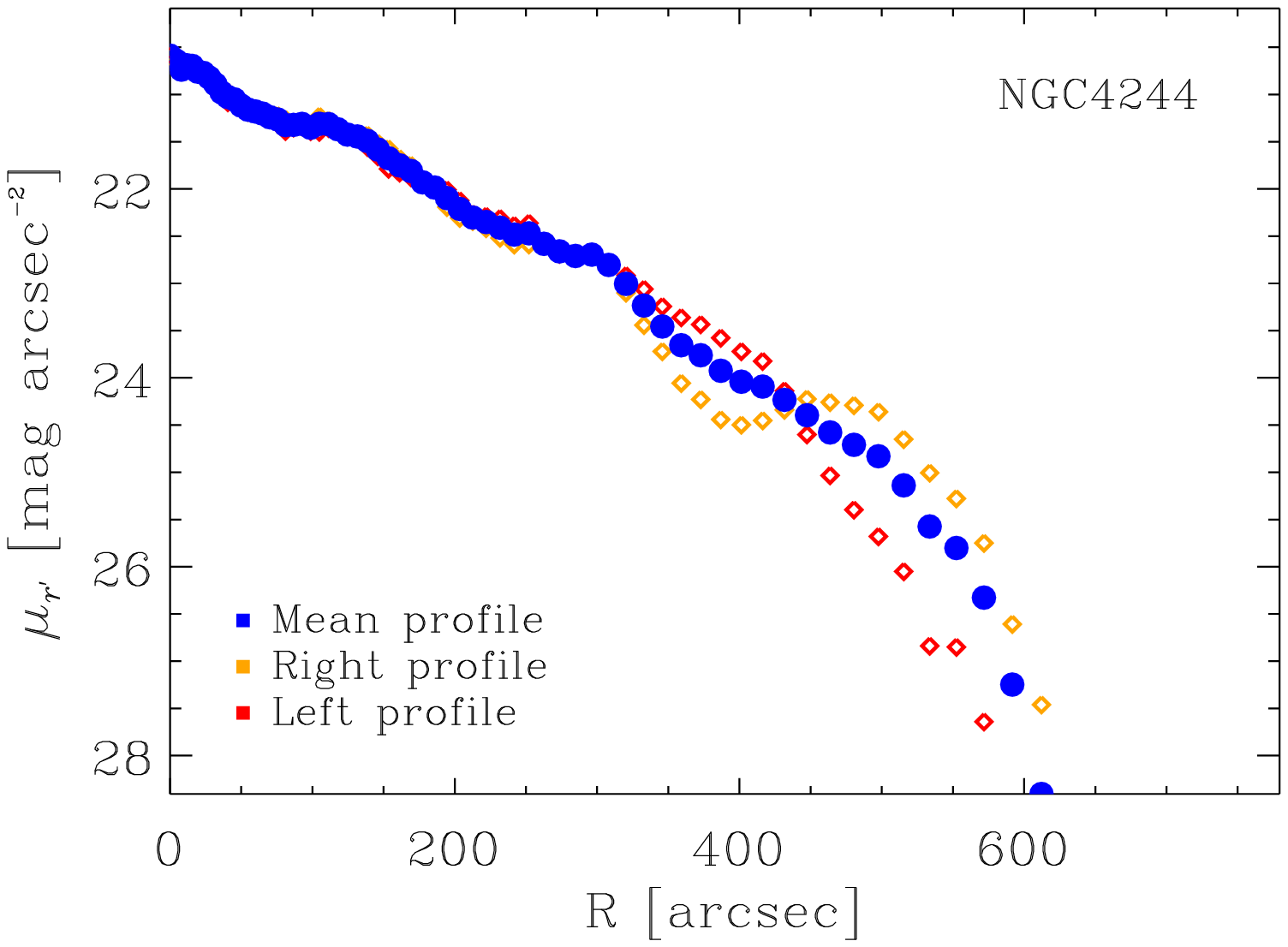}
\end{center}
\caption{Surface brightness profile for the galaxies NGC 5907 (upper panel) and NGC 4244 (bottom panel) in the $r'$ band. In blue it is shown the mean profile, whereas left and right profiles are represented in red and yellow respectively.}
 \label{img:leftright}
\end{figure}

Finally, a mean profile was constructed as the average of the left and right profiles. These averaged profiles are the ones used in this study. Looking at the individual right/left profiles of all our sample, it was clear that in some cases (e.g. NGC~4244) there are considerable asymmetries (up to $0.8$ $r'$-mag arcsec$^{-2}$ difference between the averaged profile and the right/left profiles at $ r = 550$ arcsec for that object) but as we wanted to explore the surface brightness profiles down to very faint regions, the combination of the two sides of the galaxies is necessary. As an example, in Fig. \ref{img:leftright}  we represent the mean, left and right surface brightness profile in the SDSS $r'$-band for a symmetric case (NGC~5907) in opposition to the more asymmetric galaxy NGC~4244.

Beyond a certain radius, the sky subtraction becomes a dominant factor at determining surface brightness profiles. Having used different sky subtractions techniques, the maximum radial extension of a reliable surface brightness profile needs to be measured separately for the SDSS and for the S$^4$G data.

\subsubsection{SDSS surface brightness limit}
To estimate how faint we can explore the SDSS profiles, we defined a certain critical surface magnitude ($\mu_\mathrm{crit}$) that sets the limit for our profiles. Following \citet{pt06}, we defined $\mu_\mathrm{crit}$ as the surface brightness where the two profiles obtained by under/over subtracting the sky by -1/+1 $\sigma_\mathrm{sky}$ start to differ by more than 0.2 mag arcsec$^{-2}$ where $\sigma_\mathrm{sky}$ is given by
\begin{displaymath}
\sigma_\mathrm{sky}^2 = \sigma_\mathrm{sky,\ left}^2 + \sigma_\mathrm{sky, \ right}^2
\end{displaymath}
with $\sigma_\mathrm{sky,\ left}$ and $\sigma_\mathrm{sky,\ right}$ the standard deviation of the residual sky measured on the left and right sides of the galaxy. On average, we could reach a surface brightness magnitude limit equal to 26.1 $\pm$ 0.1 mag arcsec$^{-2}$ in the SDSS $r'$ band. This value is a characteristic limit for the $g'$, $r'$ and $i'$ SDSS data, with the remaining SDSS images being typically shallower (e.g. $\left\langle \mu_{u'\mathrm{,lim}}\right\rangle  = 25.7$ mag arcsec$^{-2}$ ; $\left\langle \mu_{z'\mathrm{,lim}}\right\rangle  = 24.2$ mag arcsec$^{-2}$ ). 

\subsubsection{S$^4$G surface brightness limit}\label{sec:s4glim}
The limit for the S$^4$G data had to be more strictly defined because of the presence of gradients across the S$^4$G images. As discussed in \S \ref{sec:sky}, we have, for each object, five images from five different polynomial fits to the background. The limit was established where the function
\begin{displaymath}
\mu_\mathrm{crit} = \displaystyle \frac{1}{5} \displaystyle \sum_{n=0,..,4} \displaystyle \vert \mu_n - \left\langle  \mu \right\rangle  \vert 
\end{displaymath}
became greater than 0.2 mag arcsec$^{-2}$. In other words, the limit for the S$^4$G profiles was placed where the mean difference between the five profiles and the mean one was greater than 0.2 mag arcsec$^{-2}$.  A typical value of 26.1 $\pm$ 0.2 mag arcsec$^{-2}$ was found for this limit in the S$^4$G 3.6$\mu m$ band.

The limit defined in this a way is not necessarily the same for the left and right sides of a galaxy. The mean profile is in fact the average of two profiles until the surface magnitude limit of the ``shallower'' side. Beyond that, the mean profile is only that of the ``deeper'' side of the galaxy. 

\subsection{Color profiles}\label{sec:color}
From the surface brightness profiles one can directly obtain the radial color profiles of an object. However, the interpretation of the edge-on color profiles is not straightforward. On the one hand, the presence of dust in the galactic plane will generate redder colors and change the actual shape of the color profile compared to that expected in its absence. On the other hand, effects related to the edge-on projection (integration of light from different radii along the line of sight and a bigger optical depth because of the presence of dust) are difficult to handle and to account for in an edge-on color profile.

In this work, we have focused only on those profiles that could reach the lowest surface brightness values (i.e., the color profiles derived from the deepest surface brightness profiles). In particular, we calculated for each galaxy the ($g'-r'$), ($r'-i'$) and ($r'-3.6 \mu m$) color profiles.

\subsection{Stellar surface mass density profiles}
The final quantity that we obtained from the images is the stellar surface mass density ($\Sigma_\lambda$). We can relate $\Sigma_\lambda$ with the surface brightness at a certain wavelength ($\mu_\lambda$) and with the mass to luminosity ratio $(M/L)_\lambda$ at that given $\lambda$ (see Bakos et al. 2008) using the following expression:
\begin{displaymath}
\log \Sigma_\lambda = \log(M/L)_\lambda - 0.4(\mu_\lambda - m_{\mathrm{abs} \ \odot, \lambda}) + 8.629
\end{displaymath}
where $\Sigma_\lambda$ is given in $M_\odot$  pc$^{-2}$. Having $\mu_{r'}$ from the photometry, one just needs to know the value $(M/L)_{r'}$ along the galaxy to get the $\Sigma_\lambda$ profile. We used the relation proposed by \citet{bell03} to estimate $(M/L)_{r'}$ as follows
\begin{displaymath}
\log(M/L)_{r'} = [a_{r'} + b_{r'}\times(g'-r')]-0.15
\end{displaymath}
where $a_{r'}$ and $b_{r'}$ are $-0.306$ and $1.097$ respectively. We have then the stellar surface mass density as a function of all known variables.

As mentioned in \S \ref{sec:color}, the interpretation of edge-on color profile is not as direct as in the face-on case. The dust attenuation and the error propagation in this kind of profiles are very important and thus, one has to be aware of all these effects while interpreting the surface mass density profiles.

\subsection{Presentation of the results}\label{sec:pres}
Shortly bellow we will show some examples of the profiles extracted in this work. For the rest of the galaxies, we refer to the reader to Appendix \ref{sec:sample}. Upper left panels in these figures show the six bands surface brightness profiles used in this paper. The dashed vertical lines mark those radii where a change in the exponential behavior happens defined as described in \S\ref{sec:classi}.

On the bottom left panel we represent the color profiles. The color profile are shown down to the distance where data points have an error less than or equal to 0.2 mag, with the error calculated as the quadratic sum of the errors in the individual surface brightness profiles.

The bottom right panel is occupied by a $g'$-band image of the galaxy, with vertical lines marking the characteristic radii, in the same way as in the surface brightness profiles. The angular distance from the center of the galaxy to those lines is shown at the top of the image.

Finally, the upper right panel shows the stellar surface mass density profile. In the same panel is over-plotted the 3.6$\mu m$ surface brightness profile, that is expected to be a good tracer of the stellar mass density. In this panel we have not limited the extent of the profiles and therefore the farthest points have to be considered with caution.

\section{Analysis} \label{sec:ana}
\subsection{Break radius and scalelengths}\label{sec:measu}
\begin{figure}
\begin{center}
\includegraphics[width=215pt]{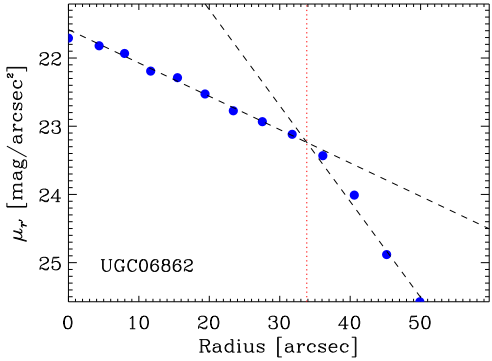}
\end{center}
\caption{Fit of the $r'$-band surface brightness profile of UGC 06862. The fit of the two characteristic regions is shown as a black dashed line, while the radius where the change in the slope occurs is marked with a red dotted line. We have defined this radius as the intersection point between the two linear fits.}
 \label{img:fit}
\end{figure}

In order to quantify the radial distance where the change in exponential scalelength occurs, we have linearly fitted the $r'$-band averaged surface brightness profile of each galaxy. We have visually selected all the disk regions showing a differentiated exponential behavior and then we have made an independent linear fit over each one of those regions. The characteristic radius where the exponential scalelength changes is set at the intersection point between two straight lines derived from the fit of two adjacent regions. Fig. \ref{img:fit} shows, as an example, the fit for UGC 06862. The black dashed lines represent the linear fit of each part of the disk. The red dotted line marks the intersection point between the two fits. Table \ref{tab:regi} lists the regions of the disk where the fits were made for each individual galaxy. In addition, Table \ref{tab:res} lists where all characteristic radii occur in each galaxy in the sample.

Apart from identifying different characteristic radii, we have measured $r_\mathrm{max}$ as the radial distance where a hint of a new change in the exponential behavior is present but there were not enough points to perform a reliable fit because it occurs at a surface magnitude too close to the surface brightness limit. If no such hint is detected, $r_\mathrm{max}$ represents then a lower limit for a potential new characteristic radius detection. 

\subsection{Overall behavior}\label{sec:over}
To exemplify the findings of our work, we will focus on the interpretation of those objects that, in particular, better reflect the general behavior of the surface brightness profiles of our sample. In any case, the same information is available for all the galaxies in Appendix \ref{sec:sample}.

\begin{figure*}
\begin{center}
\includegraphics[width=400pt]{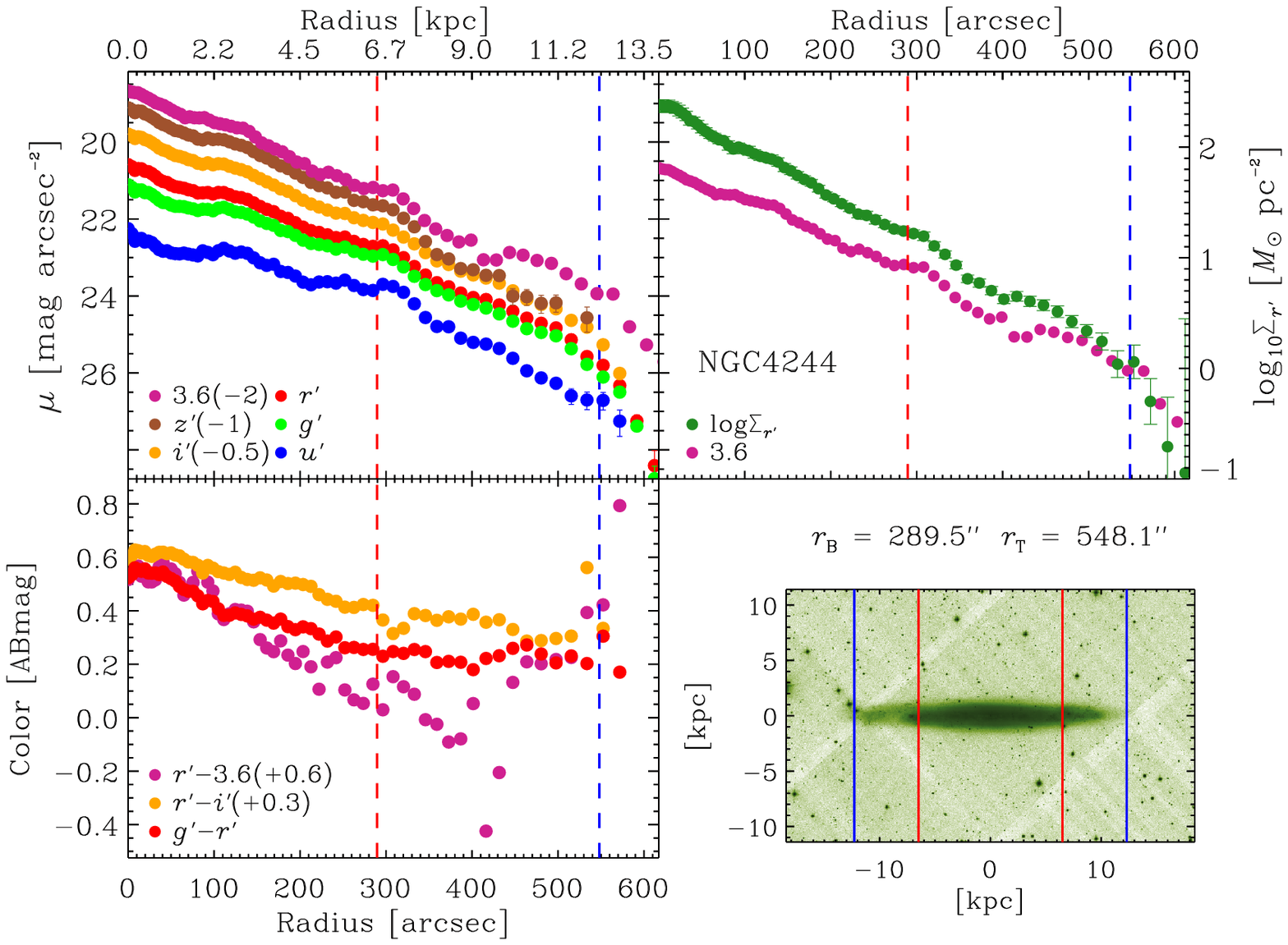}
\end{center}
\caption{Surface brightness, color and stellar surface mass density profiles of NGC 4244. The upper left panel shows six surface brightness profiles from the different photometric bands used in the current paper. Dashed vertical lines mark where a change in the exponential behavior happens. On the upper right panel are simultaneously plotted the stellar surface mass density profile and the 3.6$\mu m$ surface brightness profile. The bottom left panel shows the three deeper color profiles from the available photometric bands. Lastly, the bottom right panel is occupied by a $g'$-band image of the galaxy. For this object the surface brightness distribution shows two clearly differentiated regions: one inner region before the red line and a second outer region after the blue line. The position of the two breaks is shown on the bottom right panel. The last characteristic radius is near to the visible edge of the galaxy and it is reflected as a drop in the stellar surface mass density profile. On the other hand, the first break is clearly within the galactic disk and does not seem to affect dramatically the behavior of the mass distribution.}
\label{img:4244}
\end{figure*}

A paradigmatic example of the behavior found in the surface brightness profiles is found in NGC~4244 (Fig. \ref{img:4244}). In the inner parts of the galaxy, the shape of the surface brightness profile is dominated by an exponential decline but, reaching a break radius $r \sim 290"$ (red vertical line in Fig. \ref{img:4244}), the slope becomes more pronounced. If we keep moving farther away, we find a second change in the exponential behavior (blue line) happening at $r \sim 550"$. In the image of the galaxy shown in the bottom right panel, we can see how the first break radius (red lines) marks an inner disk, while the second break radius is very close to the visual edge of the galaxy for this exposure. It is worth noting here that the asymmetry of the brightness distribution in this object (see \S\ref{sec:asy}) makes the break radii different for the two sides of the galaxy. This has been already noticed in several studies \citep{jong07,seb11,hol12}. While \citet{hol12} noted a bright star-forming knot on either side of this galaxy but at slightly different radii, \citet{seb11} stated that this asymmetry ``can be explained by a combination of the galaxy not being perfectly edge-on and a certain degree of opacity of the disk''. \citet{jong07} proposed this asymmetry as the cause of the difference between their measurements for the break radial distance of the ``shorter'' galactic side, and the results of \citet{van81} and \citet{fry99}, who placed the break radius at around 570'' as a consequence of averaging both sides of the galaxy. It is worth noting that this asymmetry makes the breaks smoother that in the case of more symmetric galaxies.

Lastly, in the stellar surface mass density plot, the second radius is associated with a steeper drop in both profiles than the first break, which has only a minor feature in Fig. \ref{img:4244} like that in the surface brightness profiles. 

\begin{figure*}
\begin{center}
\includegraphics[width=400pt]{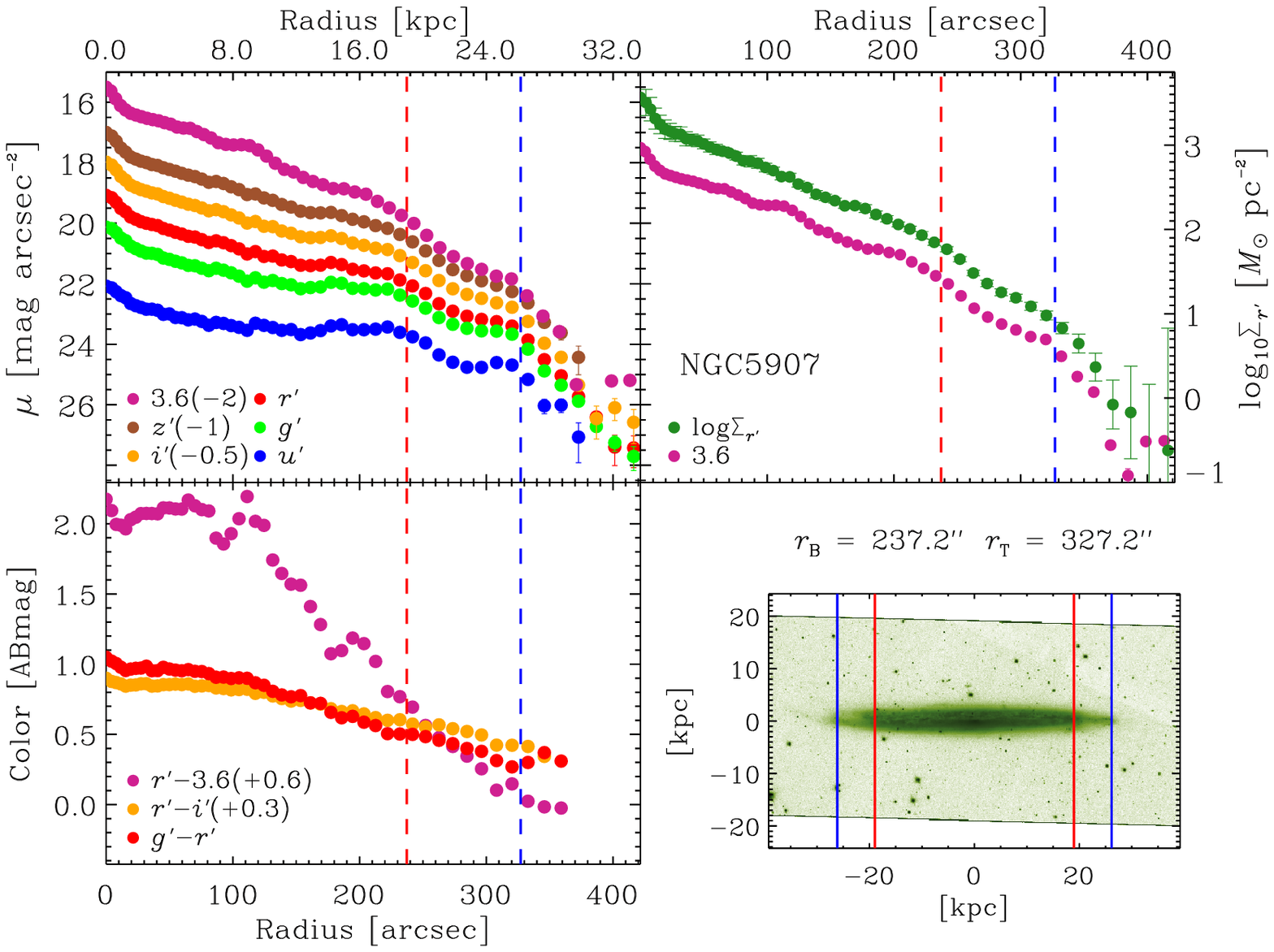}
\end{center}
\caption{As in Fig. \ref{img:4244}, now for NGC 5907. The first break radius is placed at around 240''. The farther break radius can be found at $\sim 330''$. Contrary to what we observe in NGC 4244, there is a big difference between the values of the optical color profiles and the ($r'-3.6\mu m$) color profile that could be related to a stronger dust attenuation of the inner parts of this galaxy compared to NGC 4244. }
\label{img:5907}
\end{figure*}

Another representative example (in this case a symmetric galaxy) is NGC~5907 (Fig. \ref{img:5907}). The surface brightness profiles resemble a typical TII, with the inner break around 240". As in the case of NGC~4244, there is a second change in the exponential behavior at a radius $r \sim 330"$.

A clear difference with NGC~4244 are the high values of the ($r'-3.6\mu m$) color in the central region of the galaxy. Knowing that the IR band is less sensitive to dust attenuation, this red value of the ($r'-3.6\mu m$) color could be explained by the attenuation of dust in the optical bands. This is a very common behavior among our sample of galaxies since it is clearly present in, at least, 12 objects (see Appendix \ref{sec:sample}). Paying attention to the stellar surface mass density plot, one can see again how the change in the slope is much pronounced at the second break radius compared to the change at the inner break radius. This time, being the galaxy very symmetric, this change is clearly visible.

\subsection{Feature classification}\label{sec:classi}
Many of our galaxies (18 out of 34) show two break radii in their profiles. The characteristics of the breaks seems to be different (see \S\ref{sec:over}) depending whether they are in the inner or in the outer regions of the disk. For this reason we attempt to establish an observational classification for these features. Assuming that both breaks have a different physical origin, we have studied their characteristics in those cases when it was easy to distinguish one type from the other, i.e., when both breaks appear simultaneously in the same galaxy. Having characterized the type of the break, that allowed us to extent the classification to those galaxies where only one break was present in the surface brightness profile.

\begin{figure}
\begin{center}
\includegraphics[width=217pt]{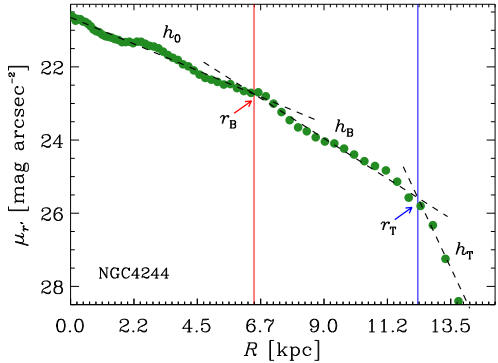}
\end{center}
\caption{Exemplification of the meaning of the parameters $h_0$, $h_\mathrm{B}$, $h_\mathrm{T}$, $r_\mathrm{B}$ and $r_\mathrm{T}$ using the \break $r'$-band surface brightness profile (NGC~4244).}
\label{img:fea}
\end{figure}

We labeled the first radius as the break radius ($r_\mathrm{B}$) and the second radius as the truncation radius ($r_\mathrm{T}$). In order to define an empirical criterion, we have investigated the distributions of these breaks and truncations as a function of two different parameters: $h_0/h_\mathrm{feature}$ (with $h_0$ the innermost scalelength and $h_\mathrm{feature}$ the scalelength after the break ($h_\mathrm{B}$) or the truncation ($h_\mathrm{T}$)) and $r_\mathrm{feature}/h_\mathrm{feature}$ (with $r_\mathrm{feature}$ representing $r_\mathrm{B}$ or $r_\mathrm{T}$, i.e., the radial distance from the galactic center to the break or the truncation position). The meaning of each parameter is exemplified in Fig.~\ref{img:fea}.

\begin{figure*}
\begin{center}
\includegraphics[width=230pt]{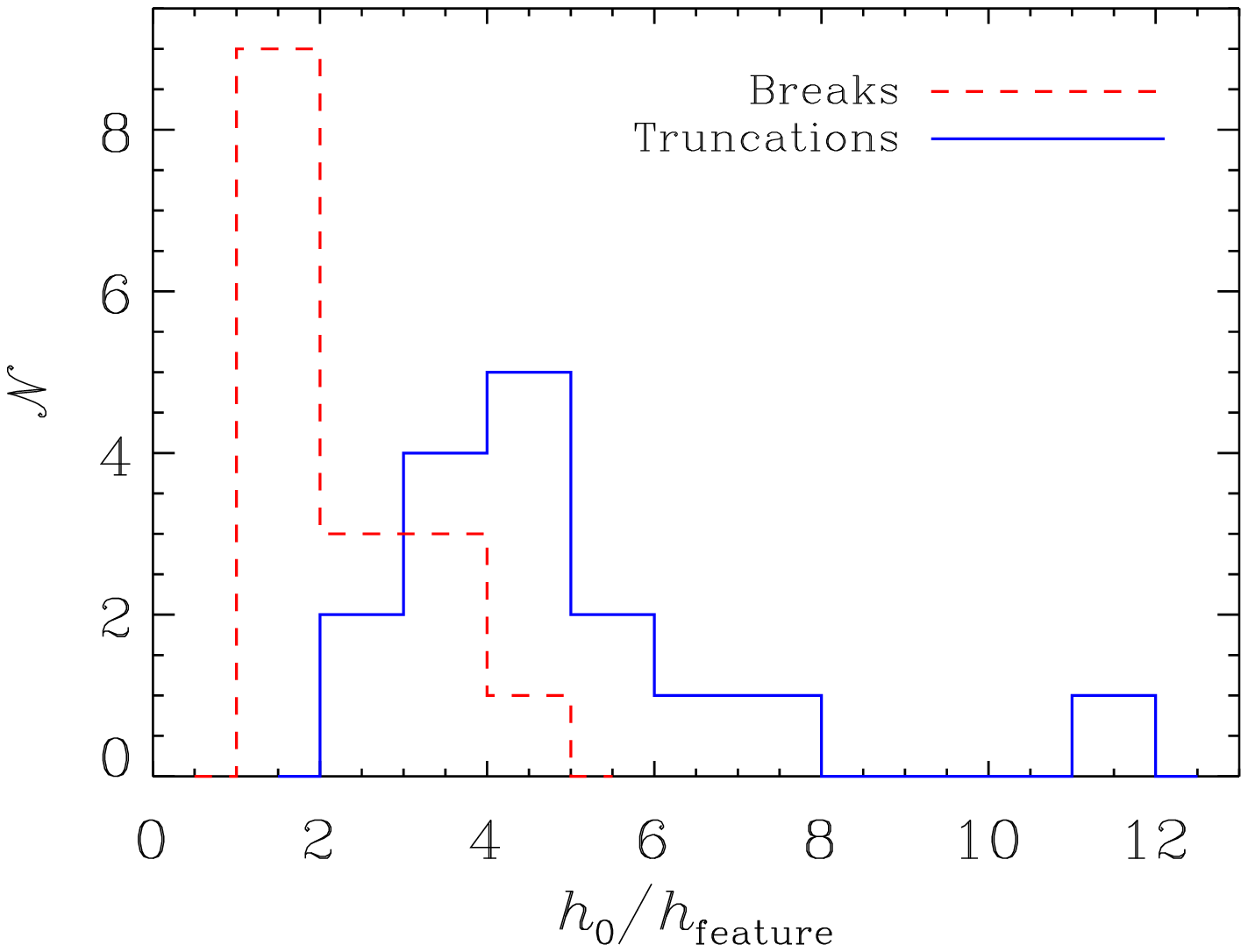}
\includegraphics[width=230pt]{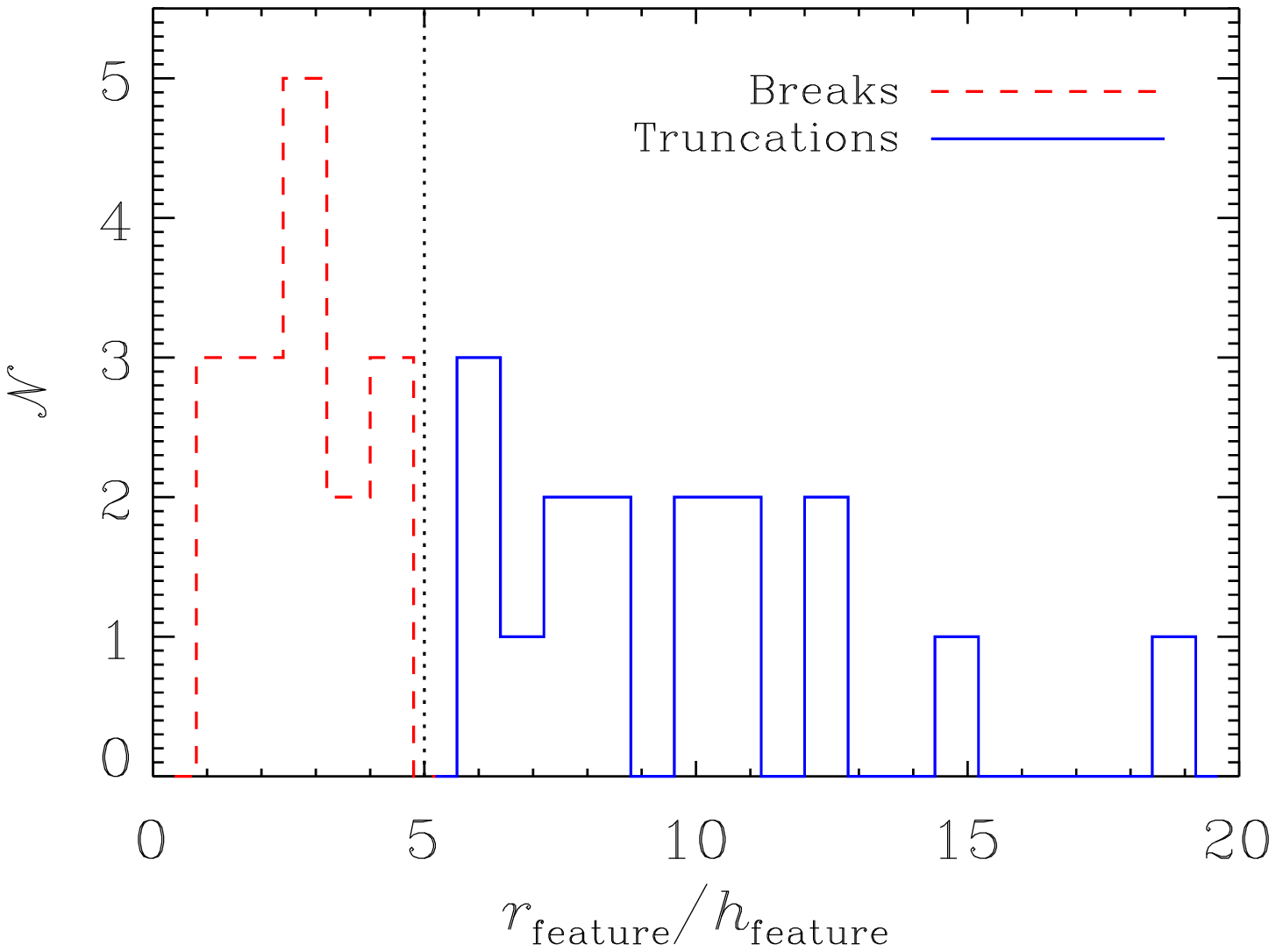}
\end{center}
\caption{Histograms representing the distribution of breaks and truncation versus the $h_0/h_\mathrm{feature}$ (left) and the $r_\mathrm{feature}/h_\mathrm{feature}$ (right) parameters where $h_0$ is the innermost exponential scalelength (between the galactic center and the first break) and $h_\mathrm{feature}$ is the exponential scalelength after the first break (\textit{break}) or after the second break (\textit{truncation}), measured in the SDSS $r'$-band. $r_\mathrm{feature}$ represents the distance from the center of the galaxy where the break or the truncation is measured. The degeneracy shown in the left panel between both features is broken in the right panel.}
 \label{img:space}
\end{figure*}

Fig. \ref{img:space} shows the histograms for the breaks and truncation vs each one of the above defined ratios, measured in the SDSS $r'$-band surface brightness profiles. The breaks distribution is shown in red while the truncations are represented by a blue histogram. It is clear from Fig. \ref{img:space} that breaks and truncations are partially degenerated when using the $h_0/h_\mathrm{feature}$ parameter but both characteristic radii are fully differentiated using the $r_\mathrm{feature}/h_\mathrm{feature}$ parameter. We find $r_\mathrm{feature}/h_\mathrm{feature} = 5$ as a typical value defining the break-truncation boundary.

We define, then, a \textbf{break} on the surface brightness profile as a slope change (from $h_\mathrm{0}$ to $h_\mathrm{B}$), occurring at a radial distance $r_\mathrm{B}$, if the ratio between $r_\mathrm{B}/h_\mathrm{B}$ is less than 5. Conversely, \textbf{truncations} are changes in the slope ($h_\mathrm{T}$) of the surface brightness profiles happening at a radius $r_\mathrm{T}$, with a ratio $r_\mathrm{T}/h_\mathrm{T}$ greater than 5. Applying this recipe to the whole sample of galaxies, we can analyze separately both features and study whether they are actually different phenomena. We note that in \citet{seb12} they use a slightly different notation in which ``break'' describes all changes in slope, with Type II breaks being termed ``truncation'' and Type III breaks being called ``antitruncations''. Our galaxies have been traced out further than in most previous studies, and this leads to our recognition of a second break in some cases. Breaks with $r_\mathrm{B}/h_\mathrm{B} > 5$ are called truncation in keeping with previous nomenclature even if there is no clear evidence for a sharp end to the disk. In fact, most of our truncations are transitions to a far-outer region that may be as exponential as the inner regions, although with a steeper decline. Studies extending outer disk surface photometry to greater radii with star counts do not find a sudden drop-off either \citep{rich08,saha10,gross11,bar12,rad12}.

\begin{figure*}
\begin{center}
\includegraphics[width=400pt]{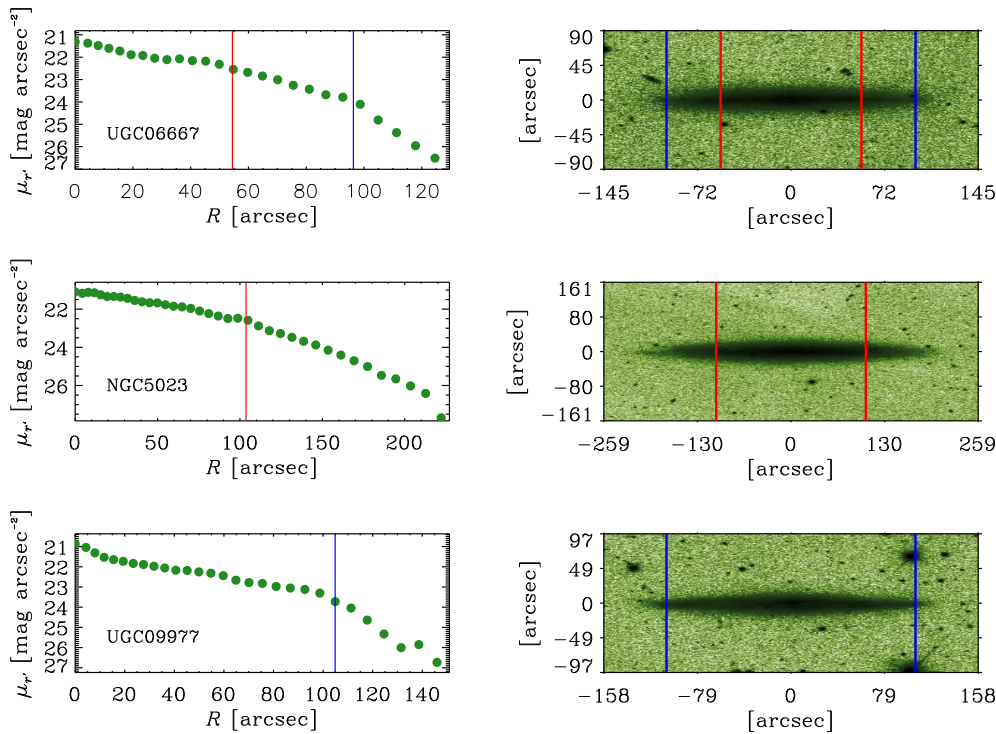}
\end{center}
\caption{Upper panel: UGC 6667 showing both a break (red line) and a truncation (blue line). Middle panel: The galaxy NGC 5023 with a break marked in red. There is a hint of a truncation around $r \sim 210$ arcsec but as there were not enough points above the surface brightness limit, we did not perform a fit over that region. Lower panel: Galaxy UGC 9977 where we only detected a truncation. The right column shows the position of the breaks and truncations over the galaxy images.}
 \label{img:types}
\end{figure*}

In Fig. \ref{img:types} we show three examples of a galaxy with break \textit{and} truncation (top), a galaxy where only the break has been detected (middle) and a galaxy with just a truncation (bottom) using our criteria. The red and blue lines mark the break and truncation radii (respectively) in both the surface brightness profiles and in the galaxy images. It its clear from Fig. \ref{img:types} that the break occurs closer to the galactic centre compared to the truncacion which appears near the very end of the optical disk. Also, the change in the exponential scalelength seems to be stronger after the truncation than after the break.

We show in Table \ref{tab:res} the summary of all the measurements and the derived quantities for breaks and truncations.

\subsection{Inner break analysis}
Looking at the relative frequencies of the different profile types around the inner break radius, 82$\pm$16\%\footnote{The errors are given by $\Delta T_\mathrm{i} = \sqrt{N_\mathrm{T_i}}/N_\mathrm{S}$ where $N_\mathrm{T_i}$ is the number of $\mathrm{T_i}$ profiles in a sample of $N_\mathrm{S}$ elements. } (28 cases) of our galaxies can be classified as TII (i.e., after the break there is a downbending on their surface brightness distribution). For the TI and TIII we found frequencies of 6$\pm$4\% (2 cases) and 12$\pm$6\% (4 cases) respectively. These values are different than those found by \citet{pt06} (10\%/60\%/30\% for TI/TII/TIII) but two factors must be taken into account. On one hand, having 34 galaxies, the statistical analysis is not as robust as in the case of \citet{pt06}. On the other hand, the study of \citet{pt06} includes morphological types from Sb to Sdm while our sample only collects Sc or later spiral types. It is well established \citep{pt06,guti11} that there exists a trend in the sense that TII becomes dominant in late-type galaxies. If the statistics of \citet{pt06} are re-calculated for the same morphological range as used in the current paper, we find that the relative frequencies become compatible (6$\pm$4\% vs 12$\pm$7\% for TI, 82$\pm$16\% vs 76$\pm$17\% for TII and 12$\pm$6\% vs 12$\pm$7\% for TIII)

The mean surface brightness value for the TII inner breaks is 22.5 $\pm$ 0.1 mag arcsec$^{-2}$ in the $r'$ band. For the others bands employed in the current paper, the mean surface brightness at the break radius are listed in Table \ref{tab:surf}. We find a mean radius equal to 7.9 $\pm$ 0.9 kpc which is in agreement with the result of \citet{pt06} who found a typical radius of 9 $\pm$ 3 kpc. The break scalelength $h_\mathrm{B}$ has a mean value of $2.7 \pm 0.3$ kpc.

\begin{figure*}
\begin{center}
\includegraphics[width=400pt]{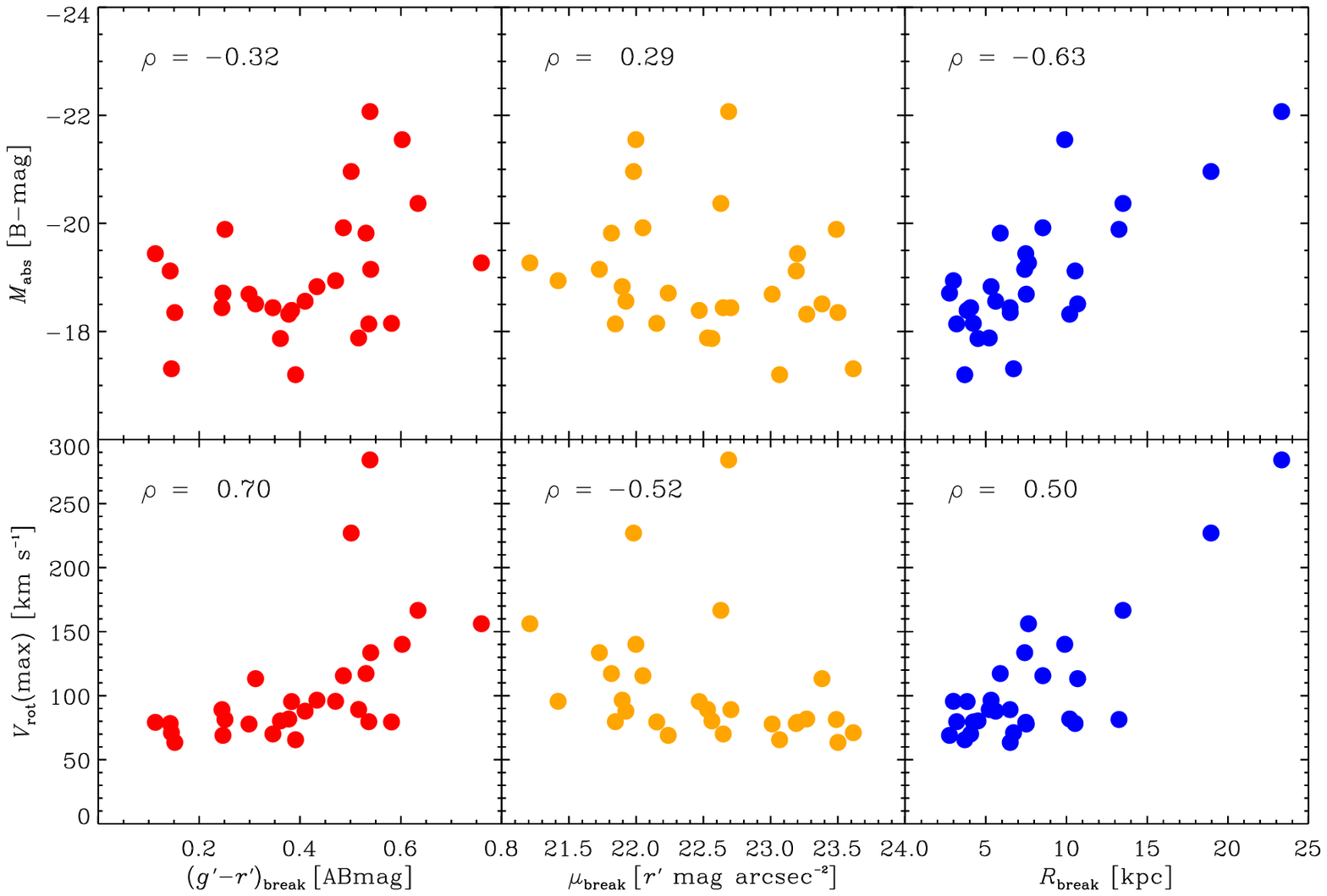} 
\end{center}
\caption{\textit{Breaks:} correlations between ($g'-r'$) (left), $\mu_{r'}$ (center) and $r_\mathrm{B}$ (right) with the $B$-band absolute magnitude $M_\mathrm{B}$ (top) and with the maximum rotational velocity (bottom). The Spearman's rank correlation coefficient ($\rho$) is over-plotted.}
\label{img:break}
\end{figure*}

Fig. \ref{img:break} shows the values for the $r'$-band surface brightness, the ($g'-r'$) color and the radial distance \{$\mu_{r'}$, ($g'-r'$), $r_\mathrm{B}$\} measured at the break radius (only TII profiles), plotted against the $M_\mathrm{B}$ and the maximum rotational velocity of the galaxy. In each plot it is over-plotted the Spearman's rank correlation coefficient\footnote{The Spearman's rank correlation coefficient varies between -1 and 1. The closer its absolute value is to one, the stronger the correlation.}. Significant correlations are found between the break radius and $M_\mathrm{B}$, and between the ($g'-r'$) color at the break and the maximum rotational velocity (top right and bottom left panels, respectively).This correlation between the break radius and $M_\mathrm{B}$, and the fact that our sample is limited to objects brighter than $M_\mathrm{B} = -17$ can explain why we marginally measure a mean break radius smaller than \citet{pt06}, whose galaxies were brighter than $M_\mathrm{B} = -18.4$

\subsection{Truncation analysis}
We found that all the profiles are downbending and steeper after the truncation than after the inner break, with a mean value for the scalelength $h_\mathrm{T}$ equal to $1.5 \pm 0.1$ kpc. The $r'$-band surface brightness has a mean value at the truncation radius of 24.1 $\pm$ 0.2 mag arcsec$^{-2}$ (see Table \ref{tab:surf} for the mean values in the other bands) and the averaged radial distance is $14 \pm 2$ kpc. In average, the 3$\sigma$ sky level in the $r'$-band is $\sim$ 3.5 mag arcsec$^{-2}$ dimmer than the typical surface brightness of the truncations so our results are not strongly affected by the sky subtraction process. The typical signal to noise ratio at the truncation radius is $\sim 10$ in the $r'$ band.

\begin{figure*}
\begin{center}
\includegraphics[width=400pt]{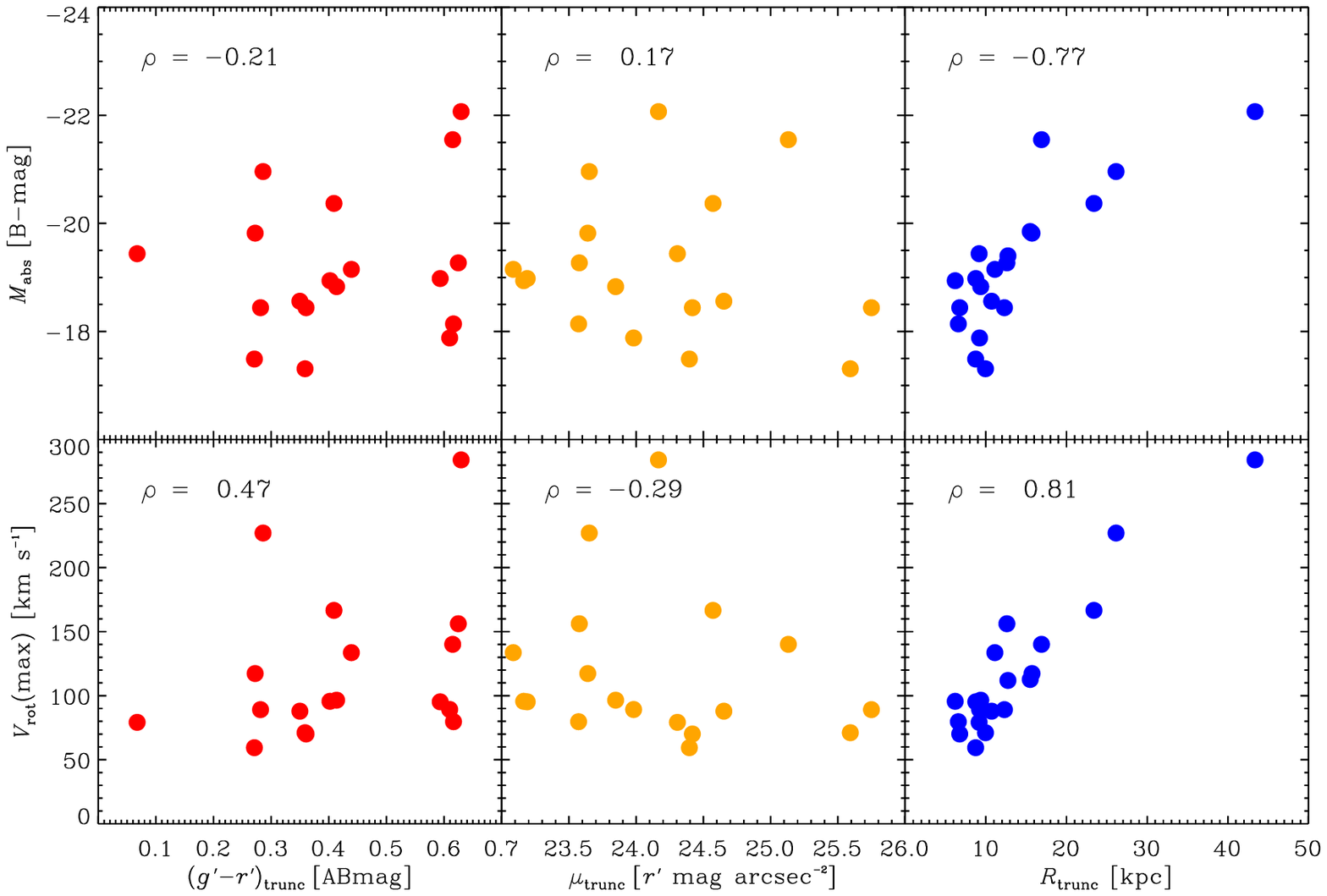} 
\end{center}
\caption{\textit{Truncations:} correlations between ($g'-r'$) (left), $\mu_{r'}$ (center) and $r_\mathrm{T}$ (right) with the $B$-band absolute magnitude $M_\mathrm{B}$ (top) and with the maximum rotational velocity (bottom). The Spearman's rank correlation coefficient ($\rho$) is over-plotted.}
\label{img:trun}
\end{figure*}

In the case of the truncations, we also looked for correlations between the same set of parameters as in Fig. \ref{img:break}, but measured obviously at the truncation radius. The results of the analysis are in Fig. \ref{img:trun}, where strong correlations appears between $M_\mathrm{B}$ and the maximum rotational velocity, and the radius where the truncation occurs.

It should be noted that the statistics derived in Figs.~\ref{img:break} and \ref{img:trun} are heavily influenced by a couple of points due to NGC~5907 and NGC~5529. The latter has a boxy/peanut bulge and a warped outer disk while NGC~5907 shows a bulge and a small warp too. If we exclude those two galaxies from the statistical analysis, the Spearman's rank correlation coefficient changes from 0.50 to 0.44 in the $V_\mathrm{rot}$(max) vs $r_\mathrm{B}$ relation and from 0.81 to 0.73 in the $V_\mathrm{rot}$(max) vs $r_\mathrm{T}$ relation.

\section{Discussion} \label{sec:discu}
\subsection{Comparison with previous studies}
Inner breaks and truncations, as defined in \S\ref{sec:classi}, have different characteristic parameters. Breaks appear closer to the galactic center compared to truncations ($r_\mathrm{B}\sim8$ kpc; $r_\mathrm{T}\sim14$~kpc) and the typical exponential scalelength is also smaller after the breaks than after the truncations ($h_\mathrm{B} \sim 3$ kpc; $h_\mathrm{T}\sim 1.5$~kpc). In addition, we have found that the ratio between the mean radius for breaks and truncations ($\langle r_\mathrm{T}\rangle  / \langle r_\mathrm{B}\rangle  =  1.8 \pm 0.5$ among our sample of galaxies) is very similar to the ratio $\langle r_\mathrm{EO}/h_0\rangle / \langle r_\mathrm{FO}/h_0\rangle  =  1.8 \pm 0.9$ between the break radii in edge-on galaxies by \citet{kruit87} and those in face-on galaxies by \citet{pt06}. Since the latter results are normalized to the innermost exponential scalelength, this quotient between edge-on and face-on measurements can be only fairly compared to our results if the innermost scalelength $h_0$ remains unchanged while looking at a galaxy in both projections. In this sense, \citet{poh07} found that their de-projected $h_0$ values match with the range given by \citet{pt06} if the influence of dust is taken into account.

It is possible, then, that the previous discrepancies found for the position of the breaks between edge-on and face-on studies can be explained as follows. Face-on studies usually measure the break radius but the truncation happens (due to the orientation) at very low surface brightness (typically $\sim$ 26.5 mag arcsec$^{-2}$), probably beyond the surface brightness limit. On the contrary, thanks to the line of sight integration, in the edge-on observations the truncation appears at higher surface brightness and it can be detected. The transition to a steeper decline after the break in the edge-on profiles can be smoothed because of this particular projection too, making it more difficult to detect this feature in edge-on galaxies than in the face-on counterpart and for this reason may have remained unnoticed in previous papers (but see \citealt{seb12}).

A natural prediction of our analysis is that truncations should be systematically observed in face-on projections if the images were deep enough. Just recently \citet{bak12} used very deep data from SDSS Stripe82 to explore the very faint regimes of disks in face-on projections by obtaining surface brightness profiles down to $\sim$ 30 $r'$-band mag arcsec$^{-2}$. Bakos \& Trujillo have not found any clear evidence for a truncation on these profiles but they have reported that the surface brightness profiles of the stellar disks show a smooth continuation into what they have identified as stellar halos. These stellar halos start to dominate the surface brightness profiles at $\sim$ 27.5. $r'$-band mag arcsec$^{-2}$ Coincidently, this means that observing truncations in face-on projections could be hindered due to the lack of contrast between the disk and the stellar halo components. The result of \citet{bak12} would, if confirmed, point out that our only opportunity of studying truncations in spiral galaxies could be limited to edge-on observations, unless an accurate parametrization and subtraction of the stellar halo component can be performed.

\subsection{\textit{Break}-\textit{truncation} scenario}
If breaks and truncations are actually two differentiated features, the two main theories (see \S\ref{sec:intro}) proposed to explain the break formation are not mutually exclusive, each playing a role at different distances from the galactic center. On the one hand, the averaged value found here for the break radius is compatible with the face-on picture of a break caused by a threshold for the star formation in the disk. On the other hand, we have found a mean truncation radius which is extremely well correlated with the maximum rotational velocity of the galaxy ($\rho = 0.81$). The above ideas are reinforced if we look at the different degree of correlations between the break/truncation radius and the maximum rotational velocity.

We can calculate the specific angular momentum of the disk to exemplify the differences between the two breaks by following the empirical expression given by \citet{nav00}:
\begin{displaymath}
j\mathrm{_{disk}} \approx 1.3 \times 10^3\left[\displaystyle\frac{V\mathrm{_{rot}(max)}}{200 \ \mathrm{km\cdot s^{-1}}}\right]^2 \mathrm{km \ s^{-1} \ h^{-1} \ kpc}
\end{displaymath}
Note that, following the equation above, the derived specific angular momentum of the disk is just a re-scaling of the maximum rotational velocity of the galaxy. Then, this specific angular momentum is showed as auxiliary information but it has not influence in further results.

\begin{figure}
\begin{center}
\includegraphics[width=217.2pt]{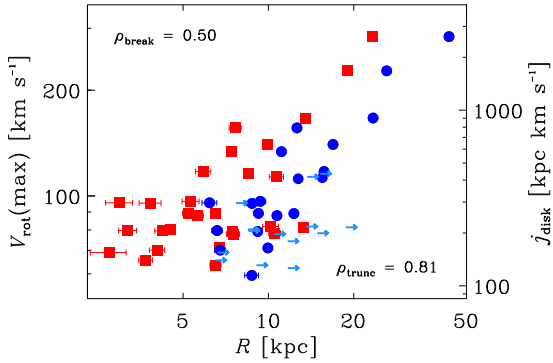} 
\end{center}
\caption{Correlations between the positions of the inner breaks (red squares) and  the radial position of the truncations (blue dots) versus the maximum rotational velocity and vs the specific angular momentum of the disk. Light blue arrows represent $r_\mathrm{max}$ as defined in \S\ref{sec:measu}. These arrows correspond to those galaxies where no truncation has been detected (so, they should be considered as a lower limit for the truncation radius). The Spearman's rank correlation coefficients for breaks (top left) and truncations (bottom right) are over-plotted.}
\label{img:corre}
\end{figure}

Fig. \ref{img:corre} shows how the break and truncation radii correlate with the maximum rotational velocity and with the specific angular momentum of the disk. The Spearman's rank coefficient reveals a very strong correlation ($\rho_\mathrm{trunc}=0.81$) between the maximum rotational velocity (and the specific angular momentum of the disk) and the truncation radius. On the contrary, the correlations between the same dynamical parameters and the break radius are significantly weaker ($\rho_\mathrm{break}=0.50$).

From Fig. \ref{img:corre} it is clear that the truncation radius correlates with the maximum rotational velocity for the whole range of values in the sample but the break radius correlates well only for those galaxies with $r_\mathrm{B} \gtrsim$ 8 kpc. Bellow this value, the break distance seems to be decoupled from the rotational velocity. In other words, for small disks, with $V_\mathrm{rot} < $  100 km s$^{-1}$, $r_\mathrm{B}$ and $V_\mathrm{rot}$ are not linked. The break radius seems to be strongly related to the maximum rotational velocity only when it happens in a galaxy with a significant angular momentum. This behavior supports the picture of truncations caused by the distribution of the galactic angular momentum and the break most likely related with a threshold in the gas density. Only when the rotational velocity is significant, the threshold in the star forming gas seems to be affected by the angular momentum distribution.

In the stellar surface mass density profile, the break has almost disappeared in many cases compared to the surface brightness profiles but the truncation barely changes (see Figs. \ref{img:4244} and \ref{img:5907}). In particular, we find a typical value for the ratio $\langle h_\mathrm{0}/h_\mathrm{B}\rangle _{\log \mathrm{\Sigma}}= 1.6 \pm 0.1$ at the break distance. Truncations on the contrary, show a ratio $\langle h_\mathrm{0}/h_\mathrm{T}\rangle _{\log \mathrm{\Sigma}}= 2.8 \pm 0.3$, associated with a quicker drop in the stellar surface mass density profile than in the break case. This is also in agreement with this \textit{break-truncation} scenario where a fast drop in the density of stars is expected after the truncation but not necessarily after the breaks, as was pointed out in \S\ref{sec:intro}.

However, two things must be noted here about the physical origin of breaks and truncations. First of all, finding a feature in a modern disk is not very clearly tied to cosmological accretion because the disk can adjust its mass distribution during its life \citep[e.g.][]{ros08}. Thus, the origin of truncations related to the angular momentum of the protogalactic cloud should be treated carefully. Also, regarding the formation of breaks, there are observational evidences that point to some level of star formation in the outskirts of spiral galaxies \citep{bush10,bar11}, arguing against a completely quenched star formation after the break radius.

Another important point regarding the break-truncation differentiation is that we have measured an averaged ratio between the scalelength after the breaks ($h_\mathrm{B}$) and after the truncations ($h_\mathrm{T}$) equal to $\langle h_\mathrm{B}/h_\mathrm{T} \rangle = 1.9 \pm 0.4$. This means that the change in the slope of the surface brightness profile is a factor of $\sim$ 2 stronger for truncations than for breaks. This result could explain why \citet{van81}, studying edge-on galaxies, found a sharp \textit{cut-off} (truncation) in their surface brightness profiles while face-on studies find a feature (break) which  corresponds to a softer \textit{break} in the radial light distribution (see Pohlen \& Trujillo 2006, \S6 in that paper).

The fact that the ratio $r_\mathrm{feature}/h_\mathrm{feature}$ is a good indicator to separate breaks from truncations (see Fig.\ref{img:space}) can be fully understood now. For the truncations, the numerator $r_\mathrm{feature}$ is larger and the denominator $h_\mathrm{feature}$ is smaller compared to the breaks. The ratio between both quantities has then, a significantly higher value for the truncations than for the breaks:

\begin{displaymath}
\left[\frac{r_\mathrm{B}(\downarrow)}{h_\mathrm{B}(\uparrow)}\right](\Downarrow) \ \leftrightsquigarrow \ \left[\frac{r_\mathrm{T}(\uparrow)}{h_\mathrm{T}(\downarrow)}\right] (\Uparrow) 
\end{displaymath}

We can conclude also that the edge-on orientation is better to study the outermost parts of galaxies because it allows us to reach the outskirts with a higher surface brightness than in the face-on case. A disadvantage of this edge-on projection is that the study of the color and stellar surface mass density profiles is less robust than in the face-on case due to the strong, yet measurable, influence of dust and line of sight integration problems (see \S\ref{sec:color}). As an example, we can compare Fig. 2 in the paper of \citet{bakos08} with Fig. \ref{img:break} and Fig. \ref{img:trun} (top-left panels) in this paper. Unlike \citet{bakos08}, we did not find any clear correlation between the ($g'-r'$) color at the break radius and the $M_\mathrm{B}$ of the galaxy.

To sum up, two differentiated features can be found in the light distribution of the disk in spiral galaxies: breaks and truncations. Face-on galaxies seem to be the perfect candidates to study breaks because both the surface brightness profile and the color profiles are available, with moderate dust influence. The advantage of integrating along the line of sight is crucial in the case of truncations, since a truncation is a feature that happens at very low surface brightness. However, the color profiles in the edge-on projection have to be interpreted with great care. Also, in the edge-on projection, it is not possible to classify the breaks depending on the morphology of the galaxy (see \citealt{pt06}, their section 4.1) unless a two-dimensional disk decomposition is made \citep{gad11}.

\section{Conclusions} \label{sec:summ}
Using SDSS and S$^4$G imaging, we have found the following important aspects regarding the behaviour of surface brightness profiles in edge-on late-type spirals:

\begin{enumerate}

\item The majority of our galaxies (82 $\pm$ 16\%) show a TII surface brightness profile as those found in photometric studies of face-on galaxies, with the break occuring at a mean radial distance from the galactic center equal to 7.9 $\pm$ 0.9 kpc.

\item Truncations, previously described in edge-on galaxies as a quick drop in the surface brightness profile, have been found in 20 of the 34 galaxies in our sample. This drop, also observed in the stellar surface mass density profile, occurs at an average radial distance of 14 $\pm$ 2 kpc.

\item For many galaxies, breaks and truncations coexist as two differentiated features in the light distribution of the disks in spiral galaxies.

\item Strong correlations exist between the truncation radius and the maximum rotational velocity, and the specific angular momentum of the disk. These correlations are, however, less strong in the case of breaks. This result reinforces the idea that breaks are more likely a phenomena related to a star formation threshold whereas truncations have a strong connection with the maximum angular momentum of the galaxy. 

\item Color and stellar surface mass density profiles are both very sensitive to the line of sight projection and to the presence of dust. Their interpretation in edge-on systems is not trivial and could be the cause of the differences between some of our results and those found in face-on works (e.g. correlations between $M_\mathrm{B}$ and break radius).

\end{enumerate}

\small{
	\footnotesize{\textit{Acknowledgments.}} We would like to thank Alexandre Vazdekis and Jes\'us Falc\'on-Barroso for their useful comments. We especially thank the referee, Prof. Piet van der Kruit, for providing very constructive and precise comments on this article.

This work has been supported by the Programa Nacional de Astronom\'ia y Astrof\'isica of the Spanish Ministry of Science and Innovation under grant AYA2010-21322-C03-02. We acknowledge financial support to the DAGAL network from the People Programme (Marie Curie Actions) of the European Union's Seventh Framework Programme FP7/2007-2013/ under REA grant agreement number PITN-GA-2011-289313. 

Funding for SDSS-III has been provided by the Alfred P. Sloan Foundation, the Participating Institutions, the National Science Foundation, and the U.S. Department of Energy Office of Science. The SDSS-III web site is http://www.sdss3.org/. SDSS-III is managed by the Astrophysical Research Consortium for the Participating Institutions of the SDSS-III Collaboration including the University of Arizona, the Brazilian Participation Group, Brookhaven National Laboratory, University of Cambridge, Carnegie Mellon University, University of Florida, the French Participation Group, the German Participation Group, Harvard University, the Instituto de Astrof\'isica de Canarias, the Michigan State/Notre Dame/JINA Participation Group, Johns Hopkins University, Lawrence Berkeley National Laboratory, Max Planck Institute for Astrophysics, Max Planck Institute for Extraterrestrial Physics, New Mexico State University, New York University, Ohio State University, Pennsylvania State University, University of Portsmouth, Princeton University, the Spanish Participation Group, University of Tokyo, University of Utah, Vanderbilt University, University of Virginia, University of Washington, and Yale University. 

This research has made use of NASA's Astrophysics Data System. We acknowledge the usage of the HyperLeda database and the  NASA/IPAC Extragalactic Database (NED), operated by the Jet Propulsion Laboratory, California Institute of Technology, under contract with the National Aeronautics and Space Administration.

}


\clearpage
\newpage

\appendix
\begin{center}
\section{Tables} \label{sec:tables} 
\end{center}

\begin{table*}
\centering
\begin{center}\begin{tabular}{l c c c c c}
\hline
\hline
Galaxy & \multicolumn{1}{c}{$M_{\mathrm{abs}}$} & Morph. & Morph. & \multicolumn{1}{c}{Distance} & \multicolumn{1}{c}{$V\mathrm{_{rot}(max)}$}\\
       & ($B$-mag) & type & code & (Mpc) & (km s$^{-1}$)             \\
\hline 
IC 1197     &  -18.56   &   Sc    &    6.0   &    25.7    &  87.9    \\
IC 2233     &  -19.44   &   SBc   &    6.4   &    13.4    &  79.2    \\
NGC 3279    &  -19.27   &   Sc    &    6.5   &    32.5    &  156.2   \\
NGC 3501    &  -19.15   &   Sc    &    5.9   &    22.9    &  133.6   \\
NGC 3592    &  -18.14   &   Sc    &    5.3   &    22.7    &  79.7    \\
NGC 4244    &  -18.44   &   Sc    &    6.1   &     4.6    &  89.1    \\
NGC 4330    &  -19.92   &   Sc    &    6.3   &    19.5    &  115.6   \\
NGC 4437    &  -21.55   &   Sc    &    6.0   &     9.8    &  140.1   \\
NGC 5023    &  -17.87   &   Sc    &    6.0   &     9.0    &  80.3    \\
NGC 5529    &  -22.07   &   Sc    &    5.3   &    49.5    &  284.1   \\
NGC 5907    &  -20.96   &   Sc    &    5.4   &    16.3    &  227.0   \\
PGC 029466  &  -18.71   &   Sc    &    6.2   &    44.8    &  69.0    \\
UGC 01839   &  -17.20   &   Scd   &    7.1   &    19.1    &  65.6    \\
UGC 01970   &  -18.98   &   Sc    &    5.9   &    33.9    &  95.2    \\
UGC 04725   &  -17.49   &   Sc    &    6.1   &    45.4    &  59.4    \\
UGC 04970   &  -18.51   &   Sc    &    5.7   &    56.7    &  113.3   \\
UGC 05347   &  -18.94   &   Scd   &    6.5   &    36.0    &  95.6    \\
UGC 06080   &  -18.69   &   Sc    &    6.5   &    36.3    &  77.9    \\
UGC 06509   &  -19.12   &   Scd   &    6.6   &    43.8    &  78.3    \\
UGC 06667   &  -17.88   &   Sc    &    5.9   &    19.8    &  89.2    \\
UGC 06791   &  -18.83   &   Sc    &    6.5   &    35.7    &  96.5    \\
UGC 06862   &  -18.32   &   Scd   &    6.7   &    62.2    &  81.8    \\
UGC 07153   &  -19.40   &   Sc    &    5.9   &    48.0    &  111.9   \\
UGC 07802   &  -18.44   &   Sc    &    6.1   &    23.2    &  70.1    \\
UGC 07991   &  -18.15   &   Scd   &    6.6   &    23.2    &  79.5    \\
UGC 08146   &  -17.31   &   Sc    &    6.4   &    18.2    &  71.1    \\
UGC 08166   &  -18.78   &   Sc    &    6.0   &    53.4    &  74.3    \\
UGC 09242   &  -19.89   &   Scd   &    6.6   &    24.0    &  81.4    \\
UGC 09249   &  -18.35   &   Scd   &    7.2   &    24.8    &  63.5    \\
UGC 09345   &  -18.39   &   Sc    &    6.4   &    36.6    &  95.4    \\
UGC 09760   &  -18.72   &   Scd   &    6.6   &    29.8    &  62.3    \\
UGC 09977   &  -19.85   &   Sc    &    5.3   &    30.5    &  112.7   \\
UGC 10288   &  -20.37   &   Sc    &    5.3   &    31.7    &  166.6   \\
UGC 12281   &  -19.82   &   Sd    &    7.5   &    33.9    &  117.3   \\
\hline 
\end{tabular}
\caption{Fundamental parameters of the final sample. The distances were collected from the NED database meanwhile the $B$-band absolute magnitude (corrected from extinction), the morphological type, the morphological code (T-type) and the maximum rotational velocity were obtained from the HyperLeda database. \label{tab:data}} 
\end{center}
\end{table*}

\begin{table*}
\centering
\begin{center}\begin{tabular}{l c c l c c}
\hline
\hline
Galaxy & \multicolumn{1}{c}{PA} & \multicolumn{1}{c}{PA} & Galaxy & \multicolumn{1}{c}{PA} & \multicolumn{1}{c}{PA}\\ 
& (HyperLeda) & (This paper) &  & (HyperLeda) & (This paper) \\
\hline
IC 1197    &    56.6    &    56.3   &       UGC 06080  &    129.4  &    127.2   \\
IC 2233    &    172.3  &    171.5   &       UGC 06509   &    79.0      &    78.9  \\ 
NGC 3279   &    152.0  &    152.0   &       UGC 06667   &    87.2   &    87.2  \\  
NGC 3501   &    28.0      &    27.1   &     UGC 06791   &    0.5    &    0.8  \\  
NGC 3592   &    117.7  &    116.5   &         UGC 06862   &    103.7  &    103.5  \\ 
NGC 4244   &    42.2   &    47.2   &        UGC 07153   &    165.8  &    166.3  \\ 
NGC 4330   &    60.0      &  57.9   &        UGC 07802   &    56.4   &    56.1  \\ 
NGC 4437   &    85.7   &    82.3   &        UGC 07991   &    172.6  &    170.5  \\ 
NGC 5023   &    28.3   &    27.8   &        UGC 08146   &    30.6   &    30.5  \\ 
NGC 5529   &    114.3  &    113.4   &        UGC 08166   &    156.5  &    156.3  \\ 
NGC 5907   &    155.5   &    154.5   &        UGC 09242   &    70.8   &    71.1  \\ 
PGC 029466 &    179.9  &    180.0   &        UGC 09249   &    85.5   &    185.5 \\ 
UGC 01839  &    43.5    &    45.8   &        UGC 09345   &    141.7  &    140.4  \\ 
UGC 01970  &    22.7   &    22.6   &        UGC 09760   &    61.2    &    55.0  \\ 
UGC 04725  &    66.2   &    67.3   &        UGC 09977   &    80.4   &    77.4  \\ 
UGC 04970  &    104.6  &    104.8   &        UGC 10288   &    90.2    &    90.0  \\ 
UGC 05347  &    17.0   &    18.2   &        UGC 12281   &    30.3   &    29.7  \\   
\hline 
\end{tabular}
\caption{Position angles (in degrees) calculated in this paper and from HyperLeda for our sample of galaxies. The PA is measured from the North to the East (between 0\degr \ and 180\degr). The mean difference between our measurements and those from HyperLeda is 0.7\degr. \label{tab:rota}} 
\end{center}
\end{table*}

\begin{table*}
\centering
\begin{center}\begin{tabular}{lcccccc}
\hline
\hline
Galaxy & \multicolumn{1}{c}{$r_1$} & \multicolumn{1}{c}{$r_2$} & $r_3$ & \multicolumn{1}{c}{$r_4$} & \multicolumn{1}{c}{$r_5$} & \multicolumn{1}{c}{$r_6$} \\ 
    & (arcsec) & (arcsec) & (arcsec) & (arcsec) & (arcsec) & (arcsec) \\ 
\hline
IC 1197    &   0.0  &   40.6  &   49.9  &   86.8  &   86.8  &   98.7  \\
IC 2233    &   0.0  &  111.3  &  111.3  &  138.6  &  145.9  &  169.2  \\
NGC 3279   &  11.7  &   45.2  &   49.9  &   86.8  &   86.8  &  104.9  \\
NGC 3501   &  11.7  &   64.8  &   64.8  &  104.9  &  104.9  &  131.5  \\
NGC 3592   &   0.0  &   40.6  &   31.8  &   64.8  &   64.8  &   81.1  \\
NGC 4244   &   0.0  &  296.3  &  308.1  &  552.4  &  552.4  &  612.2  \\
NGC 4330   &   0.0  &   98.7  &   92.6  &  161.2  &  -  &  -  \\
NGC 4437   &   0.0  &  203.5  &  212.6  &  372.8  &  372.8  &  416.3  \\
NGC 5023   &   0.0  &  111.3  &  111.3  &  212.6  &  -  &  -  \\
NGC 5529   &   8.0  &   98.7  &   92.6  &  177.4  &  177.4  &  194.5  \\
NGC 5907   &  19.4  &  222.1  &  222.1  &  320.3  &  320.3  &  401.4  \\
PGC 029466 &   0.0  &   15.5  &   19.4  &   36.1  &   -  &   -  \\
UGC 01839  &   4.4  &   45.2  &   40.6  &   64.8  &   -  &   -  \\
UGC 01970  &   4.4  &   27.5  &   27.5  &   49.9  &   54.7  &   70.1  \\
UGC 04725  &   0.0  &   19.4  &   19.4  &   36.1  &   40.6  &   49.9  \\
UGC 04970  &   0.0  &   40.6  &   40.6  &   54.7  &   -  &   -  \\
UGC 05347  &   0.0  &   23.4  &   19.4  &   36.1  &   40.6  &   54.7  \\
UGC 06080  &   0.0  &   40.6  &   45.2  &   59.7  &   -  &   -  \\
UGC 06509  &   0.0  &   49.9  &   49.9  &   75.5  &   -  &   -  \\
UGC 06667  &   0.0  &   45.2  &   45.2  &   92.6  &   98.7  &  131.5  \\
UGC 06791  &   0.0  &   31.8  &   40.6  &   54.7  &   54.7  &   64.8  \\
UGC 06862  &   4.4  &   31.8  &   36.1  &   49.9  &   -  &   -  \\
UGC 07153  &   0.0  &   49.9  &   -  &   -  &   54.7  &   70.1  \\
UGC 07802  &   4.4  &   40.6  &   40.6  &   59.7  &   59.7  &   75.5  \\
UGC 07991  &   4.4  &   31.8  &   40.6  &   75.5  &   -  &   -  \\
UGC 08146  &   0.0  &   70.1  &   75.5  &  111.3  &  111.3  &  124.6  \\
UGC 08166  &   0.0  &   27.5  &   27.5  &   40.6  &   -  &   -  \\
UGC 09242  &   0.0  &  104.9  &  124.6  &  177.4  &  -  &  -  \\
UGC 09249  &   4.4  &   49.9  &   54.7  &   75.5  &   -  &   -  \\
UGC 09345  &   8.0  &   27.5  &   31.8  &   49.9  &   -  &   -  \\
UGC 09760  &   0.0  &   49.9  &   64.8  &   81.2  &   -  &   -  \\
UGC 09977  &  11.7  &   98.7  &  -  &  -  &  111.3  &  131.5  \\
UGC 10288  &  11.7  &   92.6  &   98.7  &  138.6  &  153.5  &  169.2  \\
UGC 12281  &   0.0  &   40.6  &   40.6  &   98.7  &   98.7  &  111.3  \\ 
\hline 
\end{tabular}
\caption{Boundaries enclosing the different regions in each galaxy. The $[r_1,r_2]$ interval marks the limits between the galactic centre and the break, $[r_3,r_4]$ the limits between the break and the truncation and $[r_5,r_6]$ between the truncation and the end of the brightness profile.\label{tab:regi}}
\end{center}
\end{table*}

\begin{table*}
\centering
\begin{center}\begin{tabular}{lccccccccc}
\hline
\hline 
\multicolumn{1}{c}{Galaxy} & \multicolumn{1}{c}{Type} & \multicolumn{2}{c}{$r_\mathrm{B}$} &  \multicolumn{2}{c}{$r_\mathrm{T}$}  & \multicolumn{1}{c}{$\mu_\mathrm{B}$}& \multicolumn{1}{c}{$\mu_\mathrm{T}$}  & \multicolumn{1}{c}{($g'-r'$)$_\mathrm{B}$}  & \multicolumn{1}{c}{($g'-r'$)$_\mathrm{T}$} \\
	&   &  {(kpc)} &  (arcsec) & {(kpc)}  &  (arcsec) & {$\left(\frac{r'\mathrm{mag}}{\mathrm{arcsec^2}}\right)$} &   {$\left(\frac{r'\mathrm{mag}}{\mathrm{arcsec^2}}\right)$} &  {(ABmag)}  &  {(ABmag)} \\
\hline
IC 1197    & II &  5.6  &  45.0 & 10.7  &  86.0  & 21.9    & 24.7    &  0.41    &  0.35       \\
IC 2233    & II &  7.48 & 115.5  &  9.17 & 141.7  & 23.20    & 24.31    &  0.11    &  0.07       \\
NGC 3279   & II &   7.64   & 48.4 & 12.61 & 80.0 &  21.21   &  23.58   &   0.76   &   0.62      \\
NGC 3501   & II &   7.41 & 66.8  &  11.12 & 100.2 &  21.73   &  23.09   &   0.54   &   0.44      \\
NGC 3592   & II &   3.19   & 29.0 &   6.59 & 59.8 &  21.84   &  23.57   &   0.54   &   0.62      \\
NGC 4244   & II &   6.50 & 289.5 &  12.30 & 548.1 &  22.70   &  25.75   &   0.25   &   0.28      \\
NGC 4330   & II &   8.53 & 90.2 &  \textit{nd} &  \textit{nd}  &  22.05   &  \textit{nd}   &   0.49   &   1.32         \\
NGC 4437   & II &   9.89   & 209.0 & 16.90 & 357.2 &  22.00   &  25.13   &   0.60   &   0.61      \\
NGC 5023   & II &   4.51 & 103.7 &   \textit{nd} &   \textit{nd}   &  22.56   &  \textit{nd}   &   0.36   &   0.49         \\
NGC 5529   & II &  23.33 & 97.2 &  43.37 & 180.6 &  22.69   &  24.17   &   0.54   &   0.63      \\
NGC 5907   & II &  18.95 & 237.2 &  26.14 & 327.2 &  21.98   &  23.65   &   0.50   &   0.29      \\
PGC 029466 & II &     2.75 & 13.2 &  \textit{nd} &  \textit{nd}     &    22.24 &    \textit{nd} &     0.25 &     0.52    \\
UGC 01839  & II &    3.70  &  39.9 &  \textit{nd} &  \textit{nd} &   23.07  &   \textit{nd}  &    0.39  &    0.23        \\
UGC 01970  & III &    3.89  &  23.7 &  8.75 & 53.3 &   21.94  &   23.19  &    0.74  &    0.59    \\
UGC 04725  & III &    4.23  & 19.2 &  8.74 & 39.7 &   23.13  &   24.40  &    0.31  &    0.27    \\
UGC 04970  & II &   10.69  &  38.9 &  \textit{nd}  & \textit{nd}  &   23.38  &   \textit{nd}  &    0.31  &    0.92        \\
UGC 05347  & II &    2.99  & 17.9 &  6.20  & 37.1 &  21.42  &   23.16  &    0.47  &    0.40     \\
UGC 06080  & II &    7.51  & 42.7 &   \textit{nd} &   \textit{nd} &   23.01  &   \textit{nd}  &    0.30  &    0.38        \\
UGC 06509  & II &   10.53  & 49.7 &   \textit{nd} &   \textit{nd}  &   23.19  &   \textit{nd}  &    0.14  &    0.32        \\
UGC 06667  & II &    5.21  & 54.4 &   9.22  & 96.3 & 22.53  &   23.98  &    0.52  &    0.61     \\
UGC 06791  & II &    5.33  & 30.8 &   9.37  & 54.1 & 21.90  &   23.85  &    0.43  &    0.41     \\
UGC 06862  & II &   10.20  &  33.8 &  \textit{nd} &  \textit{nd} &   23.27  &   \textit{nd}  &    0.38  &    0.71        \\
UGC 07153  & I &   - & - & 12.74 & 54.8 &   -  &   23.49  &    0.84  &    0.23          \\
UGC 07802  & II &    4.06  & 36.1 &   6.76  & 60.2 & 22.65  &   24.42  &    0.35  &    0.36     \\
UGC 07991  & II &    4.22  &  37.5 &  \textit{nd} &  \textit{nd} &   22.15  &   \textit{nd}  &    0.58  &    0.43        \\
UGC 08146  & II &    6.72  & 76.1 &   9.96 & 112.7 &   23.61  &   25.59  &    0.14  &    0.36     \\
UGC 08166  & III &    6.85  & 26.4 &   \textit{nd} &   \textit{nd} &   23.89  &   \textit{nd}  &    0.25  &    0.14       \\
UGC 09242  & II &   13.25  &  113.9 &  \textit{nd} &  \textit{nd} &   23.49  &   \textit{nd}  &    0.25  &    0.54        \\
UGC 09249  & II &    6.51  & 54.2 &   \textit{nd} &   \textit{nd}  &   23.50  &   \textit{nd}  &    0.15  &    0.15        \\
UGC 09345  & II &    3.84  &  22.6 &  \textit{nd} &  \textit{nd}  &   22.47  &   \textit{nd}  &    0.38  &    0.40        \\
UGC 09760  & III &    8.26  &  57.2 &  \textit{nd} &  \textit{nd}  &   23.77  &   \textit{nd}  &    0.25  &    0.28       \\
UGC 09977  & I &   -  & - &  15.50 & 104.8 &   -  &   23.28  &    1.57  &    0.43         \\
UGC 10288  & II &   13.50  & 87.7 &  23.40 & 152.1 &   22.63  &   24.57  &    0.63  &    0.41     \\
UGC 12281  & II &    5.89  & 35.9 &  15.72 & 95.8 &   21.82  &   23.64  &    0.53  &    0.27     \\
\hline 
\end{tabular}
\caption{Types of profiles and measured quantities at the break distance (B subscript) and at the truncation radius (T subscript). \textit{nd} represents a non-detected truncation. The typical error for the ($g'-r'$) color profile is 0.10 AB mag at the break radius and 0.22 AB mag at the truncation radius.\label{tab:res}}
\end{center}
\end{table*}

\begin{table*}
\centering
\begin{center}\begin{tabular}{ccccc}
\hline
\hline 
   {Band} & \multicolumn{1}{c}{Break} & \multicolumn{1}{c}{$\pm \Delta$} &  {Truncation} & \multicolumn{1}{c}{$\pm \Delta$}\\ 
   {} &  {$\left(\frac{\mathrm{mag}}{\mathrm{arcsec^2}}\right)$} &  {$\left(\frac{\mathrm{mag}}{\mathrm{arcsec^2}}\right)$} &  {$\left(\frac{\mathrm{mag}}{\mathrm{arcsec^2}}\right)$} &  {$\left(\frac{\mathrm{mag}}{\mathrm{arcsec^2}}\right)$} \\
\hline
$u'$    &    24.0    &    0.1   &    25.7  &    0.2   \\
$g'$    &    22.9  &    0.1    &    24.6      &   0.2  \\ 
$r'$    &    22.5  &    0.1    &    24.2   &    0.2  \\  
$i'$   &    22.3      &  0.2 &    23.9    &    0.2  \\  
$z'$    &    22.1  &    0.2    &    23.7  &    0.2  \\ 
$3.6 \mu m$ &    22.7   &   0.2   &    24.5  &   0.2  \\ 
\hline 
\end{tabular}
\caption{Mean surface brightness at the break (second column) and at the truncation (fourth column) radius in the six photometric bands employed in this paper.\label{tab:surf}}
\end{center}
\end{table*}

\clearpage
\newpage

\section{The sample} \label{sec:sample}
In \S\ref{sec:pres} is detailed the information about how the results are presented in the following atlas.

\begin{figure*}
\begin{center}
\includegraphics[width=400pt]{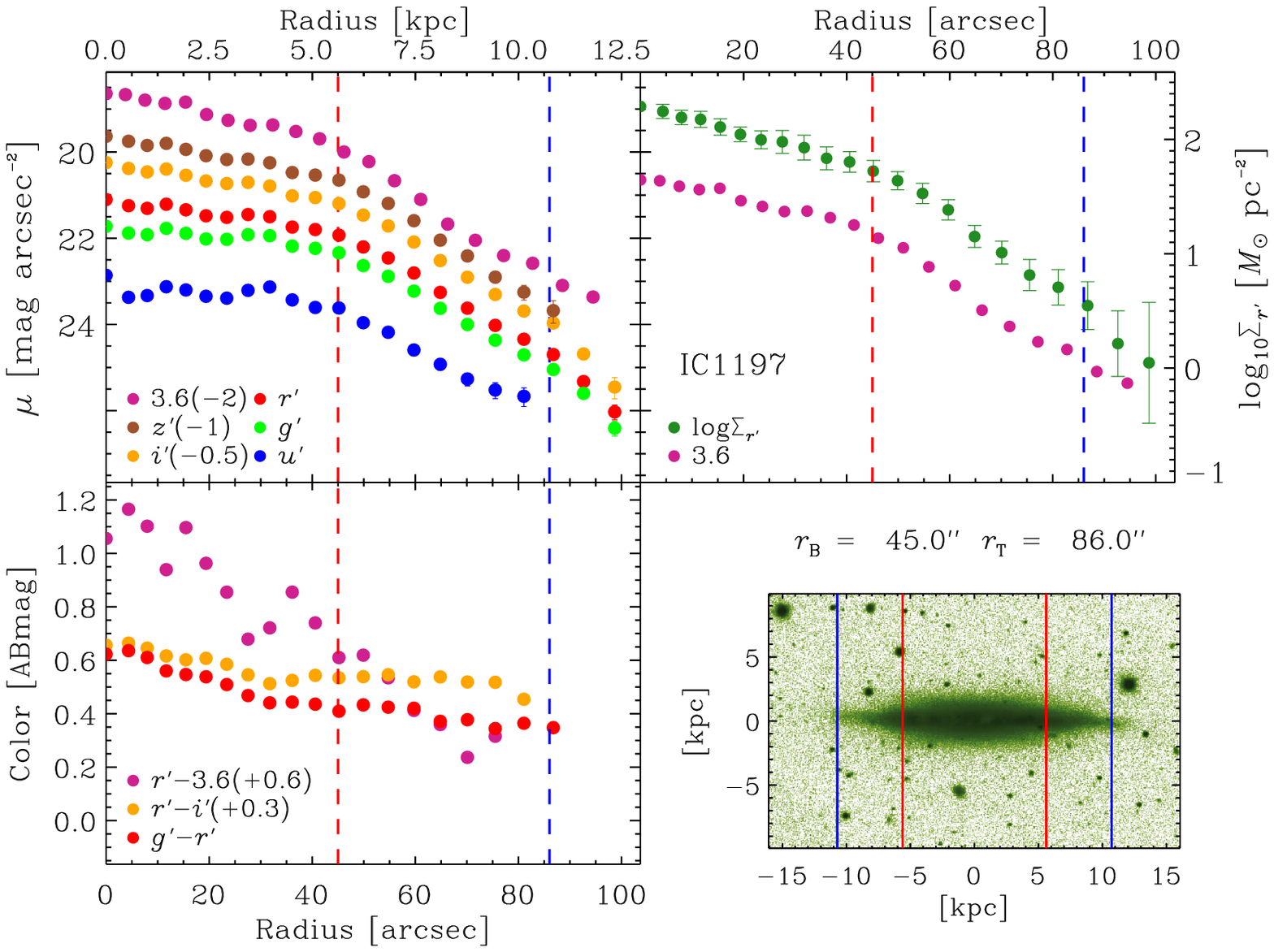}
 \end{center}
\end{figure*}

\begin{figure*}
\begin{center}
\includegraphics[width=400pt]{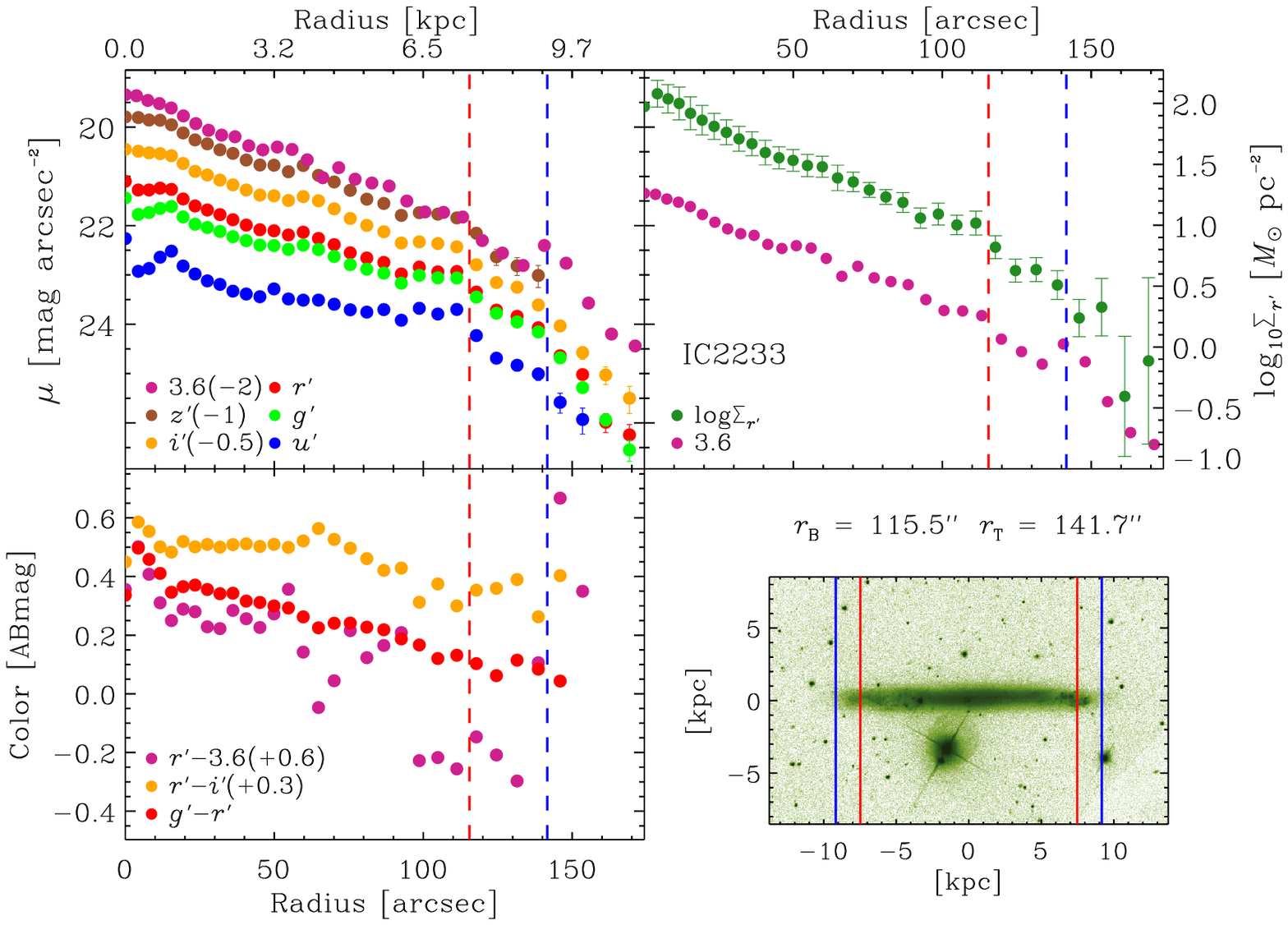}
\end{center}
\end{figure*}

\begin{figure*}
\begin{center}
\includegraphics[width=400pt]{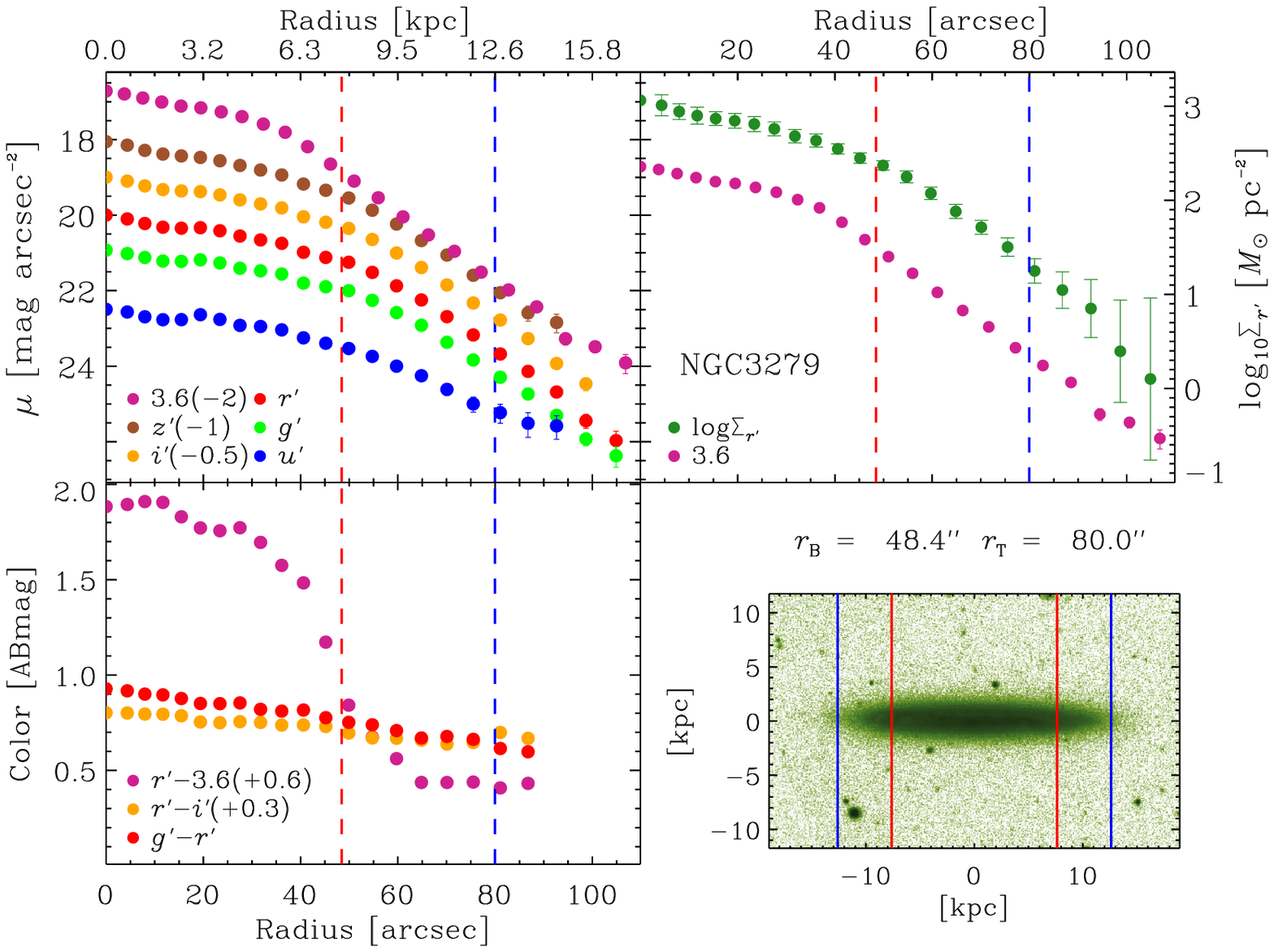}
\end{center}
\end{figure*}

\begin{figure*}
\begin{center}
\includegraphics[width=400pt]{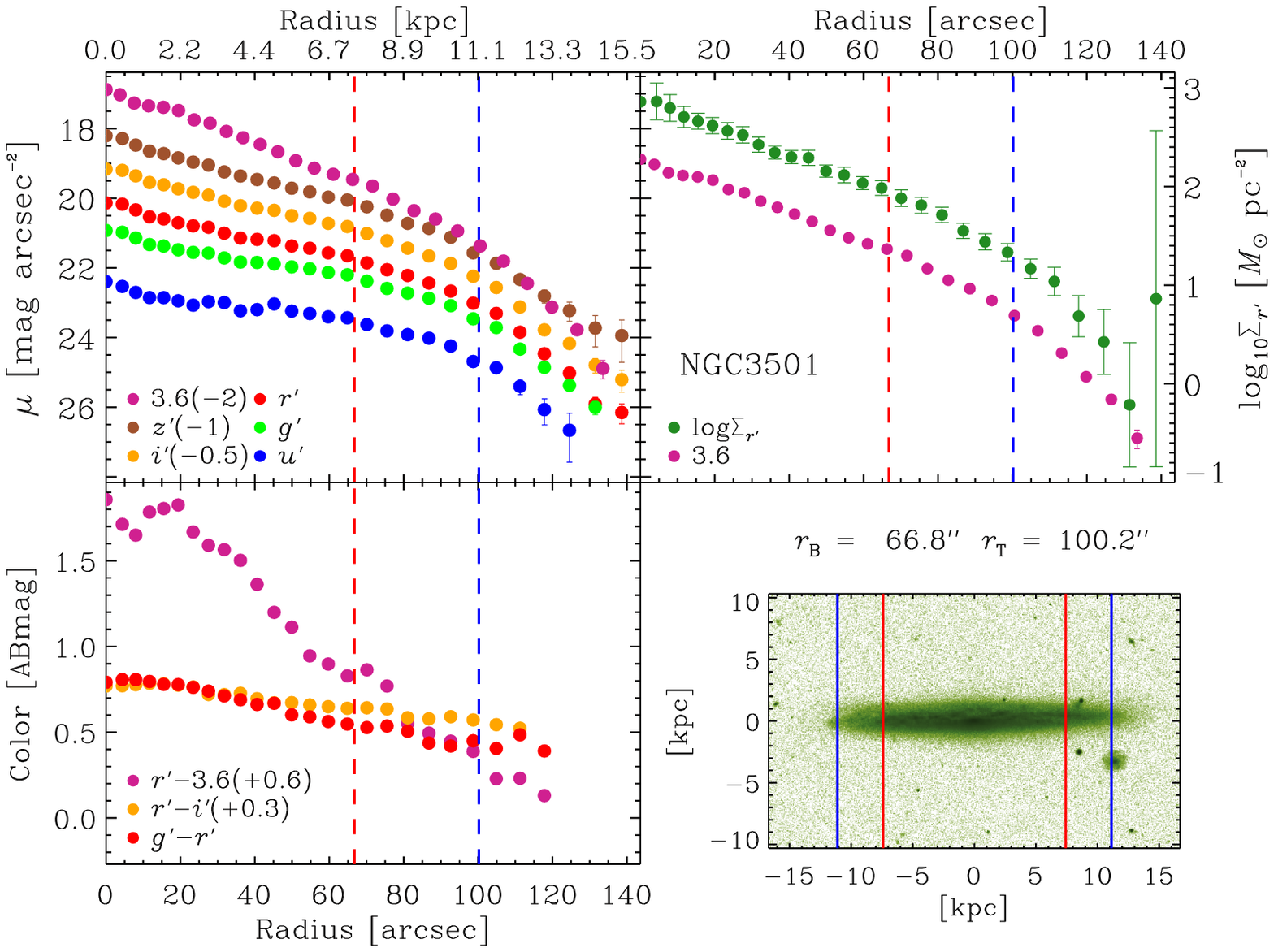}
\end{center}
\end{figure*}

\begin{figure*}
\begin{center}
\includegraphics[width=400pt]{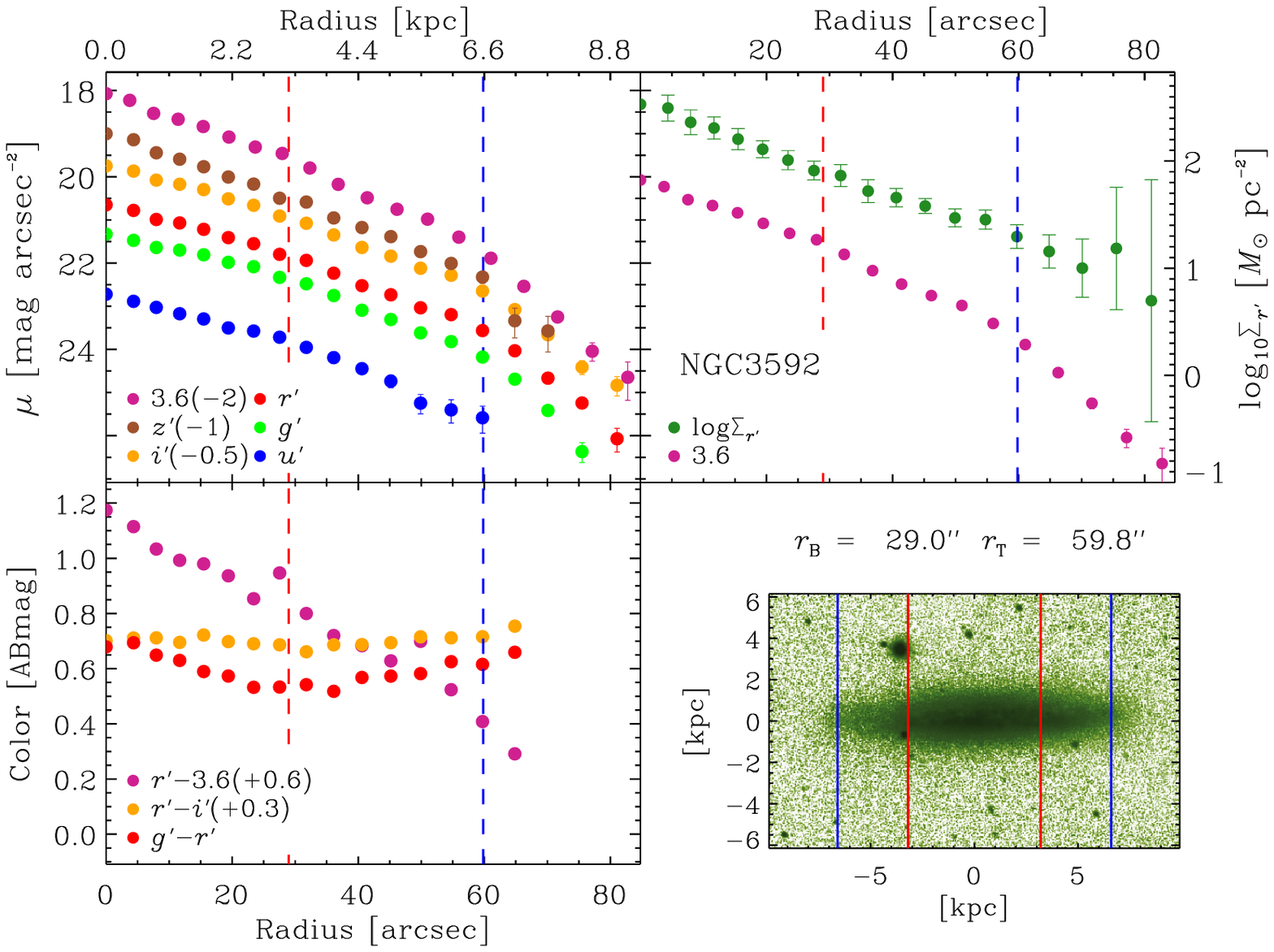}
\end{center}
\end{figure*}

\begin{figure*}
\begin{center}
\includegraphics[width=400pt]{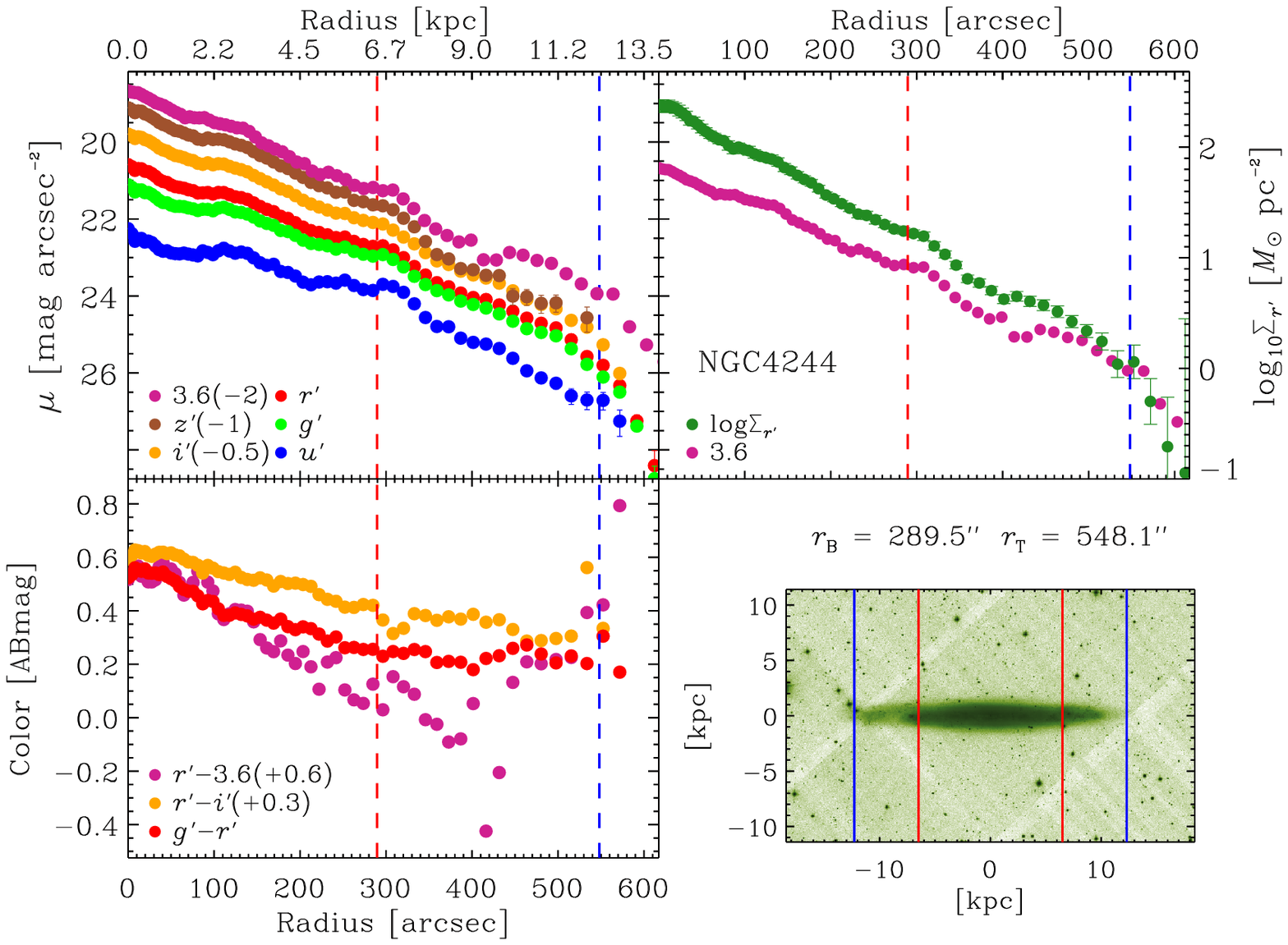}
\end{center}
\end{figure*}

\begin{figure*}
\begin{center}
\includegraphics[width=400pt]{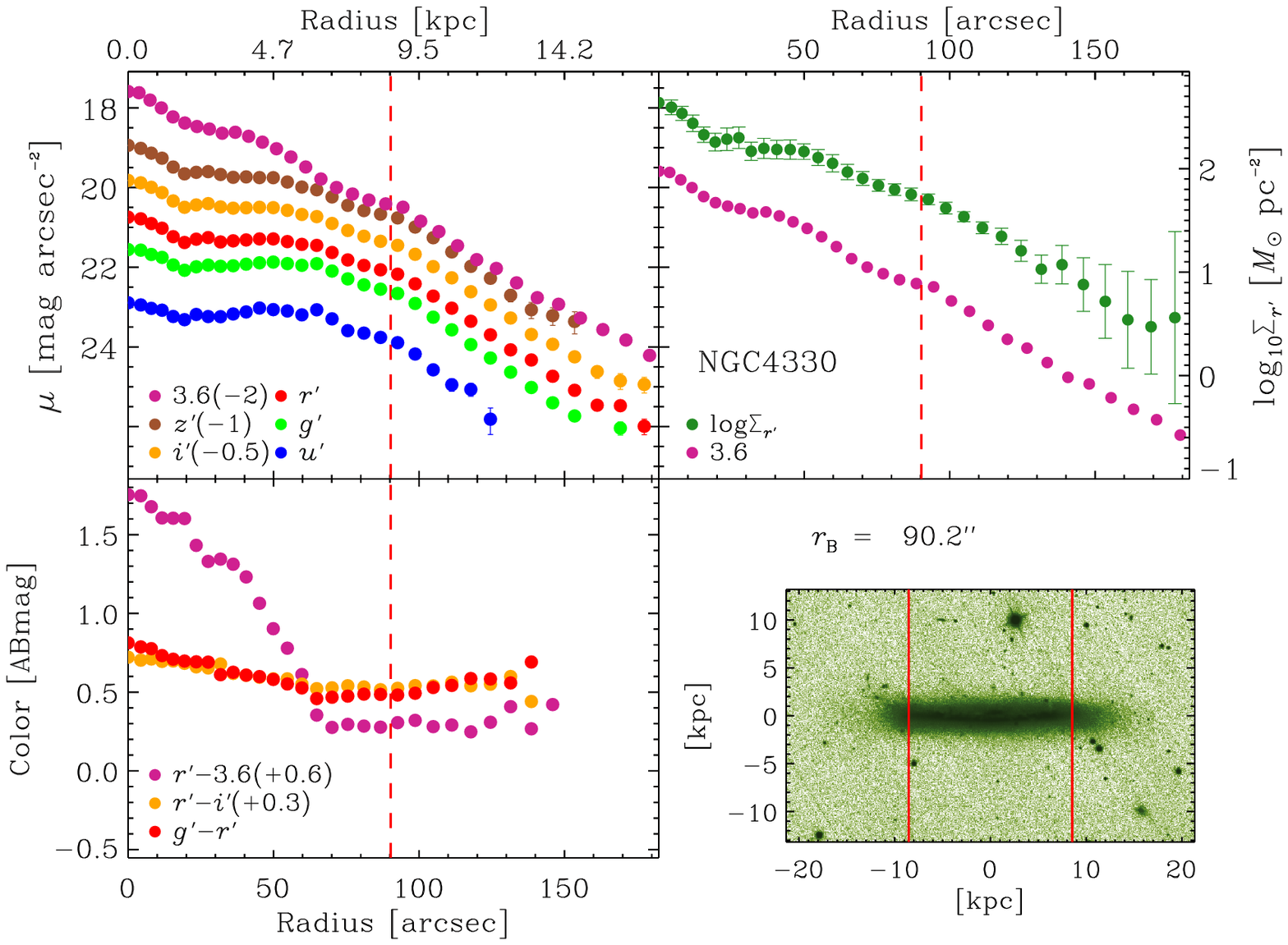}
\end{center}
\end{figure*}

\begin{figure*}
\begin{center}
\includegraphics[width=400pt]{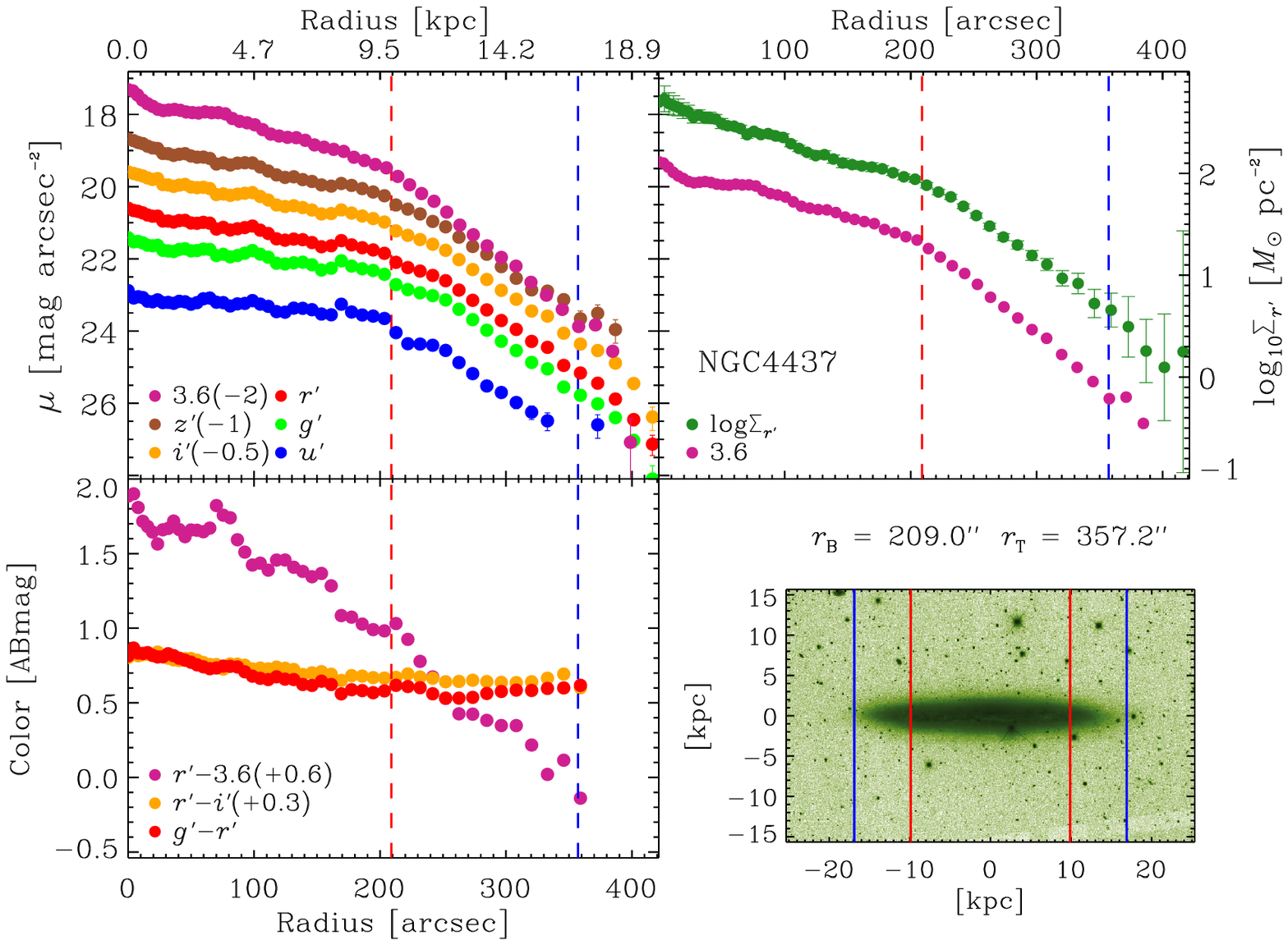}
\end{center}
\end{figure*}

\begin{figure*}
\begin{center}
\includegraphics[width=400pt]{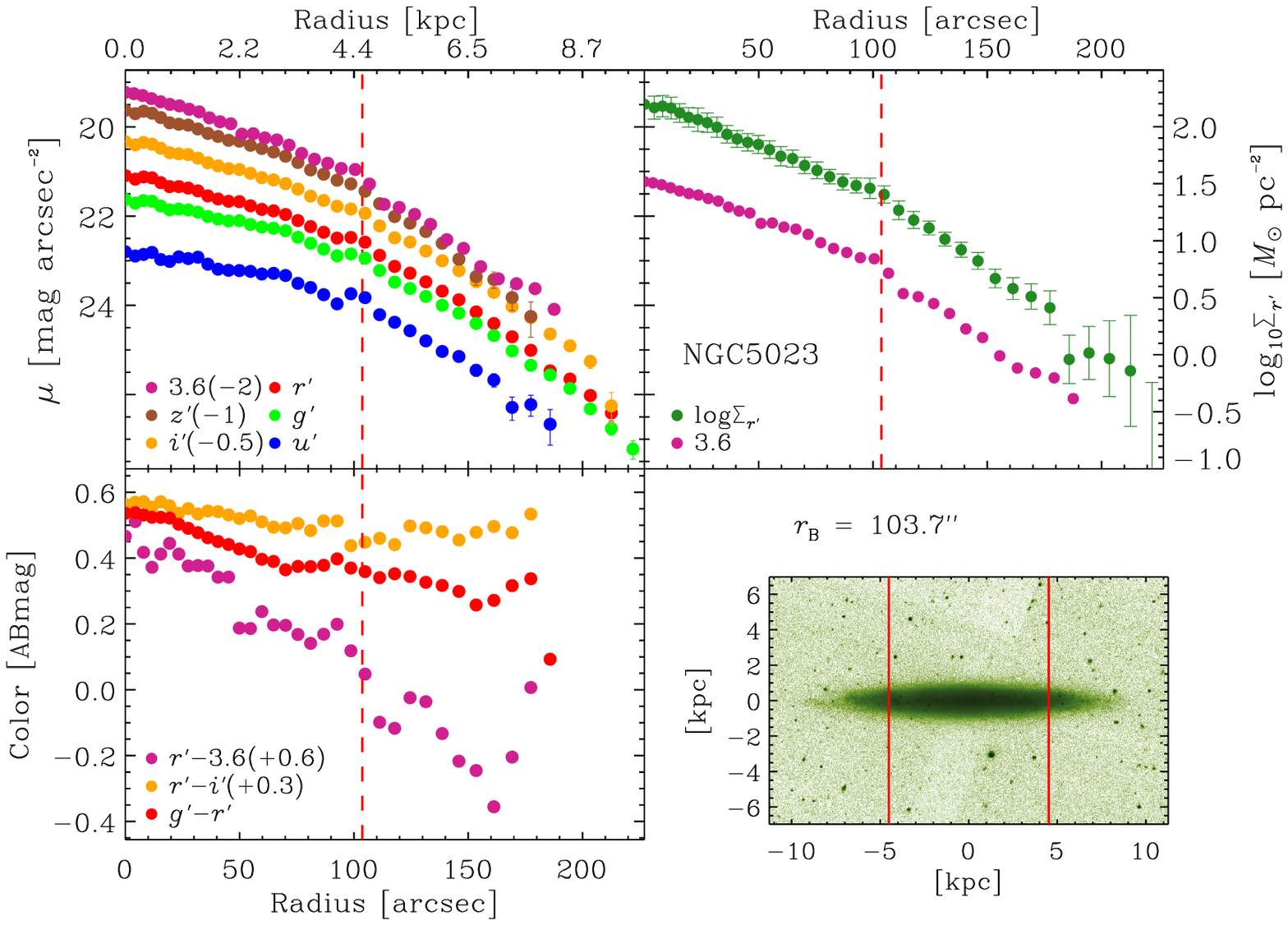}
\end{center}
\end{figure*}

\begin{figure*}
\begin{center}
\includegraphics[width=400pt]{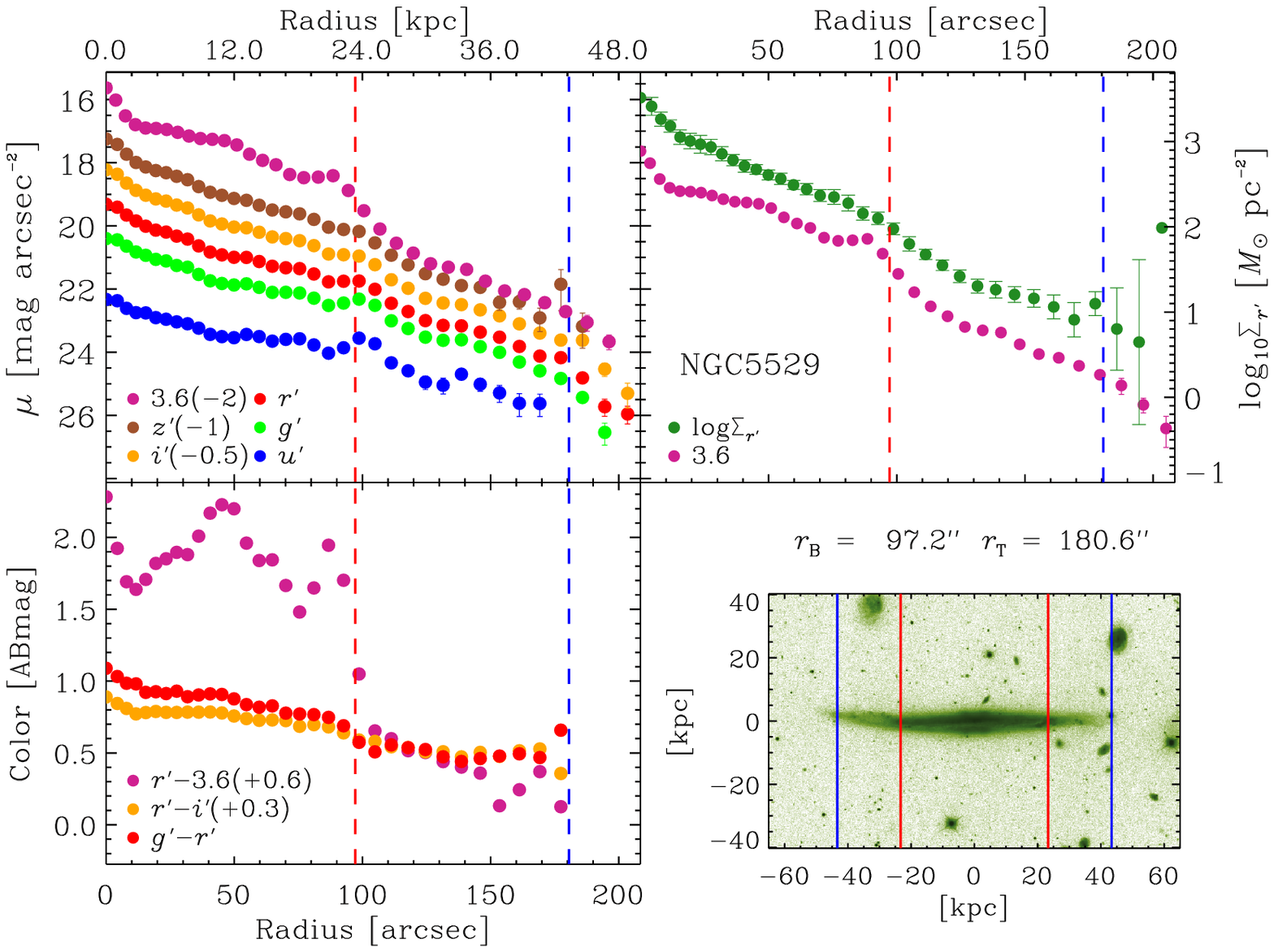}
\end{center}
\end{figure*}

\begin{figure*}
\begin{center}
\includegraphics[width=400pt]{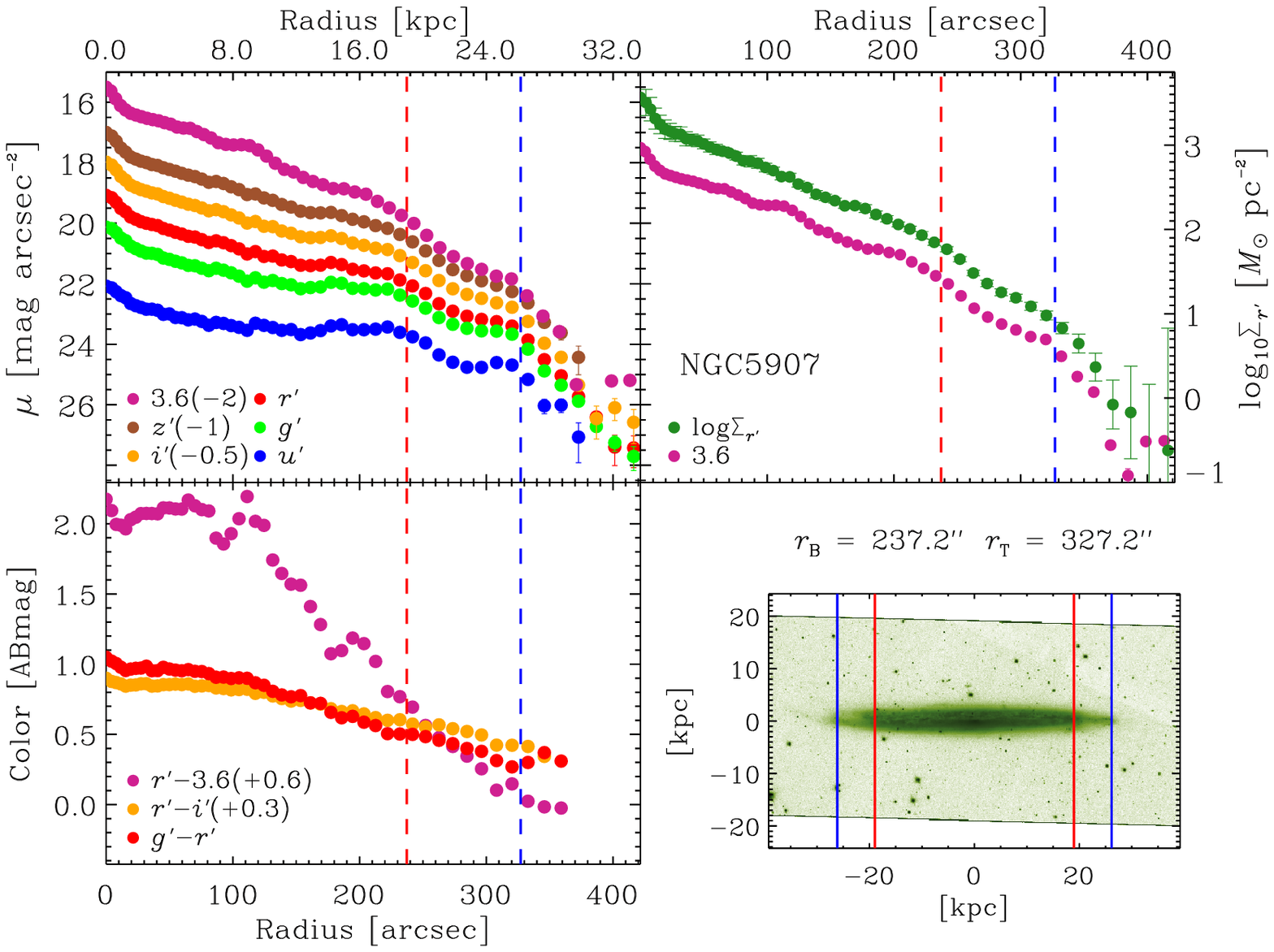}
\end{center}
\end{figure*}

\begin{figure*}
\begin{center}
\includegraphics[width=400pt]{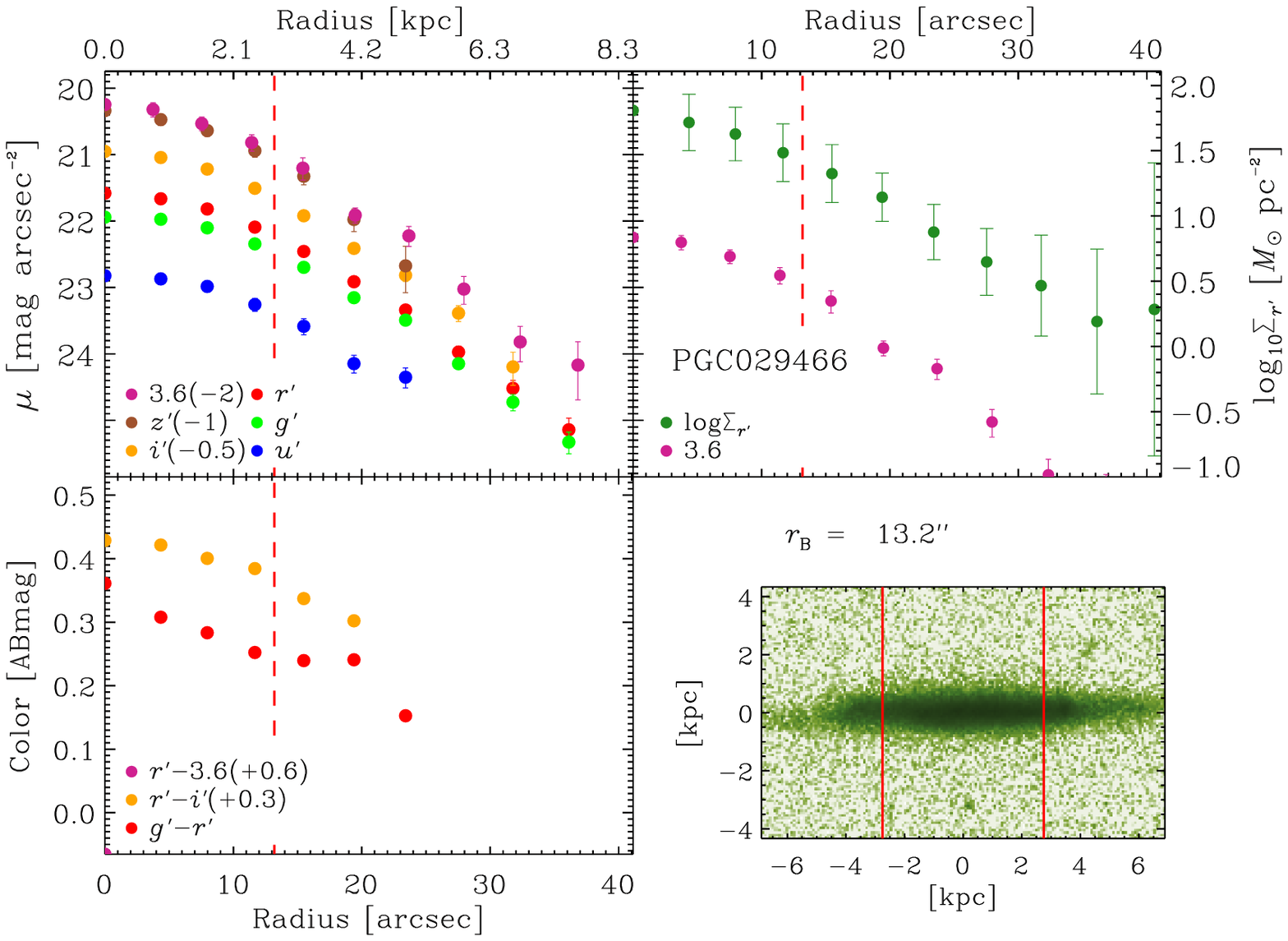}
\end{center}
\end{figure*}

\clearpage

\begin{figure*}
\begin{center}
\includegraphics[width=400pt]{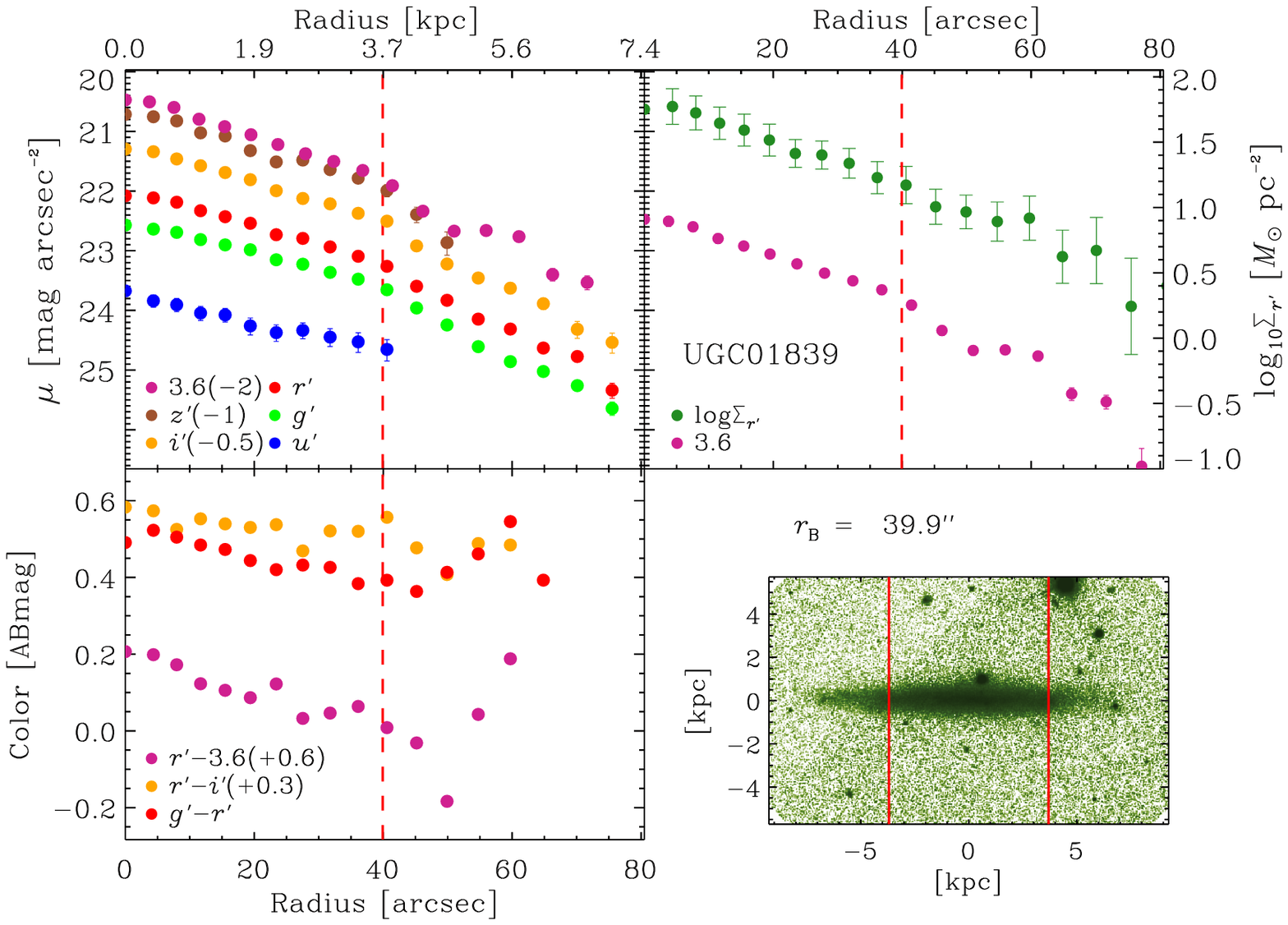}
\end{center}
\end{figure*}

\begin{figure*}
\begin{center}
\includegraphics[width=400pt]{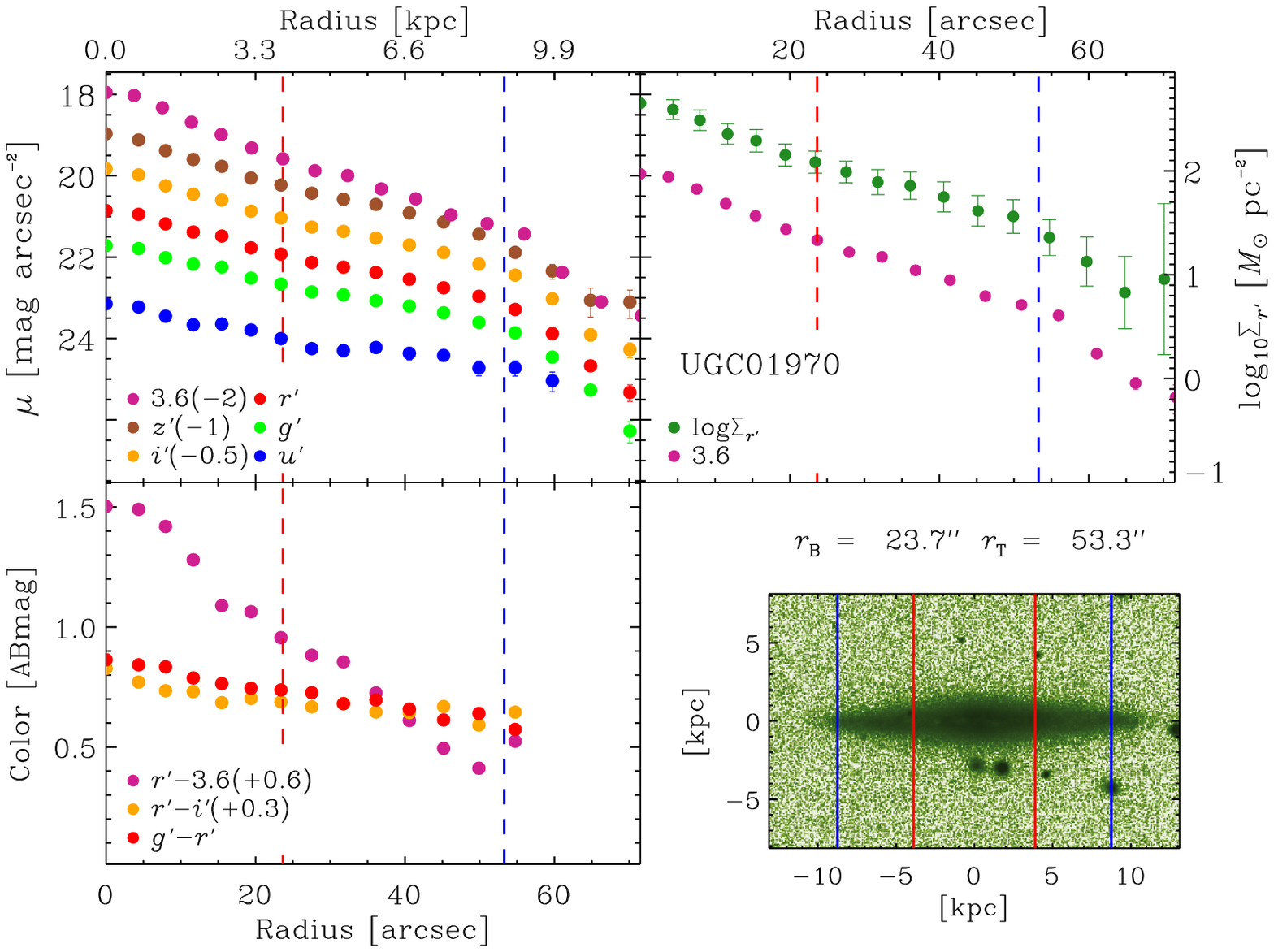}
\end{center}
\end{figure*}

\begin{figure*}
\begin{center}
\includegraphics[width=400pt]{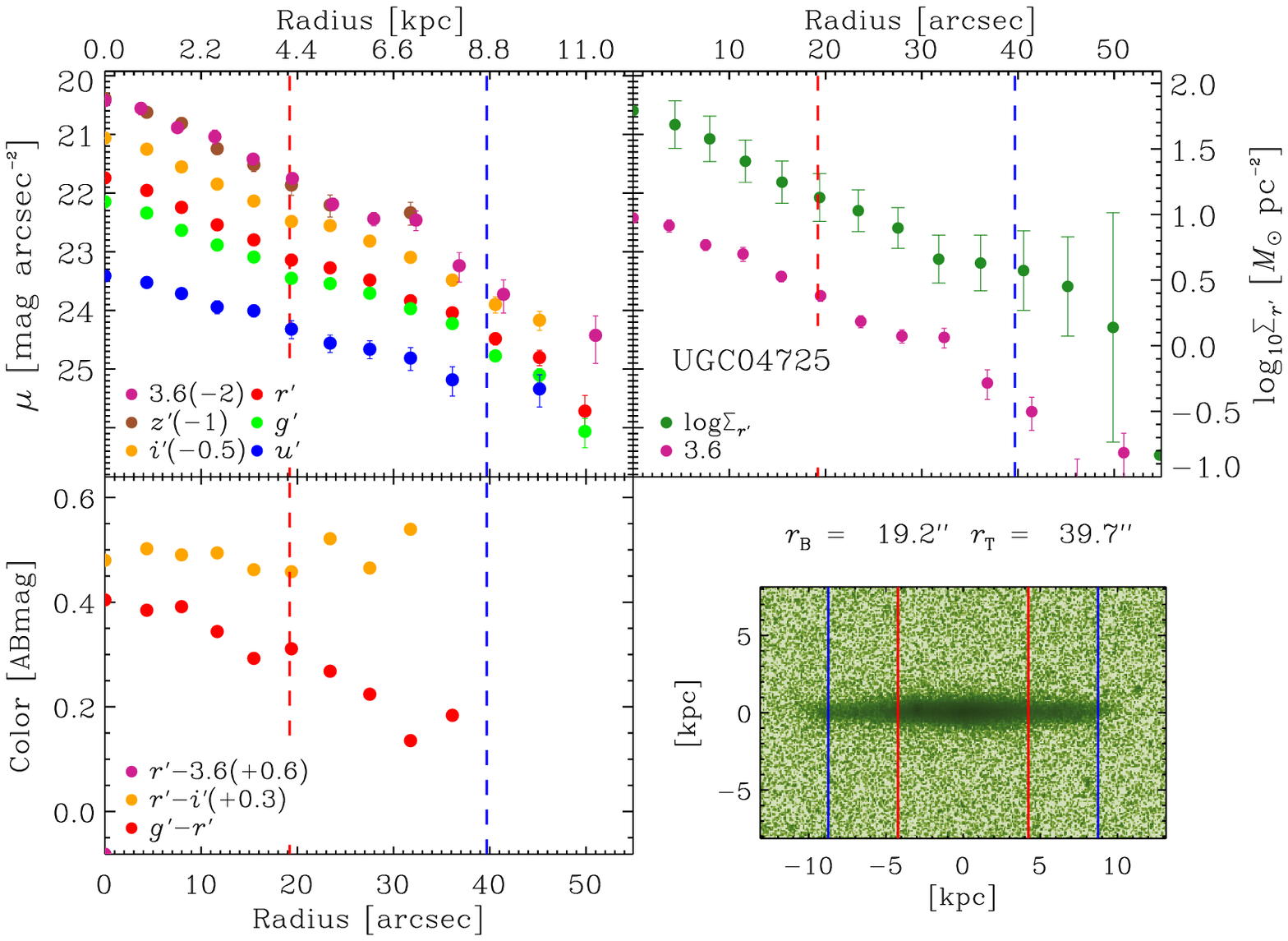}
\end{center}
\end{figure*}

\begin{figure*}
\begin{center}
\includegraphics[width=400pt]{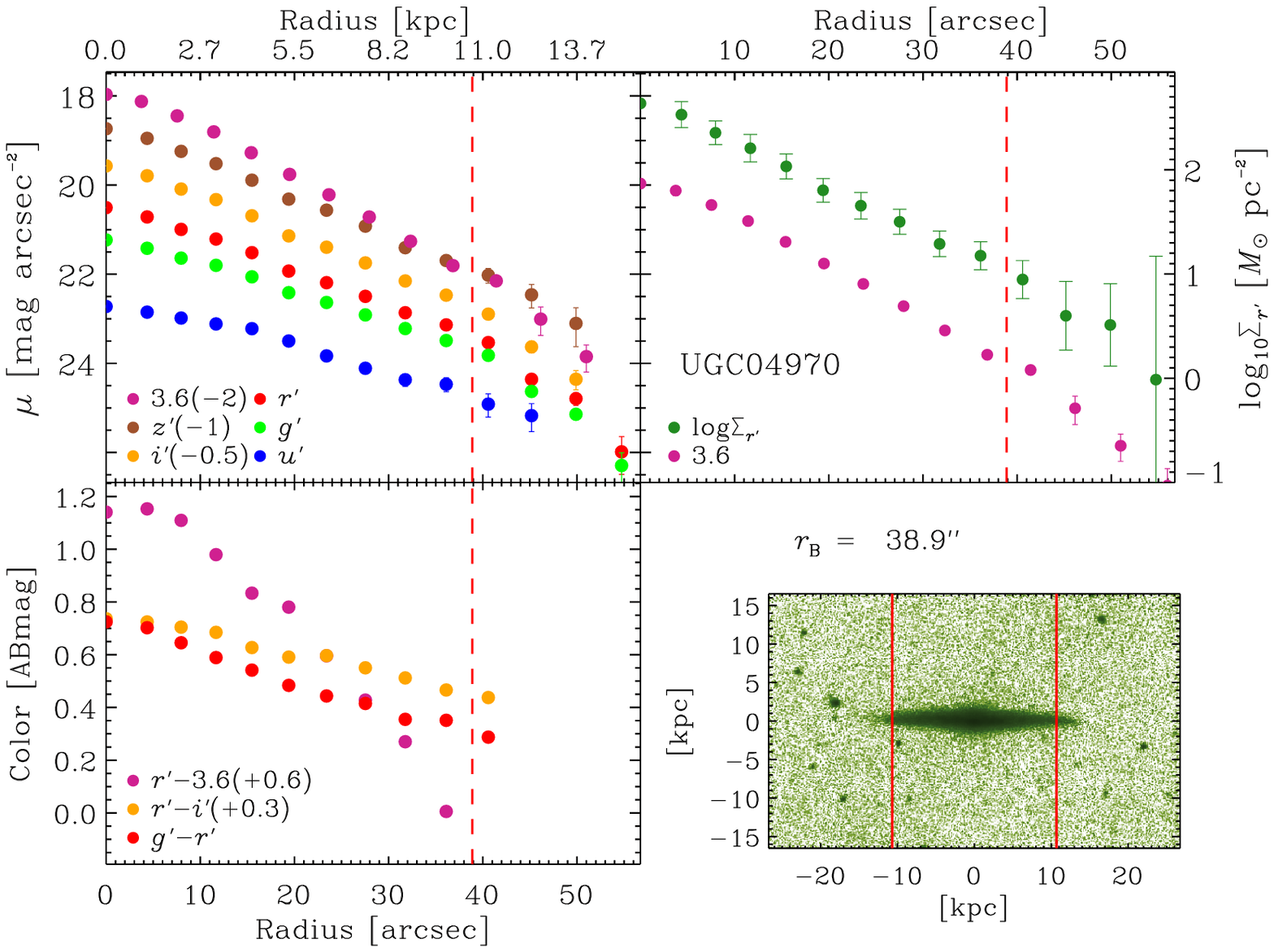}
\end{center}
\end{figure*}

\begin{figure*}
\begin{center}
\includegraphics[width=400pt]{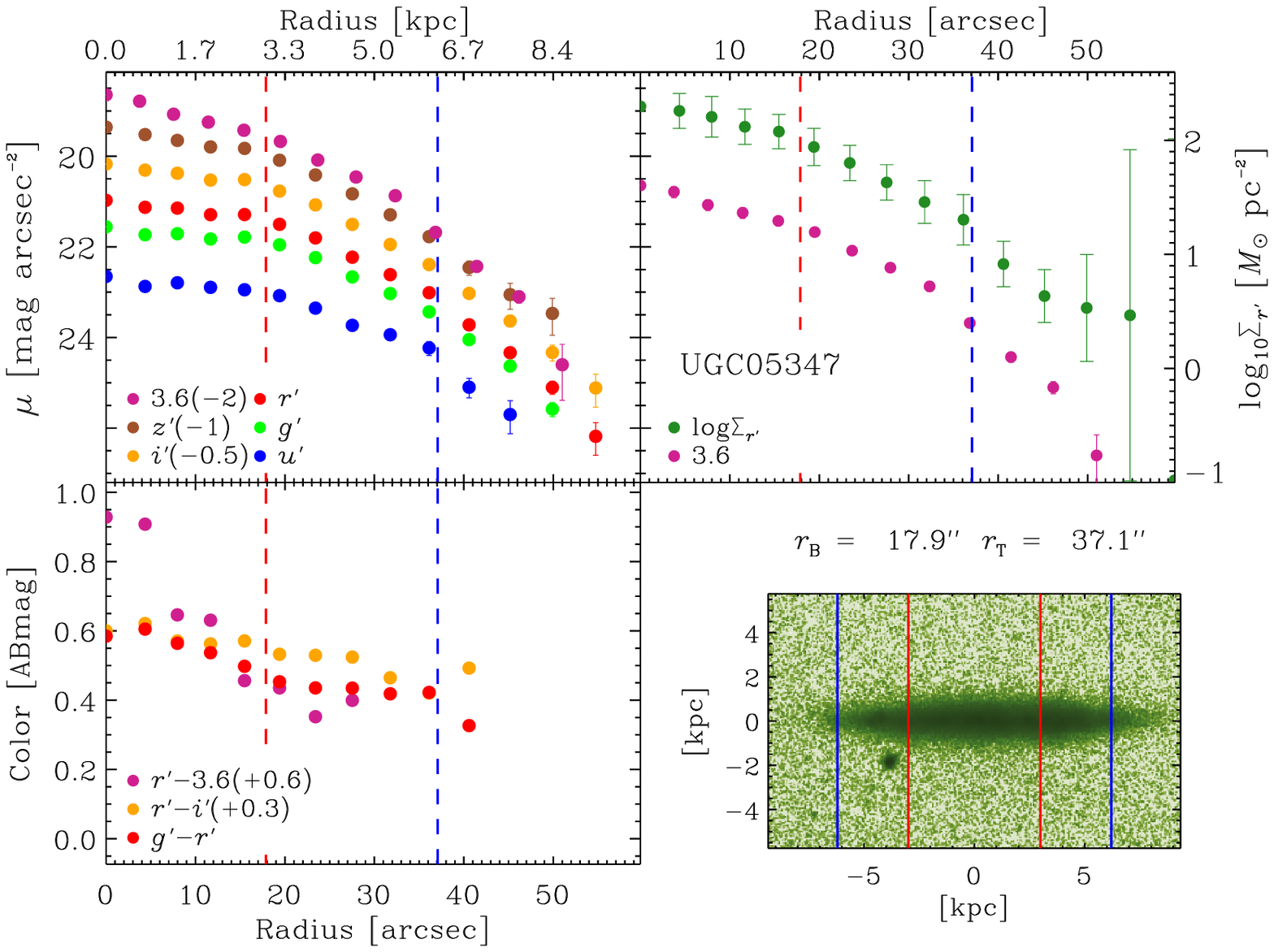}
\end{center}
\end{figure*}

\begin{figure*}
\begin{center}
\includegraphics[width=400pt]{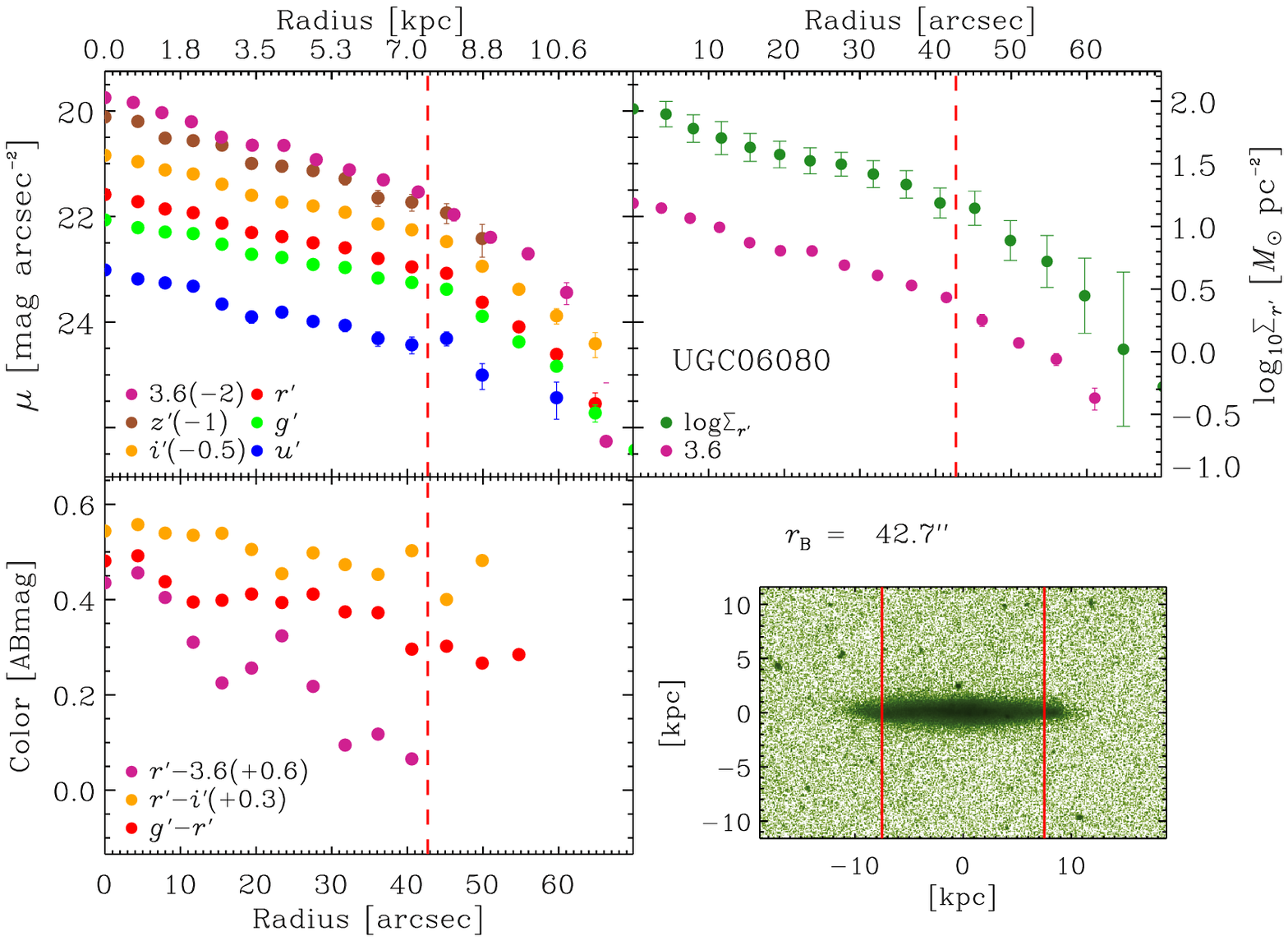}
\end{center}
\end{figure*}

\begin{figure*}
\begin{center}
\includegraphics[width=400pt]{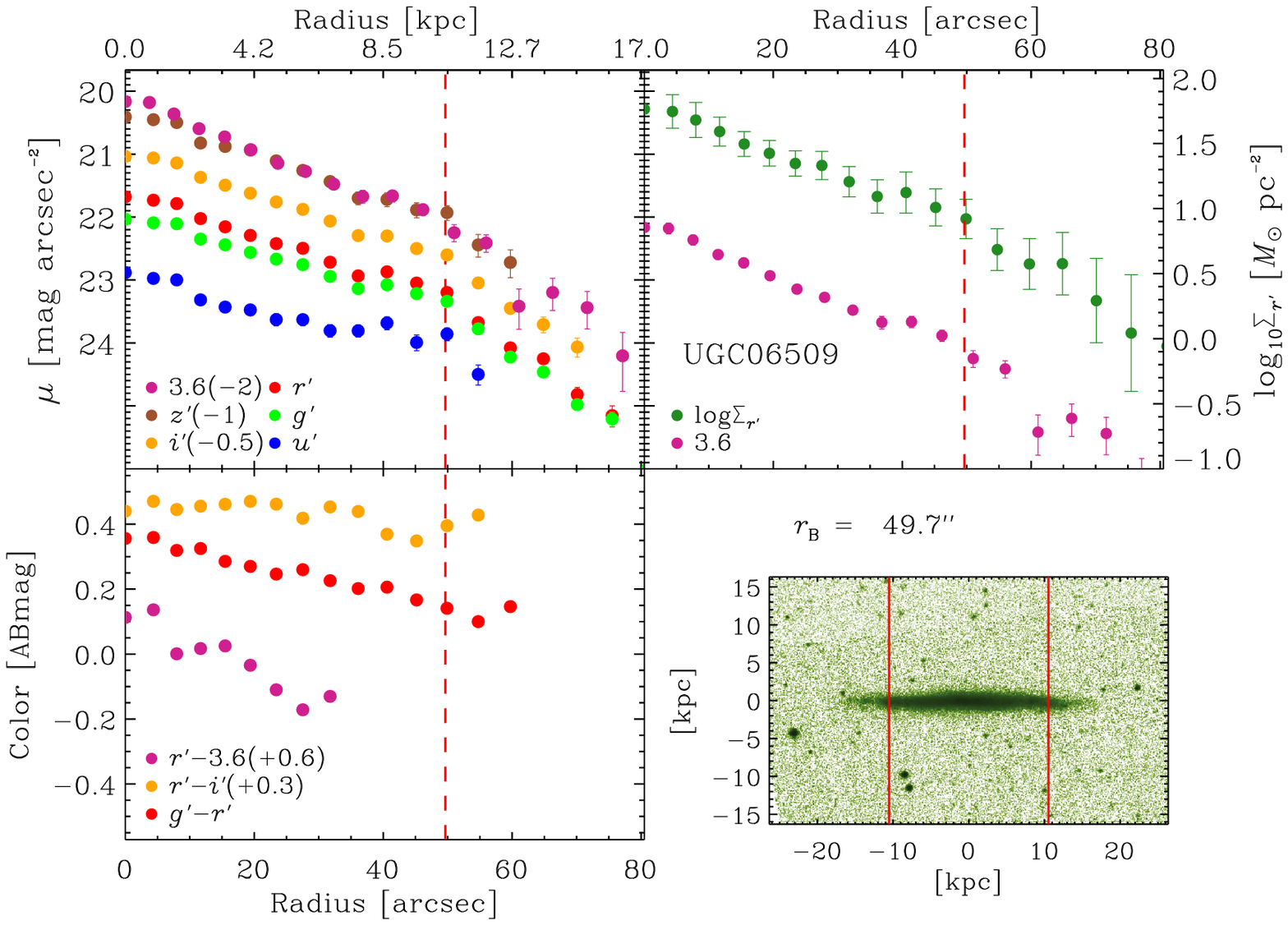}
\end{center}
\end{figure*}

\begin{figure*}
\begin{center}
\includegraphics[width=400pt]{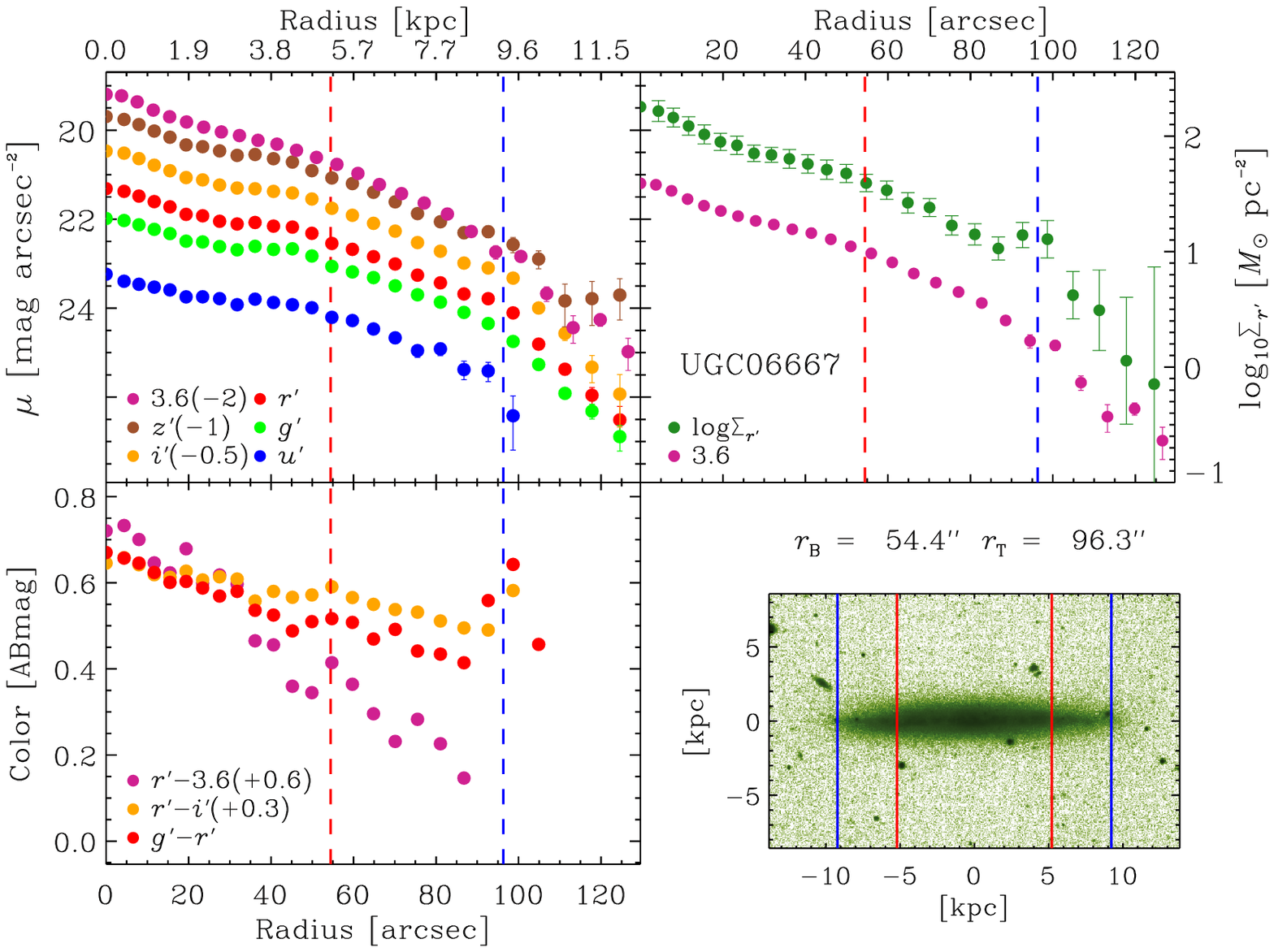}
\end{center}
\end{figure*}

\begin{figure*}
\begin{center}
\includegraphics[width=400pt]{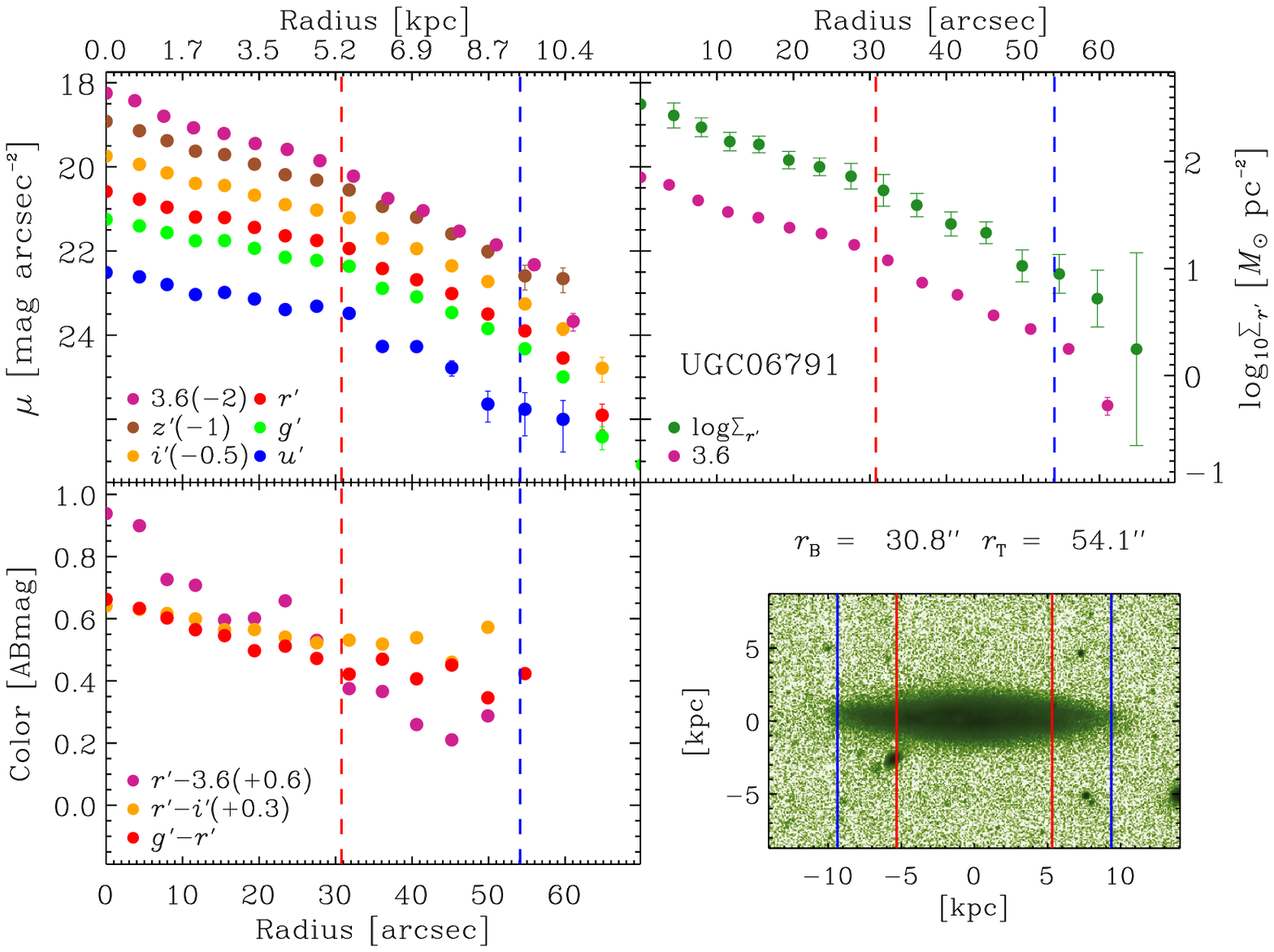}
\end{center}
\end{figure*}

\begin{figure*}
\begin{center}
\includegraphics[width=400pt]{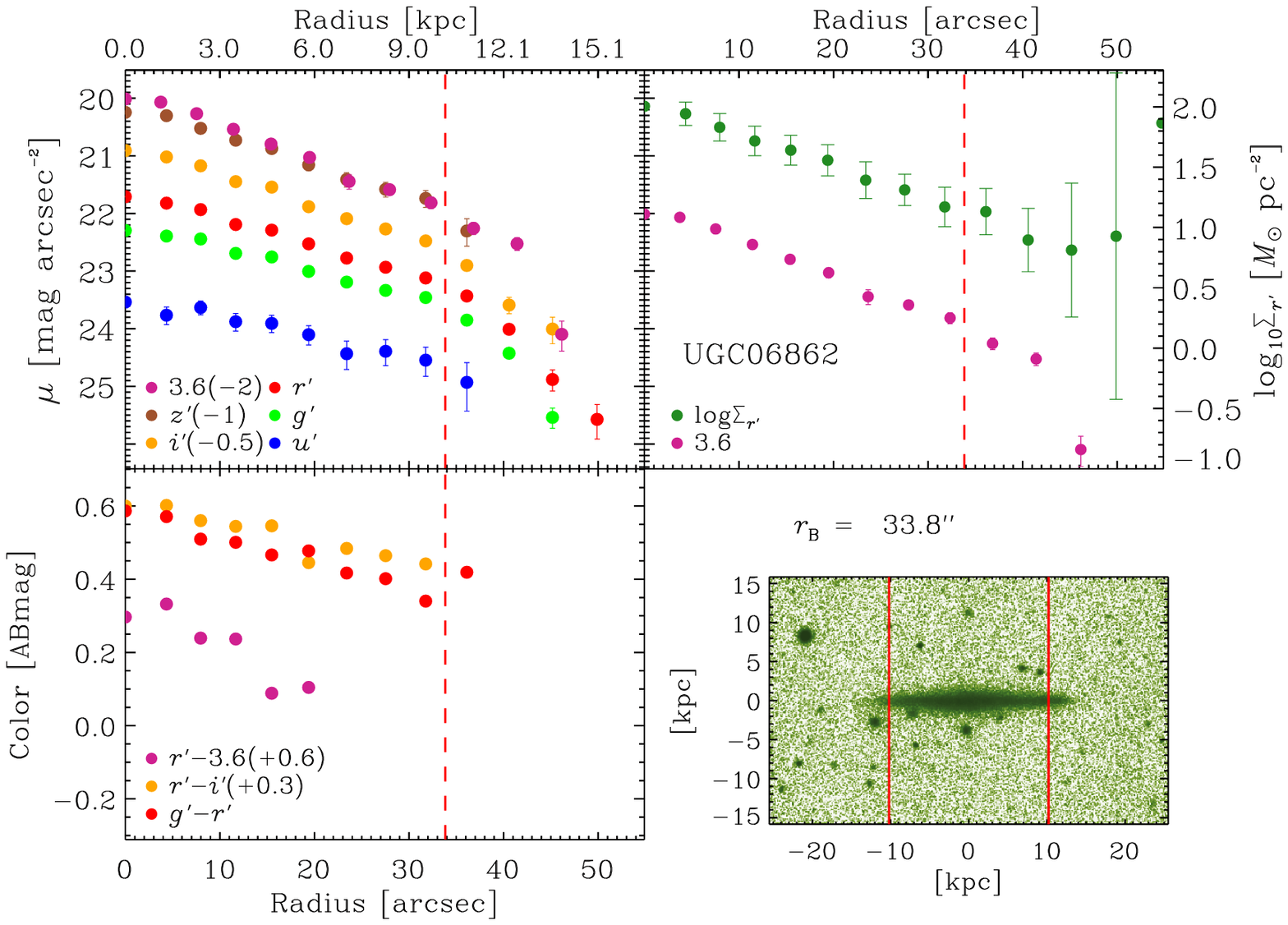}
\end{center}
\end{figure*}

\begin{figure*}
\begin{center}
\includegraphics[width=400pt]{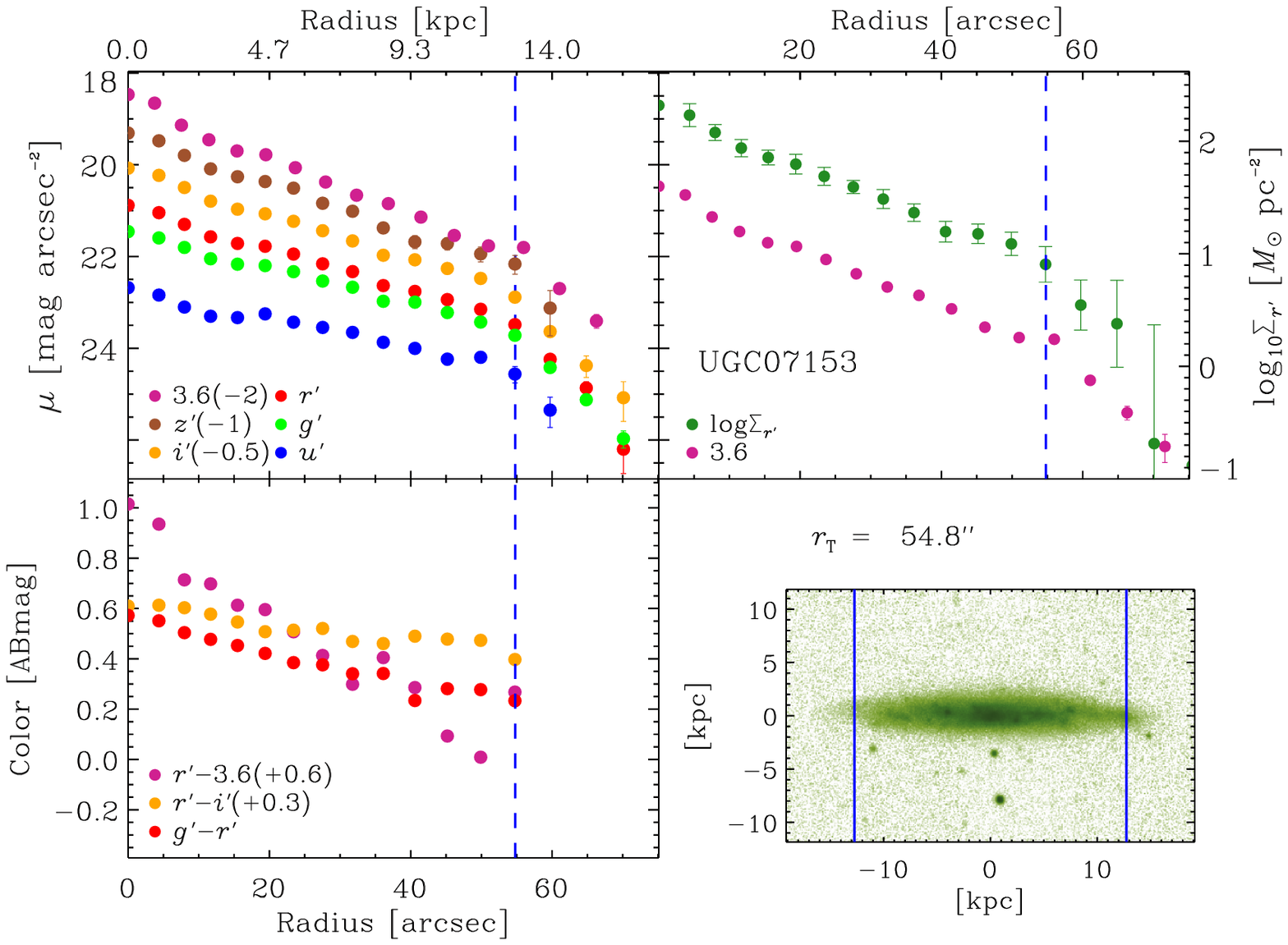}
\end{center}
\end{figure*}

\begin{figure*}
\begin{center}
\includegraphics[width=400pt]{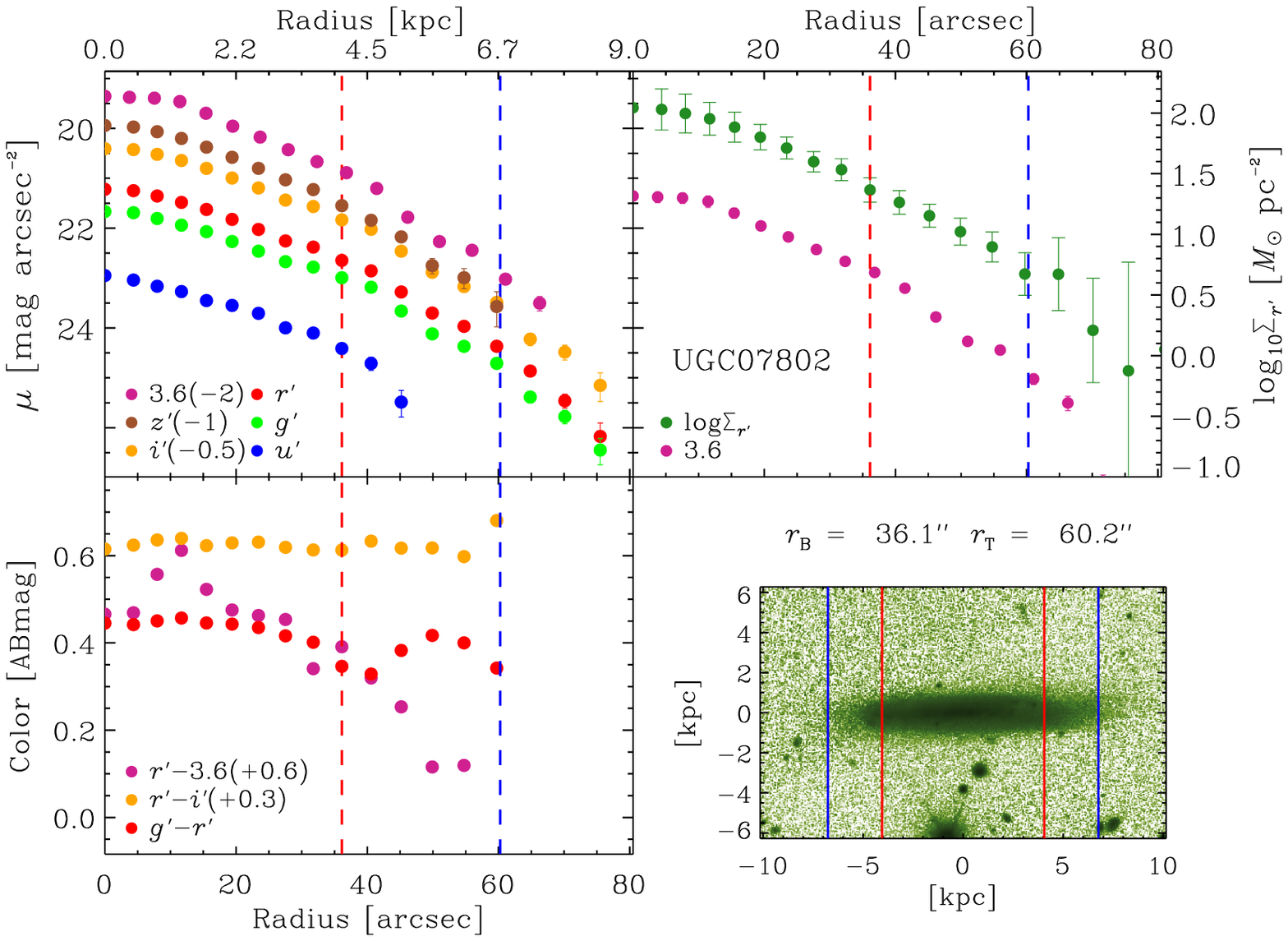}
\end{center}
\end{figure*}

\begin{figure*}
\begin{center}
\includegraphics[width=400pt]{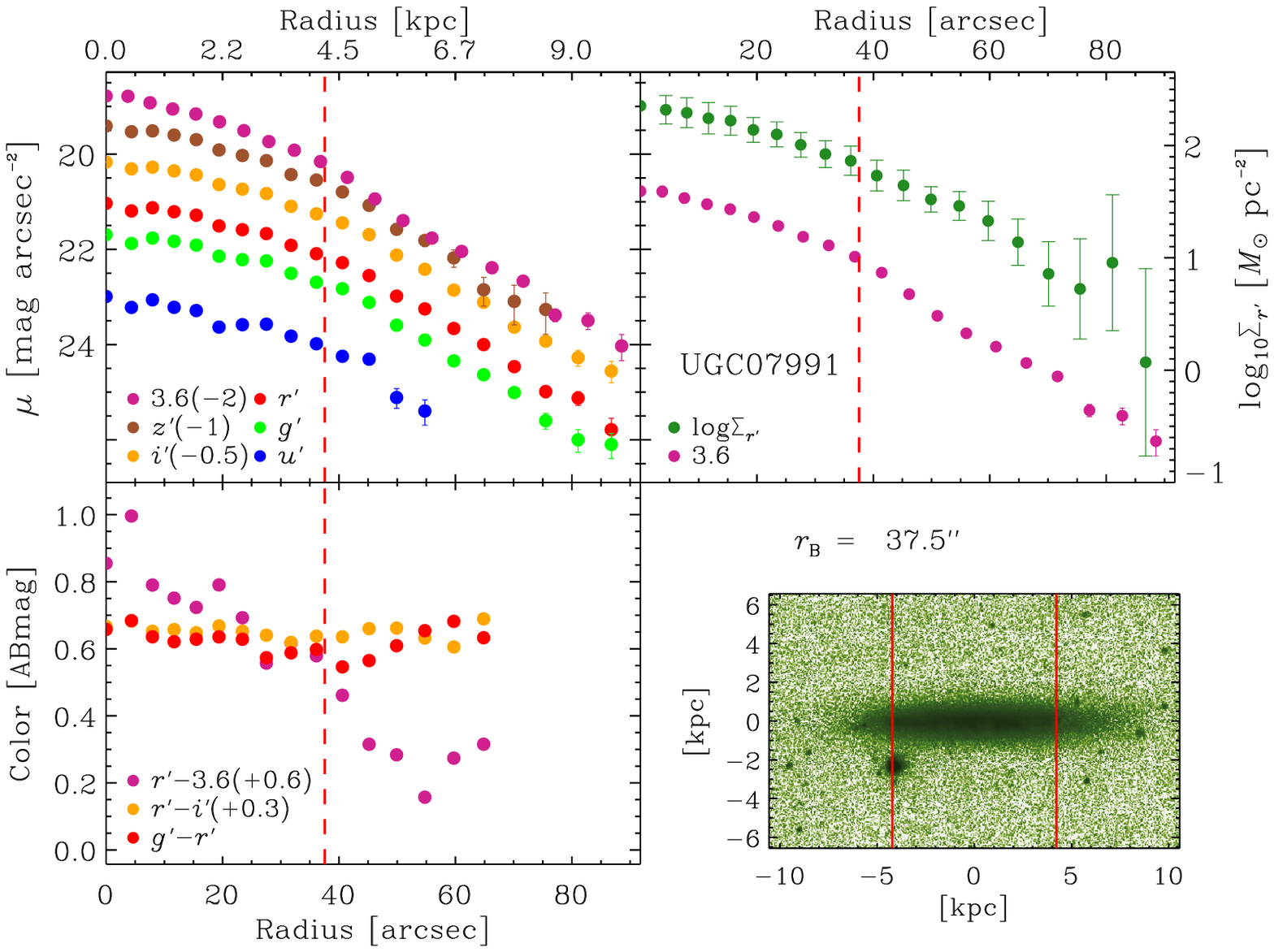}
\end{center}
\end{figure*}

\begin{figure*}
\begin{center}
\includegraphics[width=400pt]{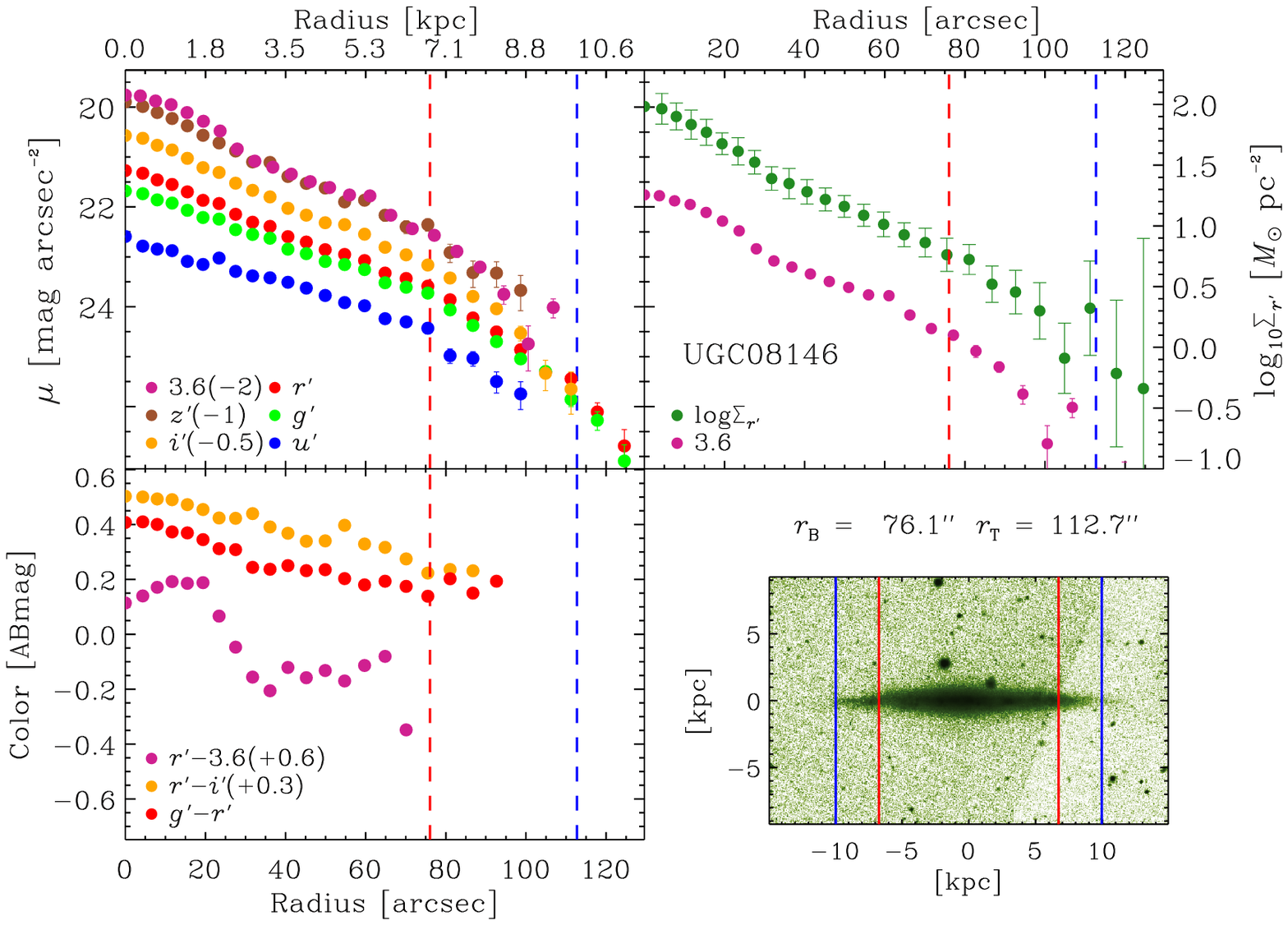}
\end{center}
\end{figure*}

\begin{figure*}
\begin{center}
\includegraphics[width=400pt]{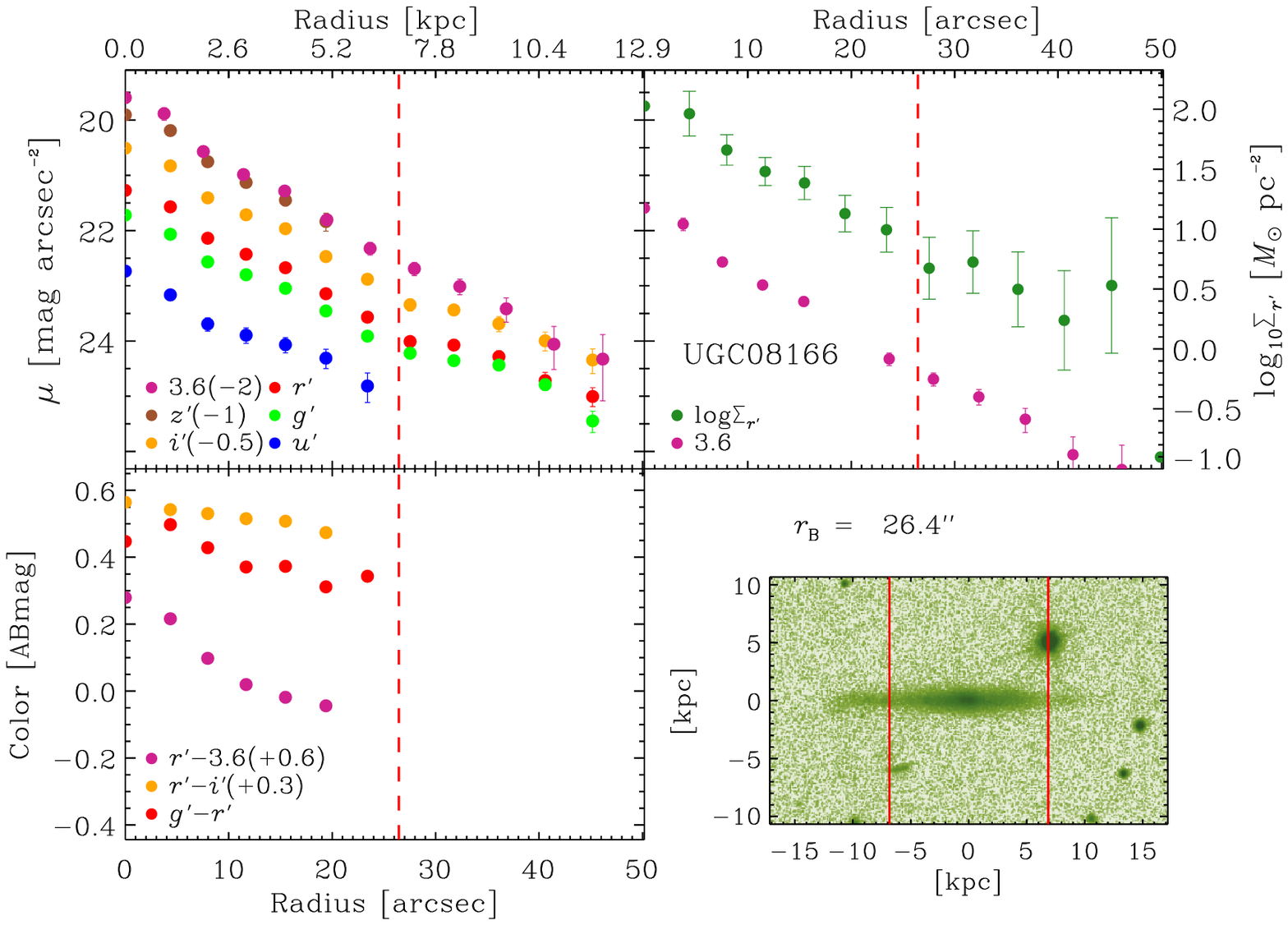}
\end{center}
\end{figure*}

\begin{figure*}
\begin{center}
\includegraphics[width=400pt]{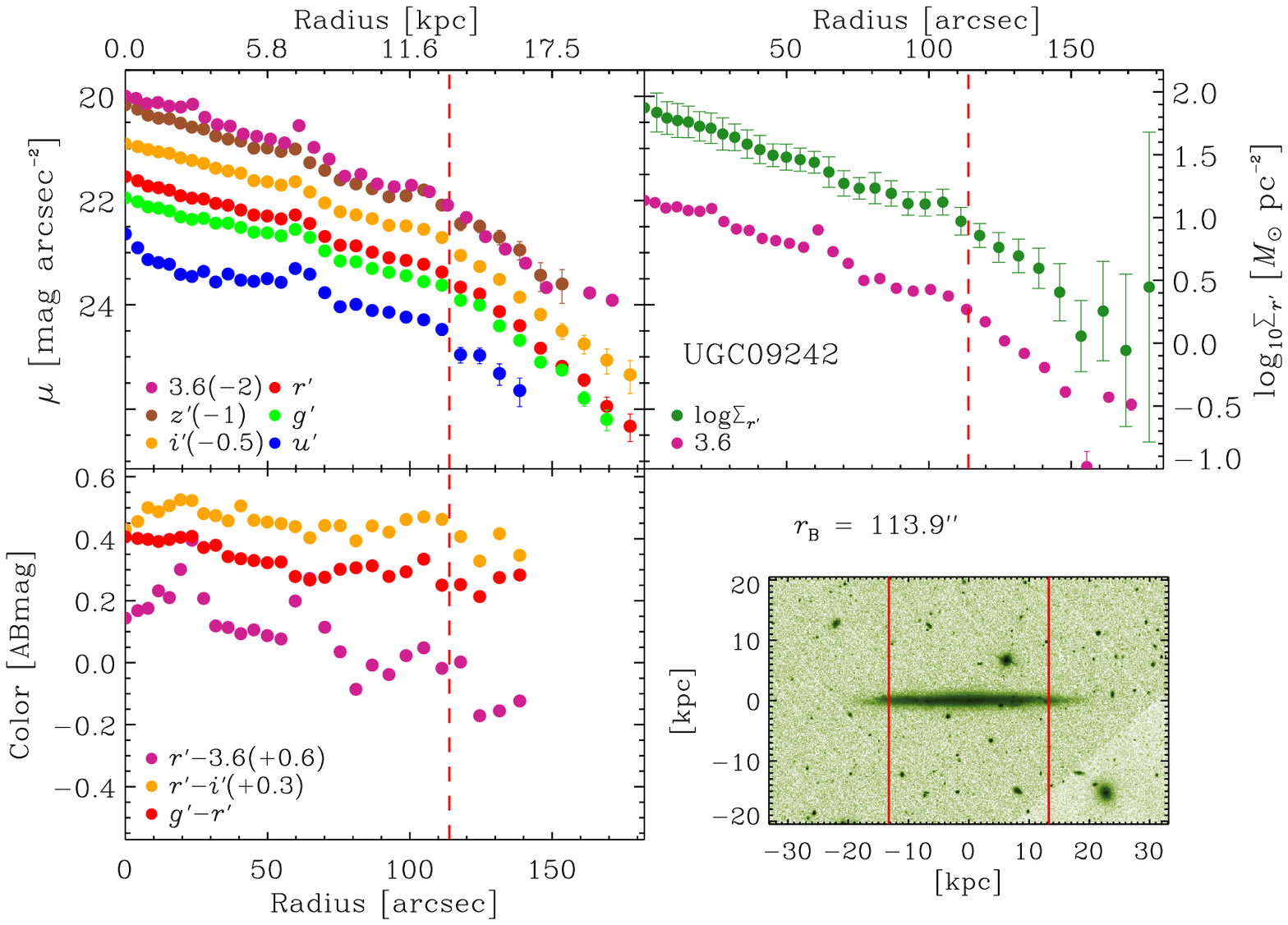}
\end{center}
\end{figure*}

\begin{figure*}
\begin{center}
\includegraphics[width=400pt]{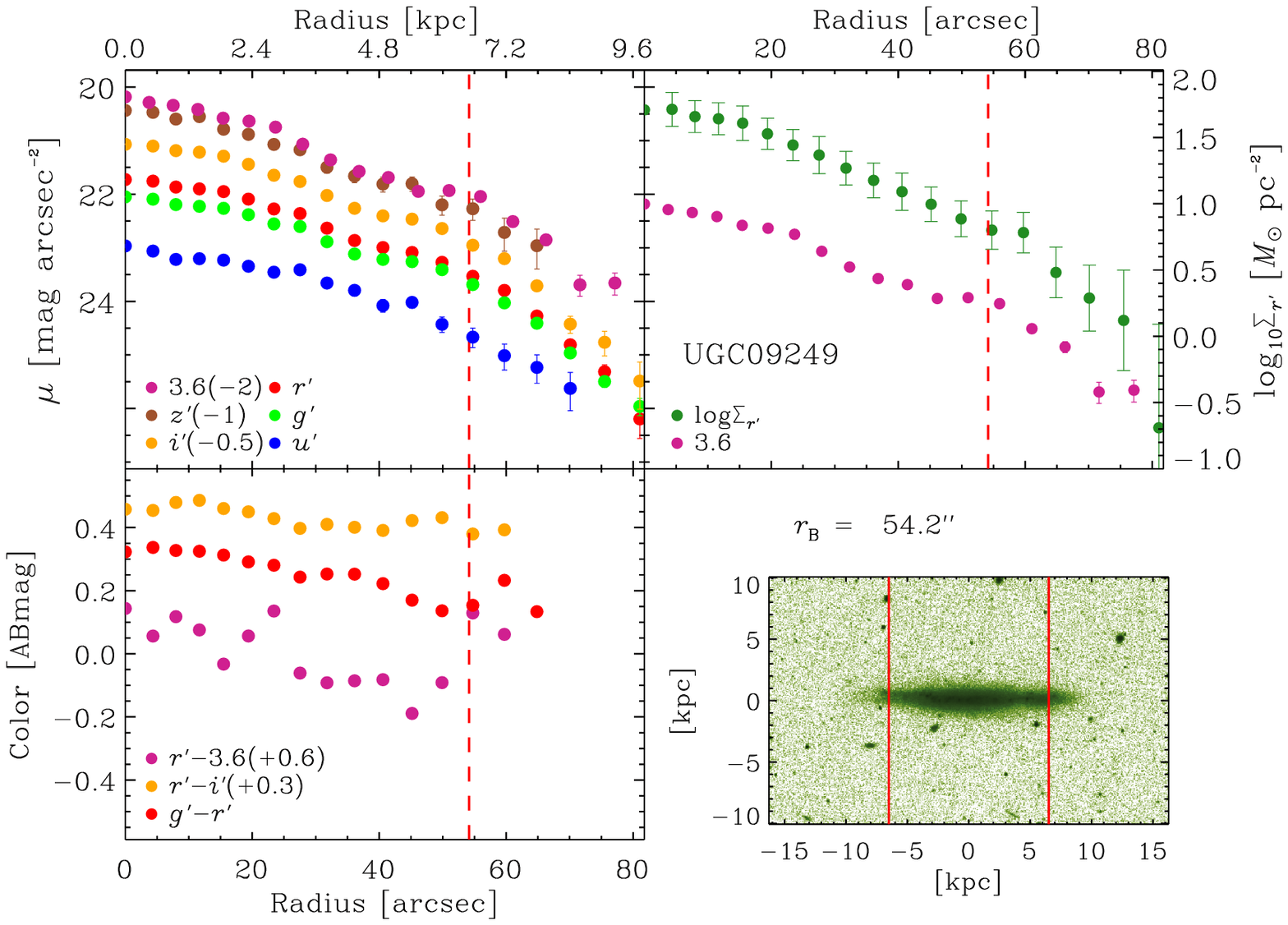}
\end{center}
\end{figure*}

\begin{figure*}
\begin{center}
\includegraphics[width=400pt]{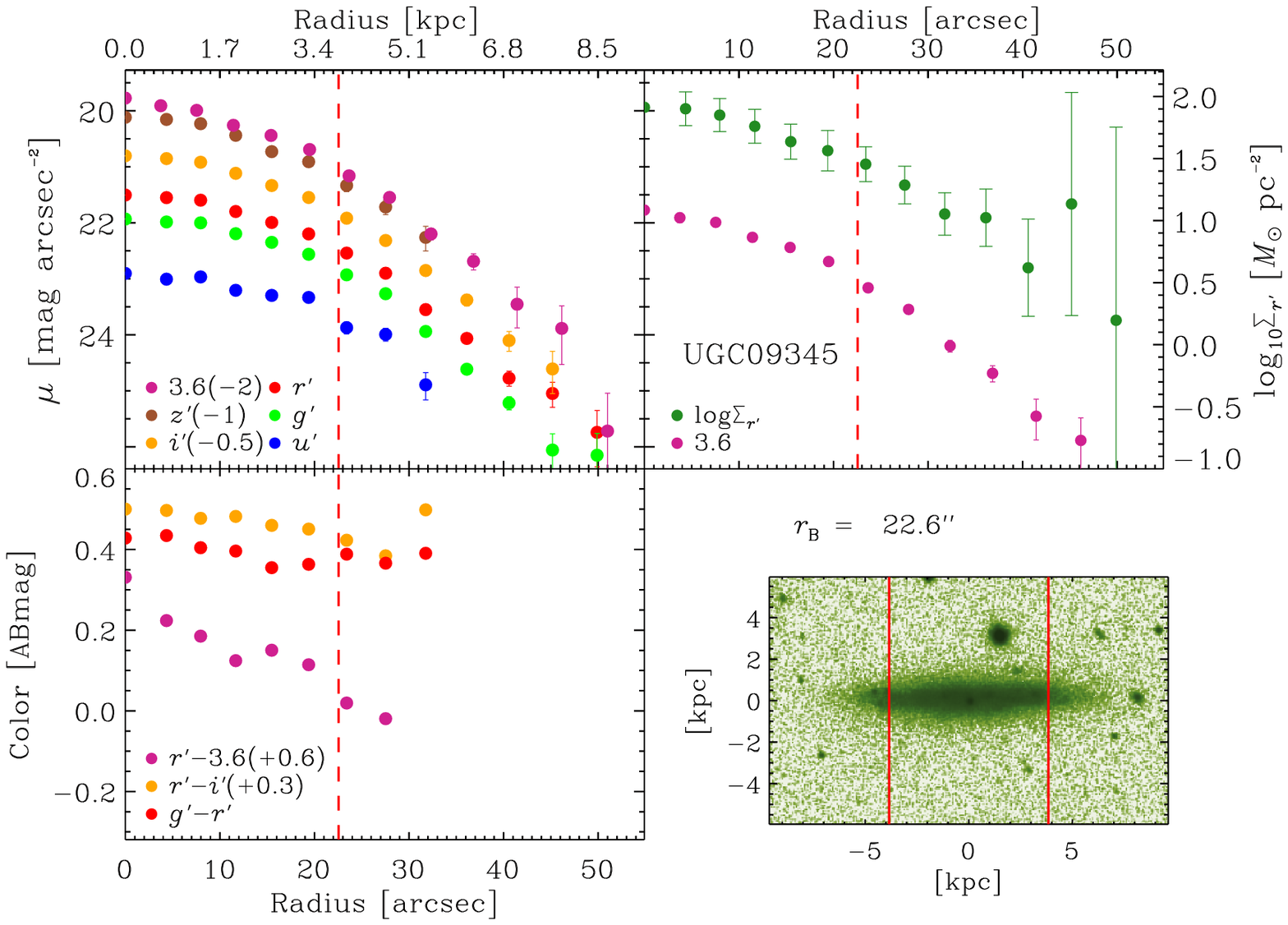}
\end{center}
\end{figure*}

\clearpage

\begin{figure*}
\begin{center}
\includegraphics[width=400pt]{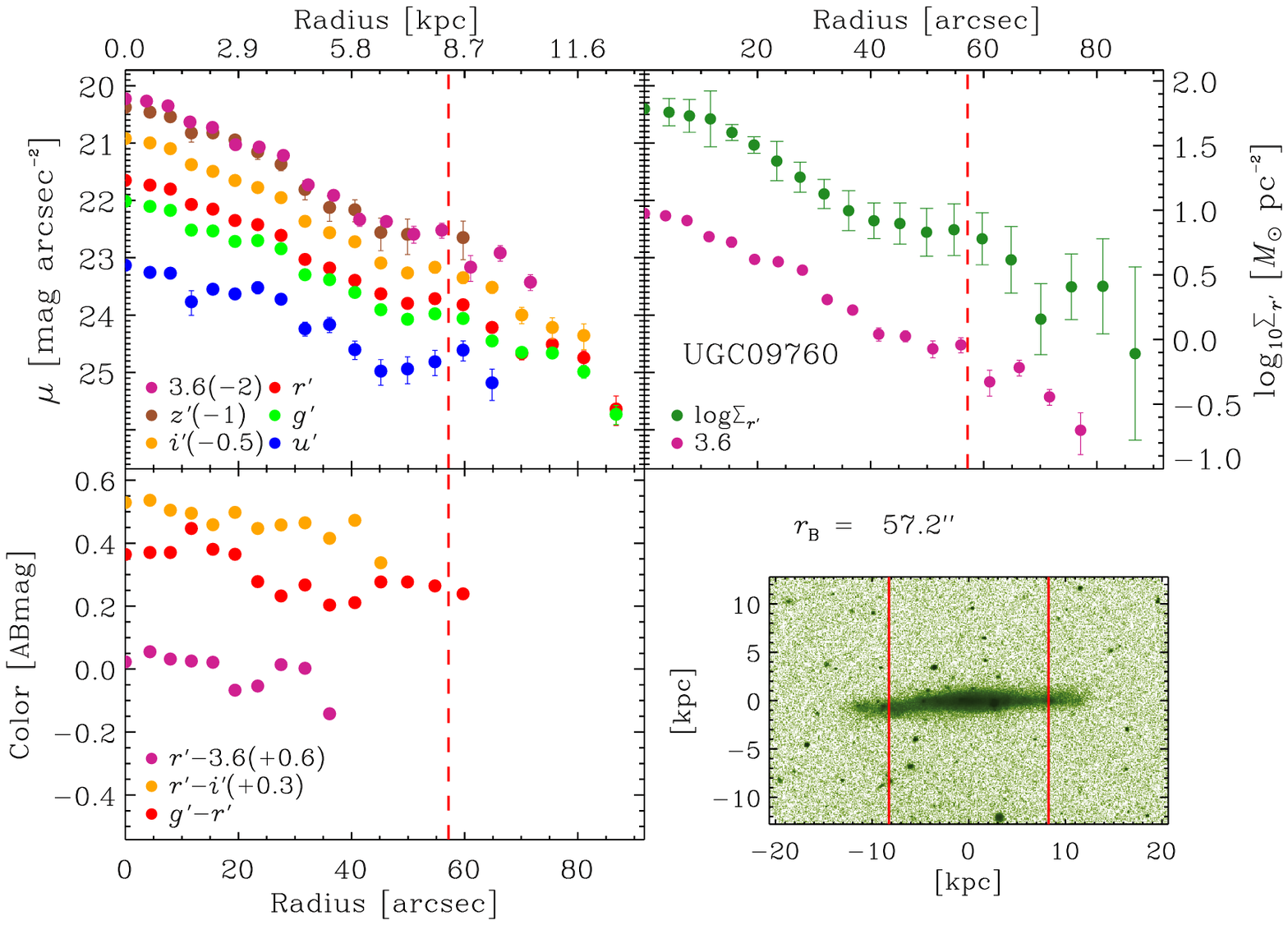}
\end{center}
\end{figure*}

\begin{figure*}
\begin{center}
\includegraphics[width=400pt]{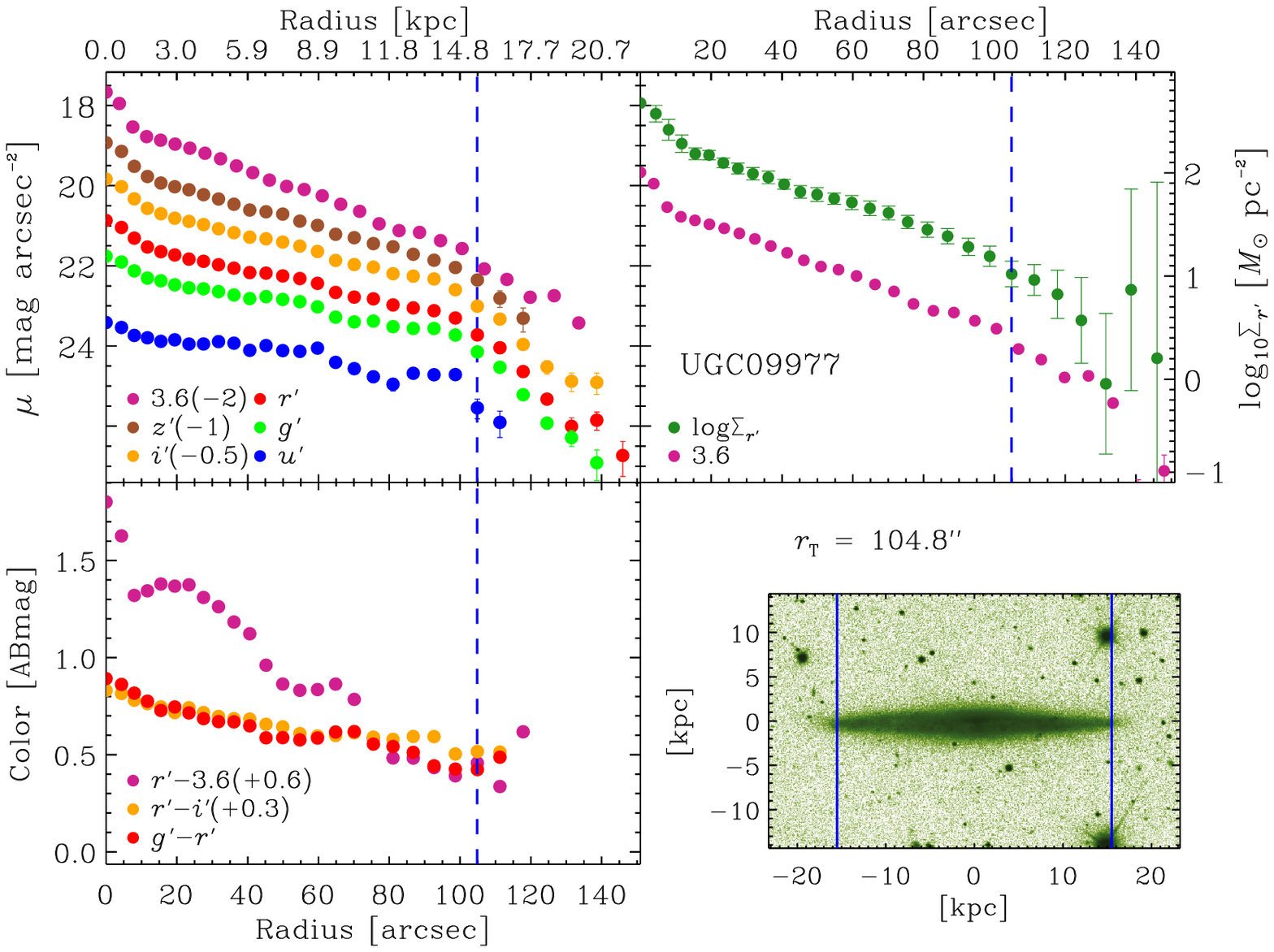}
\end{center}
\end{figure*}

\begin{figure*}
\begin{center}
\includegraphics[width=400pt]{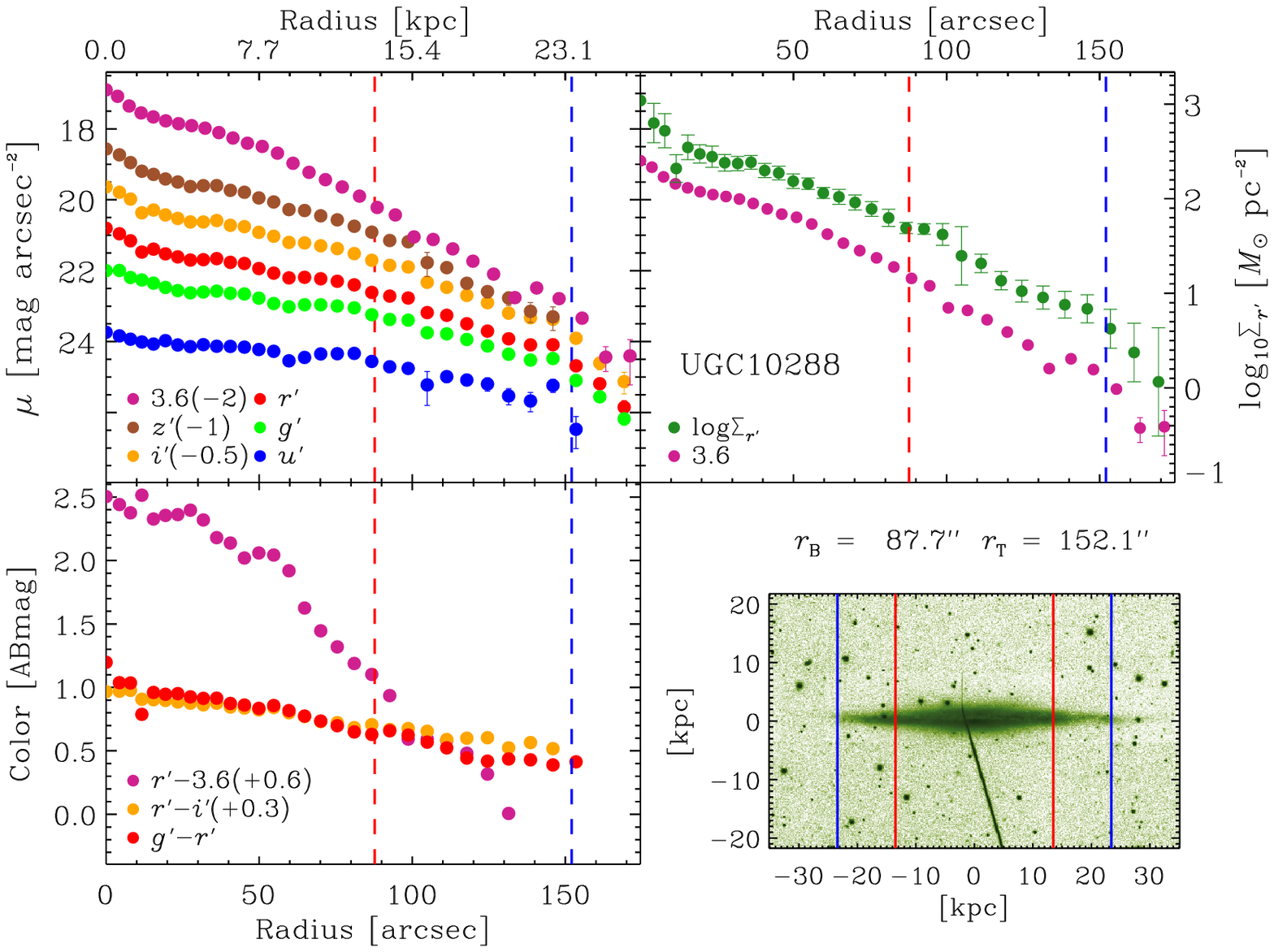}
\end{center}
\end{figure*}

\begin{figure*}
\begin{center}
\includegraphics[width=400pt]{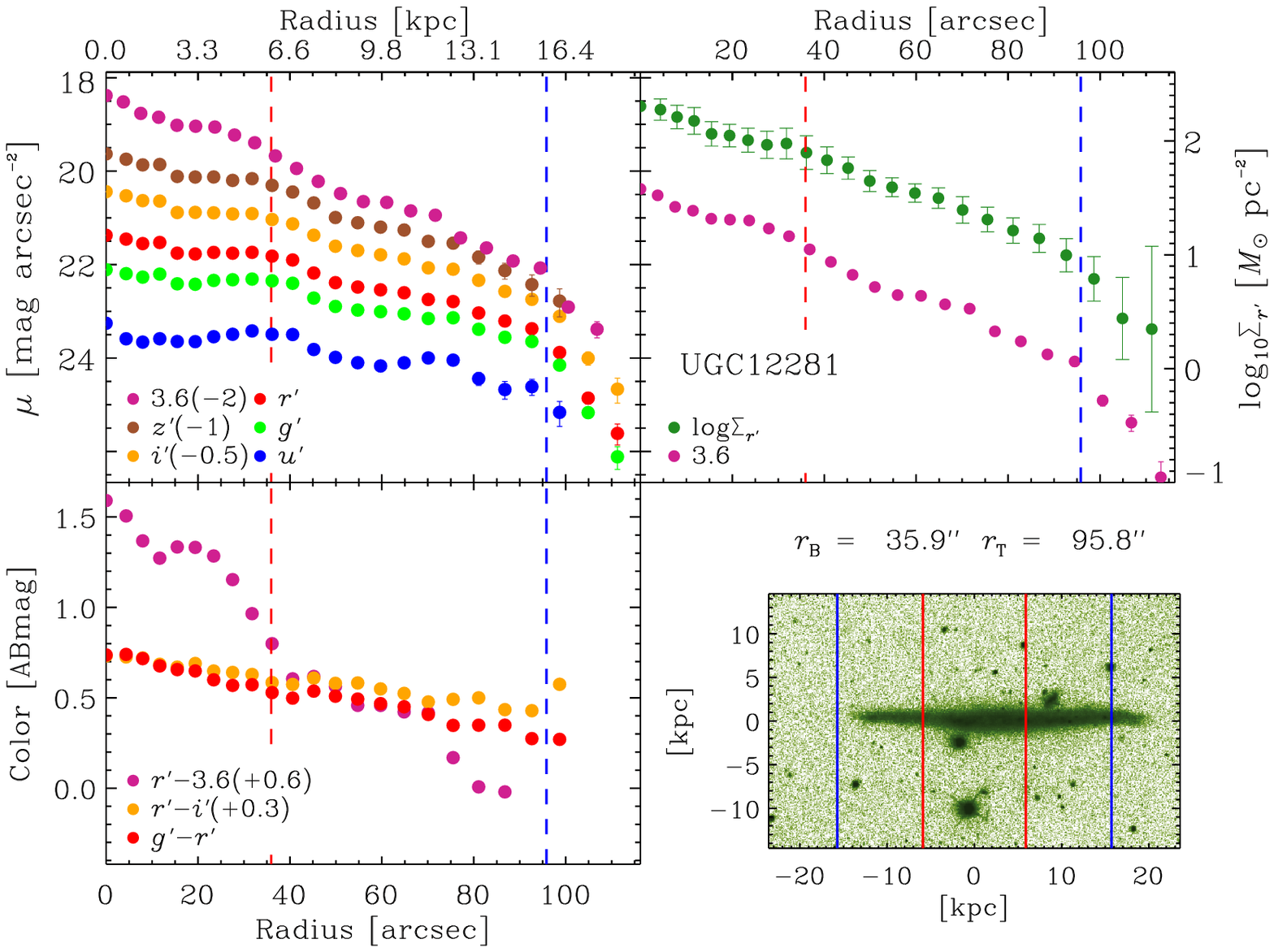} 
\end{center}
\end{figure*}

\clearpage
\label{lastpage}

\end{document}